\pdfoutput=1

\documentclass[10pt, draftclsnofoot, onecolumn]{IEEEtran}

\usepackage{setspace}

\usepackage{hyperref}
\usepackage{amsmath}
\usepackage{amsfonts}
\usepackage{amssymb}
\usepackage{amsthm}
\usepackage{epsfig}
\usepackage{color}
\usepackage{subfigure}
\usepackage{graphicx}
\usepackage{ifthen}
\usepackage{bm}
\usepackage[noend]{algpseudocode}
\usepackage{epstopdf}
\usepackage{algorithm}
\usepackage{cite}
\usepackage{etoolbox}
\usepackage[loose]{units}
\usepackage{xspace}
\usepackage{authblk}
\usepackage{multirow}
\usepackage{tabularx}
\usepackage{fancyhdr}
\usepackage{soul}
\usepackage{balance}
\makeatletter
\patchcmd{\maketitle}{\@copyrightspace}{}{}{}
\makeatother

\RequirePackage[normalem]{ulem}
\RequirePackage{color}\definecolor{RED}{rgb}{1,0,0}\definecolor{BLUE}{rgb}{0,0,1}

\newboolean{Longversion}
\setboolean{Longversion}{false}

\newboolean{TechRep}
\setboolean{TechRep}{false}

\newboolean{TechRepFinal}
\setboolean{TechRepFinal}{true}

\xspaceaddexceptions{[ ]}

\newcommand{\btodo}[1]{{\color{red}BB: #1}}
\newcommand{\gtodo}[1]{{\color{red}GG: #1}}

\newcommand{\JTKMMK}[0]{JTK-MMK\xspace}
\newcommand{\SPJCON}[0]{JTK-PSP\xspace}
\newcommand{\JTKSTA}[0]{JTK-STA\xspace}
\newcommand{\JTKMAT}[0]{JTK-MAT\xspace}
\newcommand{\JTKPSP}[0]{JTK-PSP\xspace}
\newcommand{\JTCBIP}[0]{JTC-BIP\xspace}
\newcommand{\JTCPSP}[0]{JTC-PSP\xspace}
\newcommand{\DF}[2]{[#1, #2]\xspace}

\newcommand{\bflambda}{\pmb{\lambda}}
\newcommand{\bfmu}{\pmb{\mu}}
\newcommand{\bff}{\pmb{f}}
\newcommand{\bfr}{\pmb{r}}

\newcommand{\bfL}{\pmb{L}}
\newcommand{\bfl}{\pmb{l}}

\newcommand{\maybe}[1]{}

\newtheorem{proposition}{Proposition}

\newtheorem{definition}{Definition}
\newtheorem{theorem}{Theorem}
\newtheorem{lemma}{Lemma}

\newcommand{\expect}[1]{{\mathbb E}\{#1\}}

\algrenewcommand\algorithmicthen{}
\algrenewcommand\algorithmicindent{1em}

\newcounter{problems}

    {
    } 

\newenvironment{varalgorithm}[1]
  {\algorithm\renewcommand{\thealgorithm}{#1}}
  {\endalgorithm}

\title{Joint Transmission in Cellular Networks with CoMP -- Stability and Scheduling Algorithms}

\author[1]{Guy Grebla}
\author[1]{Berk Birand}
\author[2]{Peter van de Ven}
\author[1]{Gil Zussman}

\affil[1]{Department of Electrical Engineering\\
Columbia University\\
New York, NY 10027}
\affil[2]{CWI\\
Amsterdam, The Netherlands}

\begin{document}

\maketitle

\begin{abstract}
Due to the current trend towards smaller cells, an increasing number of users of cellular networks reside at the edge between two cells; these users typically receive poor service as a result of the relatively weak signal and strong interference.
  Coordinated Multi-Point (CoMP) with Joint Transmission (JT) is a cellular
  networking technique allowing multiple Base Stations (BSs) to jointly
    transmit to a single user. This improves the users' reception quality
  and facilitates better service to cell-edge users. We consider a CoMP-enabled
  network, comprised of multiple BSs interconnected via a backhaul network. We
  formulate the OFDMA Joint Scheduling (OJS) problem of determining a
  subframe schedule and deciding if and how to use JT in order
  to maximize some utility function. We show that the
  OJS problem is NP-hard. We develop optimal and approximation algorithms for
  specific and general topologies, respectively. We consider a time dimension
  and study a queueing model with packet arrivals in which the service
  rates for each subframe are obtained by solving the OJS problem.  We prove that when the problem is
  formulated with a specific utility function and solved optimally in each
  subframe, the resulting scheduling policy is throughput-optimal.
 Via extensive simulations we show that the bulk of the gains from CoMP with
 JT can be achieved with low capacity backhaul.
Moreover,  our algorithms distribute the network resources evenly,
increasing the inter-cell users' throughput at only a slight cost to the intra-cell users.
This is the first step towards a rigorous, network-level
understanding of the impact of cross-layer scheduling algorithms on CoMP networks.
\end{abstract}

\vspace{1mm} \noindent {\bf Keywords:} Coordinated
Multi-Point (CoMP), 
Joint
Transmission, Cellular Networks, Approximation Algorithms, Scheduling, Queueing Networks.

\section{Introduction}

Cellular networks face an ever-increasing bandwidth demand, driven by the
advent of sophisticated mobile devices and new applications. Satisfying this
demand calls for improvements in the spectral utilization and reductions in
inter-cell interference. The latter is becoming more relevant as the number of
inter-cell users increases with ever-decreasing cell sizes. Such users are often unable to
receive any transmission due to the high interference.
Interference reduction can be efficiently
accomplished through multi-cell coordination, known as
Coordinated Multi-Point (CoMP) or Network-MIMO.  One particularly promising
category is CoMP with Joint Transmission (JT),
where multiple Base Stations (BSs) jointly transmit to a
single user, using the same time-frequency slots.
This technique is incorporated in the LTE-Advanced standard~\cite{lte-tr36819}.
Recently, CoMP with JT was shown to obtain substantial throughput gains in
both indoor and outdoor testbeds~\cite{irmer2011coordinated}.\footnote{There
  are two flavors of CoMP with JT: coherent~\cite{irmer2011coordinated} and
  non-coherent. We consider coherent JT but remark that all results can be
  directly extended to non-coherent JT.}






As a result of the implementation of CoMP with JT in the LTE-Advanced standard,
algorithm design and performance evaluation for these systems have recently
received increased attention in the research literature (see, e.g.,
\cite{KHHJS13,fu2014transmission,zhuang2014backhaul}). However, most existing
work is concerned with developing heuristics designed for saturation
conditions. In contrast, we consider a cellular network where new packets are generated over time, and construct a rigorous framework to develop
scheduling algorithms
for CoMP with JT that
maximize throughput. This is achieved via a cross-layer
approach, consisting of PHY (considering SINR-based transmission probabilities),
MAC (deciding on a transmission schedule), and network layer (forwarding
traffic over the backhaul).

\ifTechRep
\begin{figure}
\centering
\includegraphics[
width=0.2\columnwidth,bb=0 0 500 500]{fig/basic_example_ver3.pdf}
\caption{Example of a network supporting Joint Transmission (JT) with 3
 Base Stations (BSs).  White
 packets are available at a single BS and can be transmitted
by that BS or duplicated over the backhaul. Black packets
 are present at two BSs and can be transmitted
 jointly.\label{fig:basic_example}}
\vspace{-14pt}
\end{figure}
\fi

We consider a cellular network comprised of multiple BSs
interconnected via backhaul links.
Users are assigned a serving and a secondary BS, and packets destined for a
user can be transmitted either by the serving BS only, or jointly by the
serving and secondary BSs. The latter provides better
signal-to-interference-plus-noise ratio (SINR), but
requires a subframe in both BSs, as well as forwarding the packet from the serving to the secondary BS prior to
the transmission.
A scheduling algorithm for CoMP with JT therefore needs to balance the performance
benefits of transmitting packets using JT with the additional resources
required for doing so.

We first focus on a single subframe, and study the OFDMA Joint Scheduling
(OJS) problem of determining a schedule to maximize some utility
function, given a set of packets for each user. Such a schedule determines which packets should be forwarded over
the backhaul and which packets should be transmitted wirelessly, either using JT or by the serving BS only. 
We show that the OJS problem is
NP-hard and describe a framework for solving it efficiently by decomposing it
into problems related to knapsack and coloring.
This allows us to develop an efficient algorithm for solving the  OJS problem in bipartite backhaul
network graphs. While backhaul network graphs are not necessarily bipartite,
this result enables us to
develop approximation
algorithms for general backhaul
graphs.



We then consider the network evolution over multiple subframes. We define a
queueing model where the users are fixed, and packets for the various users are generated over time. Departures are determined by
the schedule obtained from solving the OJS problem in each
subframe.
We characterize the capacity region (i.e., the packet
arrival rates that can be sustained).
Moreover, we demonstrate that when the OJS problem is formulated with a specific
queue-length based utility function and solved optimally in each subframe, we
obtain a MaxWeight-like scheduling policy (e.g., \cite{TE92}), which we show
to be throughput optimal. This is surprising, given the difference between OJS
and the typical matching-type problems where MaxWeight scheduling performs well.
Based on the queueing model, we present extensive
simulation results to evaluate the performance of the proposed scheduling
algorithms, as well as the benefits of JT. In particular, we consider
different network topologies with an SINR-based channel model. We show that
the bulk of the gains from CoMP with JT can be achieved with low capacity
backhaul links. This result is highly relevant as the
deployments of advanced cellular wireless technologies have a strong impact on
mobile backhaul operational expenditure (OPEX), which amount to 20-40\% of
total mobile operator's OPEX due to their reliance on T1/E1 backhaul copper lines\cite{tipmongkolsilp2011evolution}. A promising alternative is wireless backhaul (e.g., satellite, microwave), which is becoming a viable technology for geographically challenging regions and 5G networks. However, since such technology has limited capacity (due to, e.g., limited wireless spectrum and poor wireless channel conditions), our result are relevant to both present and future networks.
We show that our algorithms distribute the network
resources more evenly as the backhaul capacity increases. In fact, they
increase the inter-cell users' throughput at only a slight cost to intra-cell
users.

The main contributions of this paper are two-fold:
(i) we define a rigorous model for CoMP with JT
and
develop novel scheduling algorithms with throughput guarantees for
networks with queueing dynamics; (ii) via extensive simulations, we observe the benefits of
JT and the tradeoffs related to its implementation.

The rest of the paper is organized as follows. In
Section~\ref{sec:related_work} we discuss related work. In
Section~\ref{sec:model} we present the model. In
Section~\ref{sec:hardness_results}, we introduce the OJS problem and show that it
is NP-hard. We develop approximation
algorithms for OJS in Section~\ref{sec:algorithms}.
In Section~\ref{sec:queueing_intro}, we develop and present results for a queueing model, which we
study through extensive simulation experiments in
Section~\ref{sec:simulation}. Section~\ref{sec:conclusions} provides
conclusions and directions for future research.
\ifTechRep
\else
All proofs can be found in Appendix A.
\fi

\section{Related Work}
\label{sec:related_work}
Previous work on scheduling for CoMP with JT has focused exclusively on
analyzing the performance of heuristics, and has been limited mostly to
networks that are saturated (i.e., have infinite backlog). The proposed heuristics are then evaluated via simulations or in testbeds (see, e.g.,\cite{irmer2011coordinated} and references therein). For example,~\cite{
   CYXTL11, zhang2012joint,
  JSCTXX11,
JQPBY13,
KHHJS13} present heuristics for throughput maximization, assuming a perfect
backhaul (infinite capacity and no delay), while~\cite{zhuang2014backhaul}
does the same for the finite backhaul case. In~\cite{fu2014transmission}, the
authors devise a heuristic for networks with perfect backhaul, and aim to
minimize the backhaul traffic under certain SINR constraints. To our
knowledge,~\cite{zhang2012joint,SZJM10} are the only studies that consider
unsaturated networks, where traffic is generated over time rather than assuming an infinite supply of available packets. Both of these propose heuristics, assuming a perfect
backhaul. The main contrast between our work and~\cite{SZJM10,JSCTXX11,
JQPBY13,
KHHJS13,fu2014transmission,zhuang2014backhaul,zhang2012joint} is that we derive the first scheduling policies with performance guarantees. This is done for unsaturated networks, assuming finite backhaul capacity and positive delay over the backhaul.




Models similar to the one considered in this paper have been investigated in
the context of single-cell cellular transmissions.
Packet-level scheduling algorithms for cellular networks are developed
in~\cite{
CG13-1,
CG13-2}. In \cite{andrews2007scheduling}, approximation
algorithms that provide queue stability are analyzed for a single BS.


Closely-related technologies to CoMP with JT are
network-MIMO, multi-cell MIMO, and multi-user MIMO (MU-MIMO) \cite{balanRMPC12,ZhangSKRS13,GHHSSY10,di2014spatial}. While theoretical studies (e.g.,
\cite{GHHSSY10,
cadambe2008interference,
marsch2008multicell}) show that under certain
conditions such technologies can completely cancel inter-cell interference,
achieving these gains in \emph{practical scenarios} is still
challenging~\cite{di2014spatial,ZhangSKRS13,balanRMPC12} (e.g., due to the high signal processing
complexity). The Study of scheduling schemes for these technologies is subject to further research.

\section{Network Model}
\label{sec:model}


\label{sec:system_model}

\begin{figure}
  \centering
  \subfigure[wireless network]{\label{fig:network1}\includegraphics[width=0.45\columnwidth,bb=0 0 400 400]{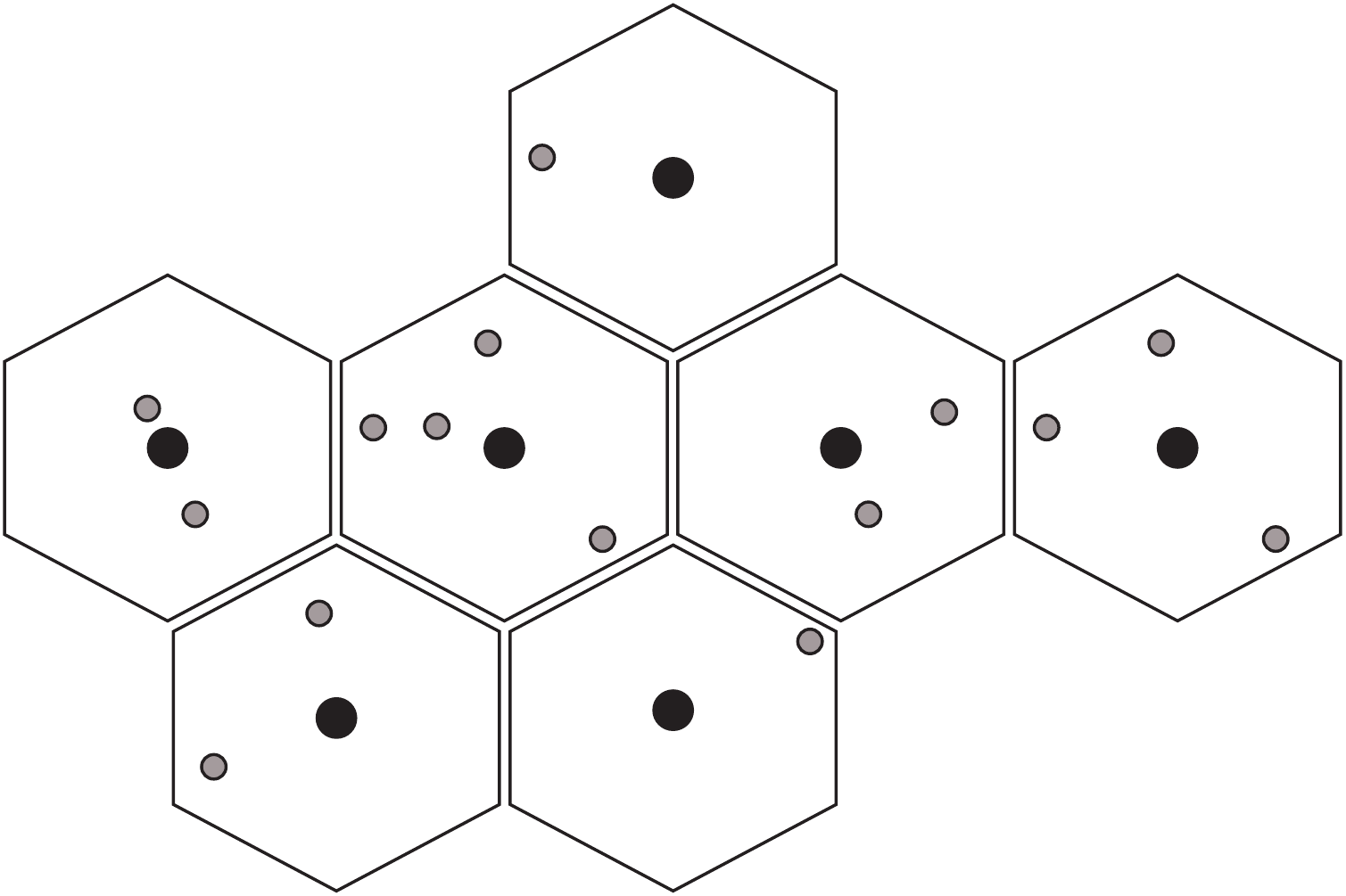}}
  \hspace{1cm}
   \subfigure[backhaul network and joint transmission graph]{\label{fig:network2}\includegraphics[width=0.45\columnwidth,bb=0 0 400 400]{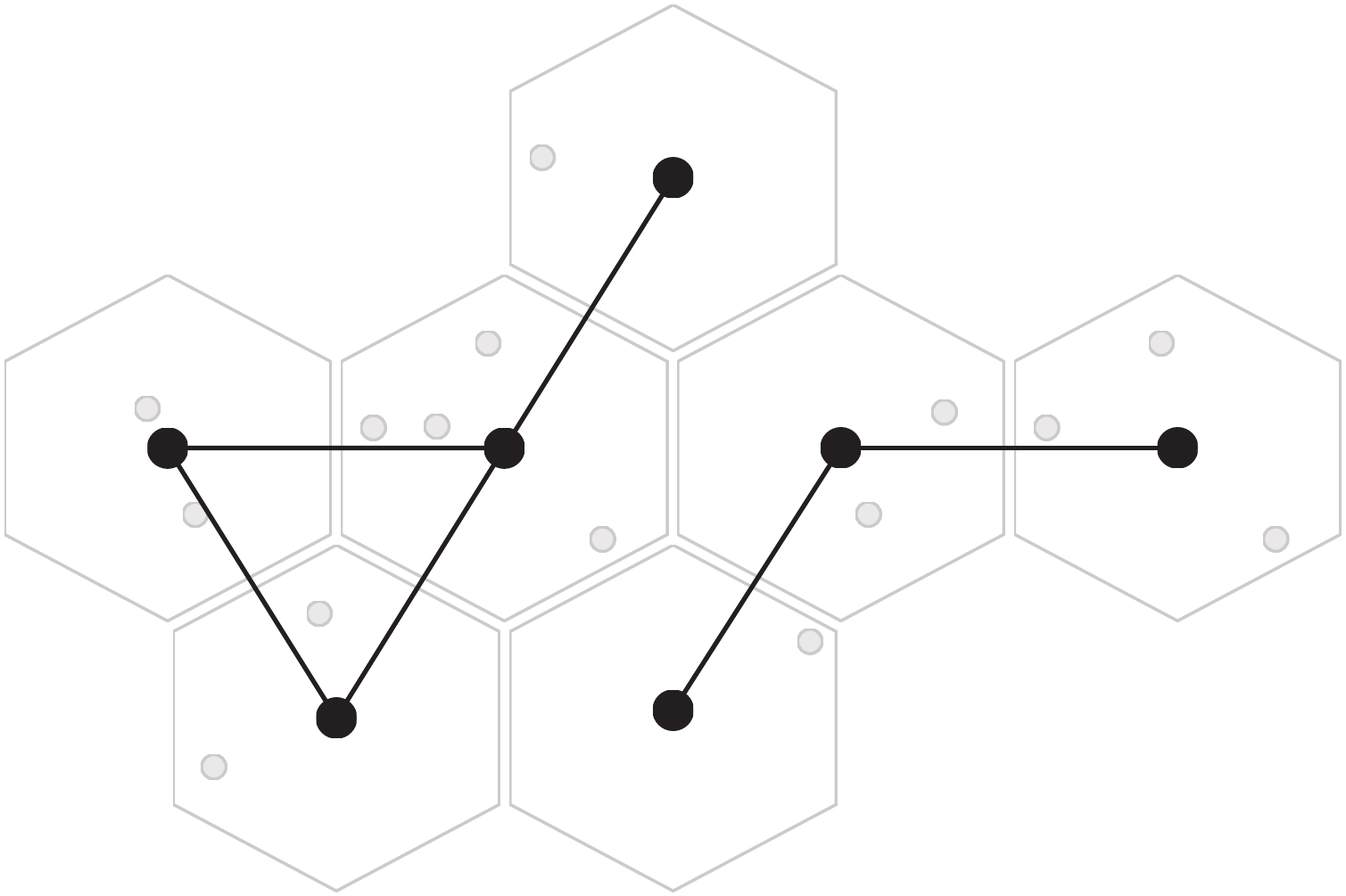}}
    \caption{
    A cellular network comprised of basestations (black circles) and users (grey circles).
    }
\end{figure}

We consider an OFDMA cellular network comprised of a set of BSs $\mathcal{B}=\{1,\ldots, B\}$ and a fixed set of stationary users
$\mathcal{N}=\{1,\ldots, N\}$, see Fig.~\ref{fig:network1}. The BSs are connected by a backhaul network
represented by a graph
$G_J=(\mathcal{B},\mathcal{C})$, where
$\mathcal{C}$ is a set of backhaul links
with $|\mathcal{C}|=C$. We refer to $G_J$ as the {\em Joint
  Transmission Graph}, as only neighboring BSs in $G_J$ can joint-transmit, see Fig.~\ref{fig:network2}.
We schedule over the downlink and assume that each backhaul link is bidirectional and that both directions share
the link capacity, but remark that all results can be readily extended to
directional backhaul links.
\begin{definition}
User $n$ is associated with up to two BSs:
\begin{itemize}
\item[-] The {\em serving BS} is denoted BS$(n)$ and is defined as the BS that
provides the highest SINR to user $n$.
\item[-] The {\em secondary BS} is
denoted $\widehat{\text{BS}}(n)$ and is defined as
the BS with highest SINR that has a backhaul link to BS$(n)$ in
$\mathcal{C}$.
\end{itemize}
Note that some users may not have a secondary BS.
\end{definition}
Packets destined for user~$n$ arrive at BS$(n)$ and are stored in a queue $Q_n$. Transmission for user~$n$ can be either single-transmitted by BS$(n)$ or joint-transmitted by BS$(n)$ and $\widehat{\text{BS}}(n)$. For a packet to be joint-transmitted, it first has to be forwarded over the backhaul to $\widehat{\text{BS}}(n)$, and stored in a queue $\widehat{Q}_n$ for joint transmissions maintained at both BS$(n)$ and $\widehat{\text{BS}}(n)$. So a packet
departs from $Q_n$ when it is single-transmitted or forwarded across the backhaul,
and a packet departs from $\widehat{Q}_n$ when it is joint-transmitted.



\begin{figure}
  \centering
  \subfigure[example network]{\label{fig:trans1}\includegraphics[width=0.3\columnwidth,bb=0 0 400 400]{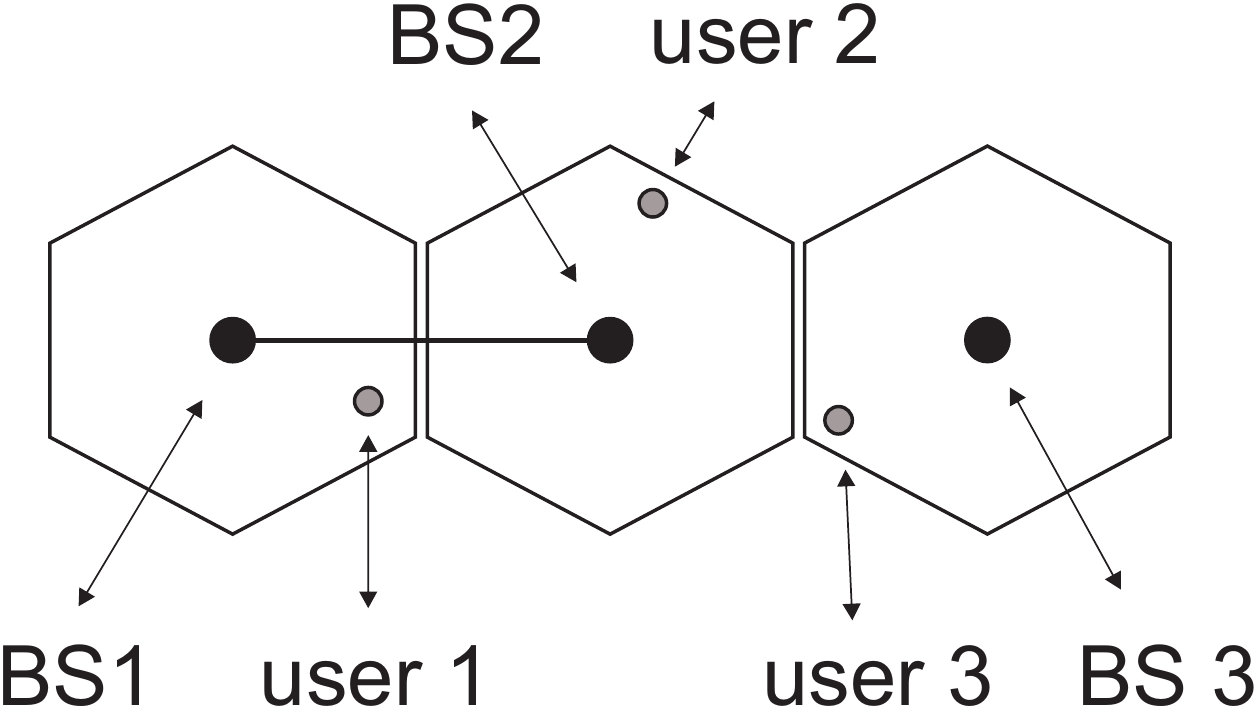}}
  \hspace{0.5cm}
   \subfigure[primary (solid line) and secondary (dashed line) BSs]{\label{fig:trans2}\includegraphics[width=0.3\columnwidth,bb=0 0 400 400]{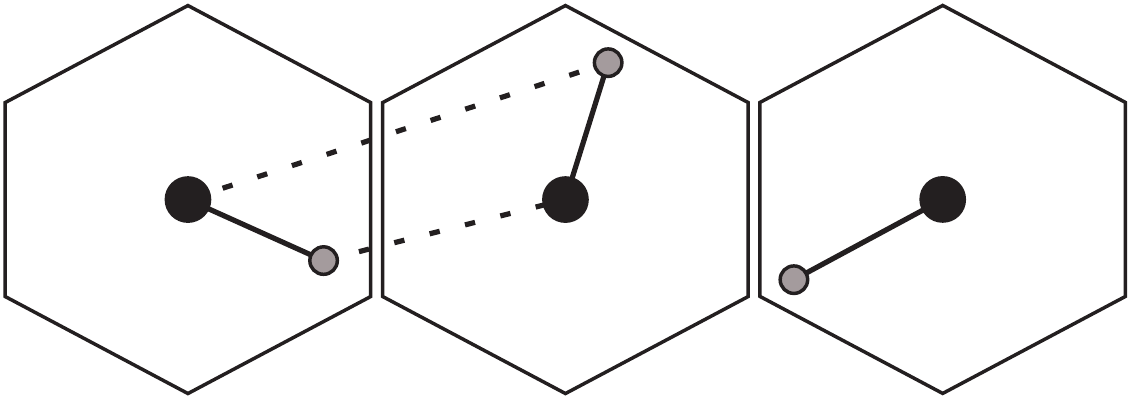}}
   \hspace{0.5cm}
   \subfigure[the storage locations of the various queues]{\label{fig:trans3}\includegraphics[width=0.3\columnwidth,bb=0 0 400 400]{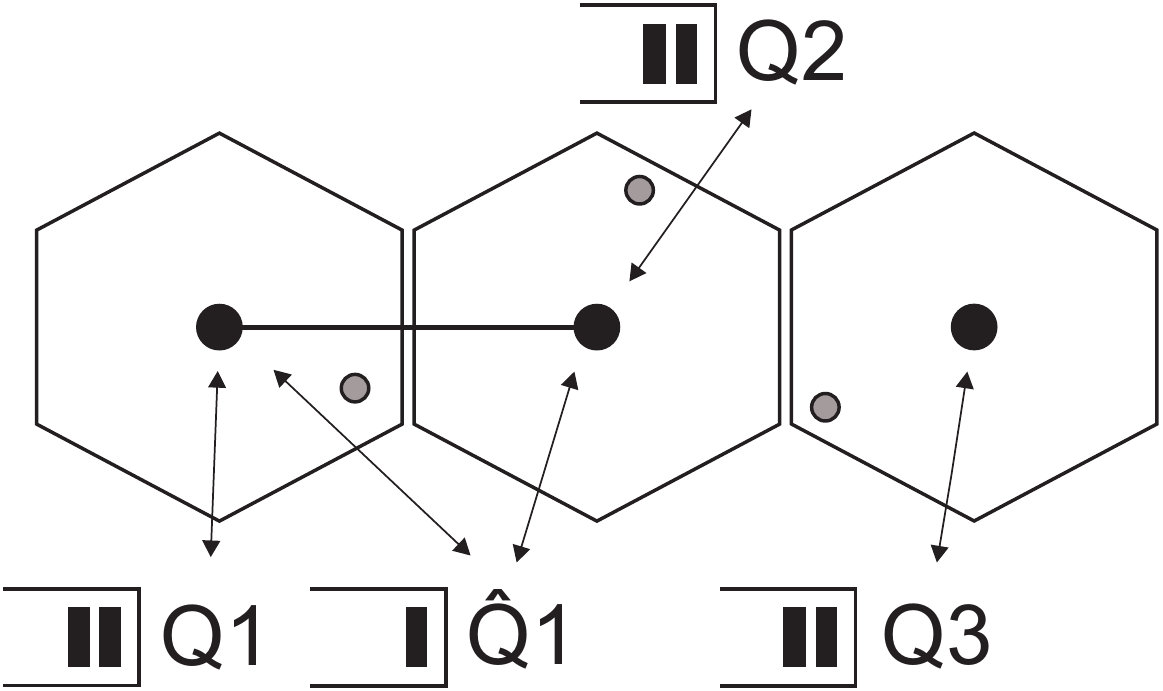}}
    \caption{
    BS allocation and queues.
    }
\end{figure}

To illustrate this, consider the network in Fig.~\ref{fig:trans1} with three
users and three BSs. In Fig.~\ref{fig:trans2} the primary and secondary BSs
of each user are marked, with a solid and dashed line, respectively. Note that
user 2's secondary BS is BS1 and not BS3, although the latter is closer, since
BS3 does not have a backhaul link to BS2. User 3 does not have a secondary BS because its primary BS does not have any backhaul connections. Fig.~\ref{fig:trans3} displays the various queues in play, and their locations.

Both wireless packet transmissions and forwarding over the backhaul lasts exactly a single subframe. Thro\-ughout, we assume that a central processing unit determines the schedule for
all BSs based on perfect knowledge on the network state.

We consider a time-slotted model indexed by $t$, $t= 0,1,\dots$, where each time slot corresponds to a single subframe. Denote $L_n(t)$ and $\widehat{L}_n(t)$  the
queue length of $Q_n$ and $\widehat{Q}_n$ at the beginning of subframe $t$,
respectively. Denote by $W_n(t)$ the number of new
packets generated for user $n$ at the beginning of subframe $t$. The $W_n(t)$ are
assumed to be i.i.d.\ over time, independent between users, and have finite second moment.  We
denote by $\mu_n^{(1)}(t)$, $\mu^{(2)}_n(t)$, and
$\mu_n^{(3)}(t)$ the number of packets transmitted towards user $n$ in
subframe $t$ using single and joint transmission, and the number of packets forwarded across the backhaul, respectively. These are determined by the resource allocation at each subframe, see Section~\ref{sec:subframe} for more details. The evolution of the queue lengths can then be written as
\begin{align}
  L_n(t\text{+}1)&= L_n(t) + W_n(t) - \mu_n^{(1)}(t)
  -\mu_n^{(3)}(t), \label{eqn:evolution_1}\\
  \widehat{L}_n(t\text{+}1)&= \widehat{L}_n(t) +
  \mu_n^{(3)}(t) - \mu_n^{(2)}(t). \label{eqn:evolution_2}
\end{align}

\subsection{Subframe model}\label{sec:subframe}

We now consider a single subframe consisting of scheduled blocks $\mathcal{S} = \{1,\dots,S\}$ for each BS, see Fig.~\ref{fig:LTE-frame}. In Sections~\ref{sec:hardness_results} and~\ref{sec:algorithms} we discuss how to allocate resources within a single subframe, which determines the $\mu_n^{(j)}$, $j=1,2,3$. The evolution of the queue lengths~\eqref{eqn:evolution_1} and~\eqref{eqn:evolution_2} over time is then discussed in Sections~\ref{sec:queueing_intro} and~\ref{sec:simulation}.

A packet $i$ which is single-transmitted requires scheduled blocks in a
subframe of BS$(n(i))$, while a joint-transmitted packet requires
scheduled blocks in the subframes of both BS$(n(i))$ and $\widehat{\text{BS}}(n(i))$. In the latter
case, the set of scheduled blocks used by BS$(n(i))$ and $\widehat{\text{BS}}(n(i))$
\emph{must have identical indices} since JT requires both BSs to transmit on
the same scheduled blocks.

\ifTechRep
\begin{figure}
 \centering
 \subfigure[]
{\label{fig:LTE-frame}\includegraphics[width=0.35\columnwidth,bb=0 0 600 100]{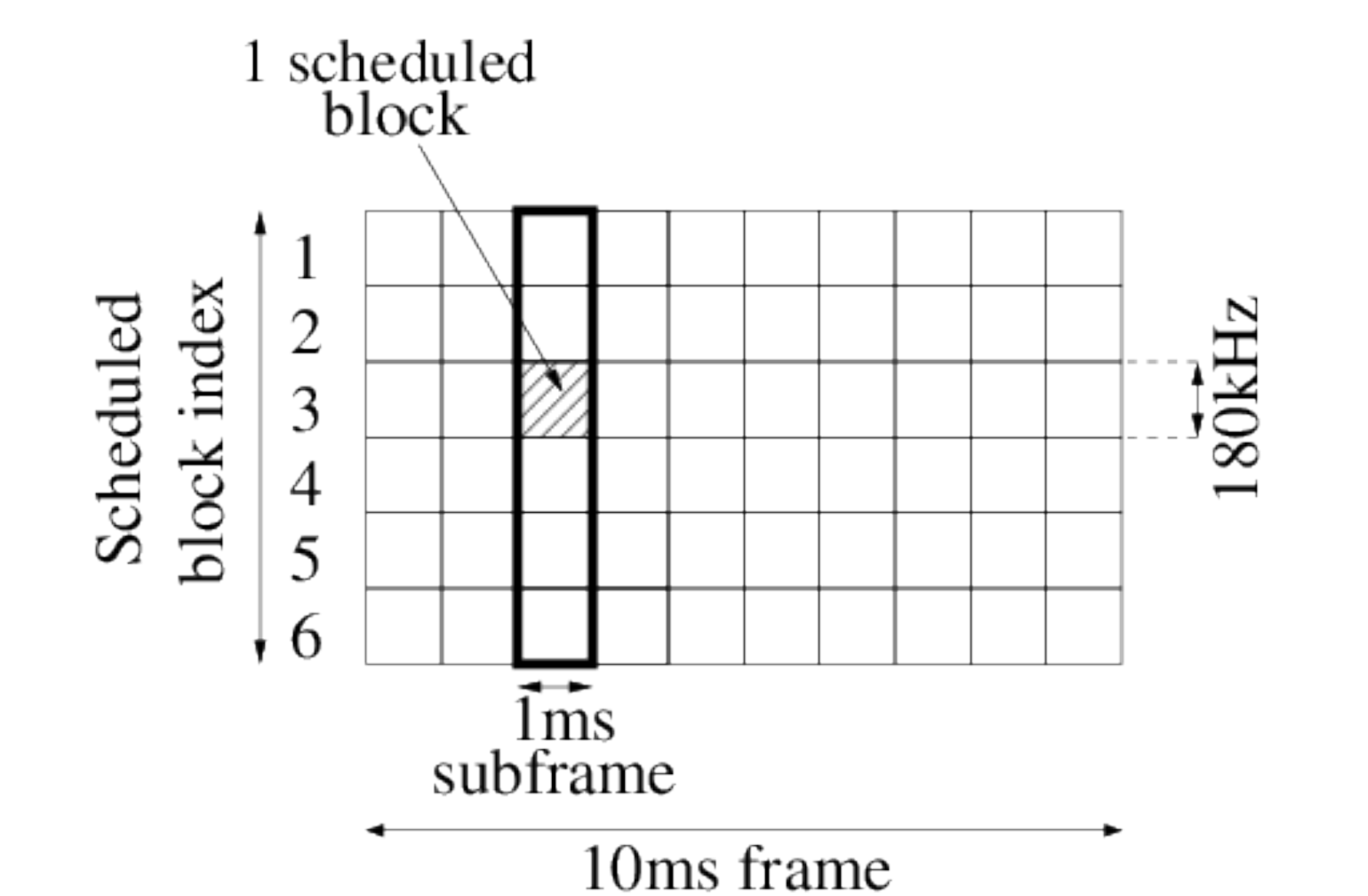}}
\hspace{3mm}
 \subfigure[]{\label{fig:GSB}\includegraphics[
   width=0.23\columnwidth,bb=0 0 500 100]{fig/GSB_v2.pdf}}
   \vspace*{-8pt}
   \caption{(a) Example of a frame (corresponding to 1.4MHz LTE). (b) A scheduled
   blocks multigraph $G_{\text{SB}}$.}
\end{figure}
\else
\begin{figure}
  \centering
  \includegraphics[width=0.4\columnwidth]{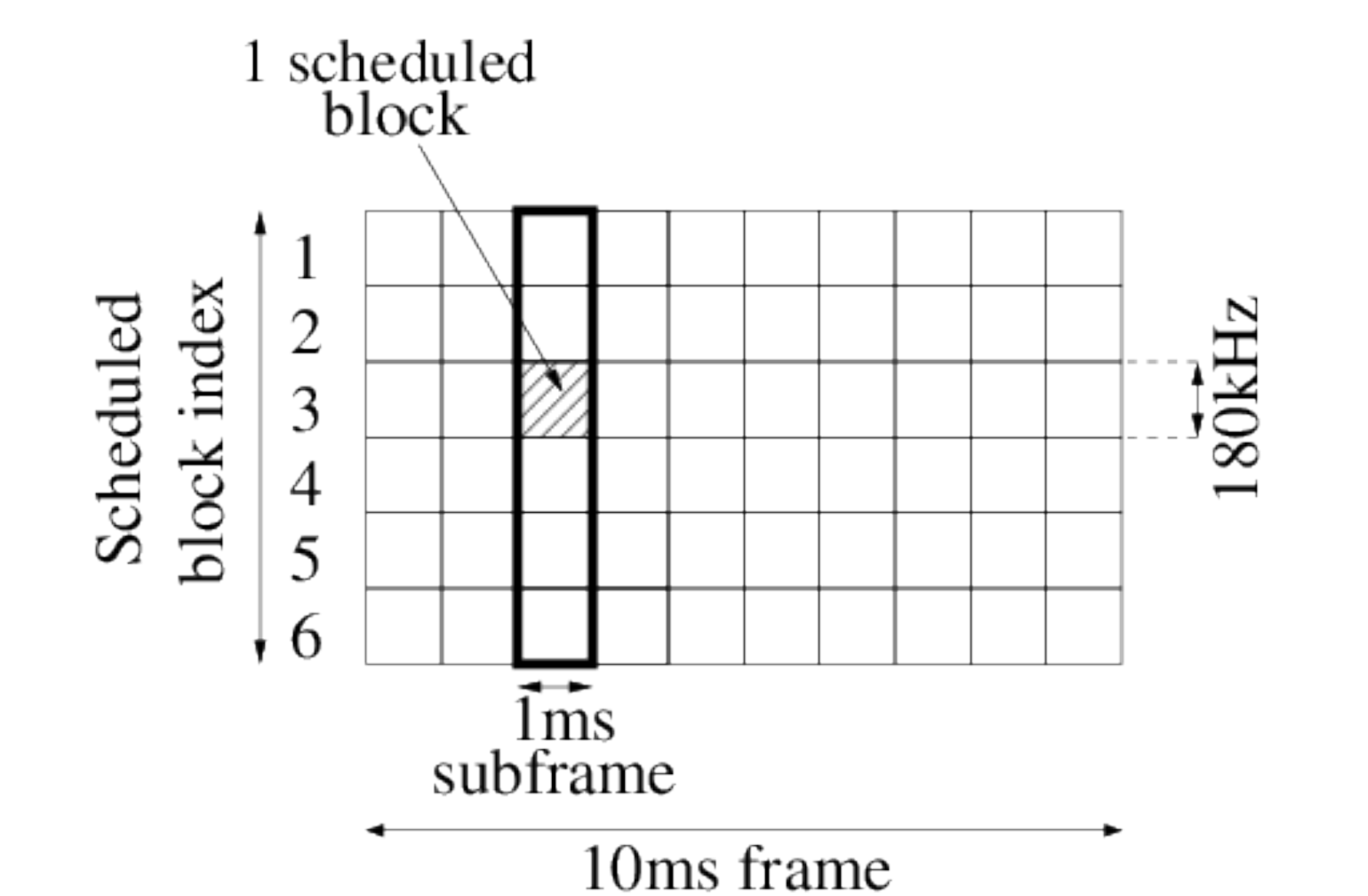}
\caption{Example of a frame (corresponding to 1.4MHz LTE).}
\label{fig:LTE-frame}
\end{figure}
\fi


A packet $i$ is characterized by the pair $(n, \beta)$, where $n$ is the
receiving user, $\beta\in \{0, 1\}$ indicates whether a packet is in $Q_n$
($\beta = 0$) or in $\widehat{Q}_n$ ($\beta = 1$).
The set of all packets is
denoted $\mathcal{I}$, $|\mathcal{I}|=I$. Given a packet $i\in \mathcal{I}$, we denote by $n(i),\
\beta(i)$  its user type and queue type, respectively.

\ifTechRep
 Each wireless transmission requires an MCS, which is selected from a set $\mathcal{M}=\{1,\ldots, M\}$ of supported MCSs.
We define the configuration set $\mathcal{R}=\{0\} \cup
\mathcal{M}$ such that $r\ge 1$ indicates wireless transmission using MCS $r$, and $r=0$
indicates a packet forwarded to another BS over the backhaul. 
Thus, a pair $(i, r)$ ($i\in \mathcal{I},\ r\in
\mathcal{R}$) defines a transmission of packet $i$ with MCS $r$ ($r \ge 1$) or packet $i$ forwarded over the backhaul ($r=0$). Note that the pair $(i,0)$ is not feasible if $\beta(i)=1$.

For each $(i, r)$, we define two properties: its size $\Gamma(i,r)$ and
its success probability $p(i,r)$. First, $\Gamma(i, 0)$ represents the size of
packet $i$ in bytes. For $r \ge 1$, the $\Gamma(r, c)$ represents the number
of scheduled blocks required for transmission with MCS $r$, which
depends on
the packet size in bytes $\Gamma(i,0)$ and the MCS $r$.

The success probability $p(i,r)$ represents the probability that packet $i$ will be successfully received by the user $n(i)$ ($r\ge 1$) or that packet $i$ forwarded over the backhaul is successfully received by $\widehat{\text{BS}}(n)$ ($r=0$). For MCS $m \in \mathcal{M}$, the success probability $p(i,m)$ depends on the packet length $\Gamma(i,0)$, the MCS $m$, the user's channel state, and whether it is single-transmitted or joint-transmitted ($\beta(i) = 0$ or $\beta(i) = 1$).

We assume that $p(i,m)$ is higher if $\beta(i) = 1$ compared to $\beta(i) =
0$, since the SINR of a user is greater when a packet is joint-transmitted.
\ifTechRep
The capacity of a
backhaul link between BSs $a$ and $b$ is $l_{ab}$
bytes.
\fi


\else

When transmitted wirelessly, packet $i$ is received successfully with
probability $p(i)$. Additionally, the success probability $p(i)$ is independent of its allocated scheduled blocks, since
  we assume that the interference in all scheduled blocks is similar (due to
  frequency reuse 1). Note that if some scheduled blocks are unused, the
  interference is lower and better performance is obtained.
We assume that $p(i)$ is higher if $\beta(i) = 1$ compared to $\beta(i) = 0$,
since when using joint transmission the two BSs configure their
  transmission parameters such that the signal combines constructively at the
  user, resulting in greater
SINR when a packet is joint-transmitted.

To simplify the presentation and due to space constraints, we make three
assumptions: (i) forwarding a packet $i$ over the backhaul is always successful;
(ii) all packets are of same length and a packet
transmission on a wireless channel requires one scheduled block from the subframe
of each of its transmitting BSs; and (iii) the
capacity of each backhaul link is $K$ packets/subframe.
\ifTechRepFinal
In Appendix~\ref{sec:extension}
\else
In the accompanying technical report~\cite{GBVZ15}, 
\fi
we show that
assumptions (i)-(iii) can be relaxed. Moreover, there we
demonstrate the applicability of our algorithms to the case where a packet
can be transmitted using one of several Modulation and Coding Schemes
(MCSs). We also remark that all the results in this paper can be
  readily applied to the setting with infinite backhaul capacity, by setting $K=S$. In our simulation study (Section~\ref{sec:simulation}) we  evaluate our algorithms for the
case where multiple MCSs are supported.
\fi

\section{The OFDMA Joint Scheduling (OJS) Problem}
\label{sec:scheduling_problem}
We now formulate the joint scheduling problem. 
\ifTechRep
Capacity constraints apply both to the subframes of the $B$ BSs
as well as the $C$ backhaul links, and we denote the total number of constraints by
$D=B+C$. We order these constraints such that the constraint $b$
corresponds to the BS $b$, $b = 1,\dots,B$,
and constraints $B+1,\dots,D$
correspond to the backhaul. Define the $D$-size vector
$\pmb{K}=(K_1,\ldots, K_D)$ such that for $1\leq d\leq B$, $K_d=S$ (number of
scheduled blocks), and for
$d\ge B+1$ that corresponds to a backhaul link between BSs $a$ and $b$,
$K_d=l_{ab}$ (number of bytes).
\fi
\ifTechRep
In order to describe the capacity used by the transmission of packet $i$, we introduce
\else
In order to describe the BSs involved in each transmission, we introduce
\fi
\begin{equation*}
h(i) = \left\{
\begin{array}{ll}
\{\text{BS}(n(i))\}    & {\rm ~if~} \beta(i) = 0,\\
\{\text{BS}(n(i)),\widehat{\text{BS}}(n(i))\}    & {\rm ~if~} \beta(i) = 1.\\
\end{array}\right.
\end{equation*}
If $\beta(i) = 0$ then $h$ returns only the serving BS, and if $\beta(i) = 1$ it returns both the serving and secondary BS. 
\ifTechRep
We define the function $w:\mathcal{I}\times \mathcal{R}\rightarrow
(\mathbb{N}_0)^D$, where $w(i, r)$ denotes the $D$-dimensional vector that
represents the capacity used for $(i,r)$. If $r \ge 1$, then $w(i, r)$ is the
all-zero vector except for $[w(i, r)]_{b} = \Gamma(i, r)$ for $b \in h(i)$. If
$r=0$, then only the entry corresponding to the appropriate backhaul link is
positive, and equal to the length in bytes of packet $i$.
\fi


\ifTechRep
The function $u:\mathcal{I} \times \mathcal{R} \mapsto \mathbb{R}_+$ represents the utility of scheduling packet $i$ according to configuration $r$. Examples include a throughput-based utility function
\begin{equation}\label{eqn:throughput_based_utility}
u_T(i,r)= p(i,r),
\end{equation}
and a fairness-based utility function
\begin{equation*}
u_F(i,r)= \log p(i,r).
\end{equation*}
In Sections~\ref{sec:queueing_intro} and~\ref{sec:simulation} we use a queue-length based utility function
\begin{equation}\label{eqn:queue_based_utility}
u_Q(i,r)= \left\{
\begin{array}{ll}
L_{n(i)} p(i,r), \quad r \ge 1,\\
\max \{L_{n(i)}- \hat{L}_{n(i)},0\},  \quad r=0,
\end{array}
\right.
\end{equation}
where $L_{n}$ and $\hat{L}_{n}$ denote the queue length of $Q_{n}$ and $\hat{Q}_{n}$,
respectively. Our model and analysis can in fact handle a wide range of
utility functions such as those used in~\cite{YRLZ13}.
\else
The function $u:\mathcal{I} \times \{0,1\} \mapsto \mathbb{R}_+$ represents the utility of
scheduling packet $i$ over the backhaul ($u(i,0)$) or wireless channel ($u(i,1)$). Examples include throughput-based utility
function $u_T$ and fairness-based utility
function $u_F$ defined by
\begin{equation}\label{eqn:throughput_based_utility}
\begin{array}{ll}
u_T(i,0)=\gamma, \quad \quad \quad &  u_T(i,1)=p(i),\\
u_F(i,0)=\gamma, &  u_F(i,1)=\log p(i),
\end{array}
\end{equation}
where $\gamma>0$ is a small constant that ensures packets are
  forwarded over the backhaul. Since here we consider a single-slot formulation, the utility of scheduling packets over the backhaul is not evident when the utility function is based only on wireless throughput; $\gamma$ compensates for this.
In Sections~\ref{sec:queueing_intro} and~\ref{sec:simulation} we use a queue-length based utility function
\begin{equation}\label{eqn:queue_based_utility}
\begin{array}{ll}
u_Q(i,0)= \max \{L_{n(i)}- \hat{L}_{n(i)},0\},\\
u_Q(i,1)= L_{n(i)} p(i),
\end{array}
 \end{equation}
where $L_{n}$ and $\hat{L}_{n}$ denote the queue length of $Q_{n}$ and
$\hat{Q}_{n}$, respectively.
Our model and
analysis can in fact handle a wide range of utility functions such as those
used in~\cite{YRLZ13}.

\fi

Based on the set of packets $\mathcal{I}$ and the utility function $u$, the centralized scheduler determines
the set of wireless transmissions to take place in
the upcoming subframe, and what packets to forward over the backhaul.

The scheduler must also determine which scheduled
blocks will be used for each packet transmission, such that for JT the
scheduled blocks of the serving BS and secondary BS are aligned (i.e., have
identical index).
\ifTechRep
The
scheduling decisions are represented using indicator variables $z_{ir} \in \{0,1\}$ and
$x_{irs}\in \{0,1\}$, where $z_{ir}$ indicates if a transmission $(i,r)$
takes place, and $x_{irs}$ indicates if scheduled block $s$ is used by a
transmission $(i, r)$.
The scheduler needs to solve the
following integer programming problem (with $\pmb{z} = (z_{ir})_{i\in\mathcal{I},r\in\mathcal{R}}$ and $\pmb{x} = (x_{irs})_{i\in\mathcal{I},r\in\mathcal{R},s \in \mathcal{S}}$).

\noindent{\bf OFDMA Joint Scheduling (OJS) Problem}:
\begin{alignat}{2}
\underset{\pmb{x},\pmb{z}}{\text{max}}&
&\:\: & \sum_{i=1}^I \sum_{r=1}^R z_{ir}
  u(i,r) =: U(\pmb{z}) \notag
\\
\text{s.t.}&
&&  \sum_{r=0}^R z_{ir} \leq 1, \quad \forall\  i\in
\mathcal{I}, \label{eq:one-conf} \\
&
&&   \sum_{i=1}^I \sum_{r=0}^R z_{ir} [w(i,r)]_d \leq K_d,
 \quad \forall
  1\leq d\leq D, \label{eq:all-capacities} \\
&
&&  \sum_{s=1}^S x_{irs}=z_{ir} \Gamma(i, r),  \quad \forall
  i \in \mathcal{I}\  \forall r\ge 1, \label{eq:sufficient} \\
&
&&   \sum_{r=1}^R \sum_{\{i:\ b\in h(i)\}} x_{irs} \leq 1, \quad \forall b\in
  \mathcal{B}\ \forall s\in \mathcal{S}, \label{eq:used-once}\\
&
&&   z_{ir} \in \{0,1\}, \quad \forall i \in \mathcal{I}\ \forall r \in \mathcal{R}, \label{eq:integer_z}\\
&
&&   x_{irs} \in \{0,1\}, \quad \forall i \in \mathcal{I}\ \forall r \in \mathcal{R}\ \forall s \in \mathcal{S}. \label{eq:integer_x}
\end{alignat}
Constraint \labelcref{eq:one-conf} ensures a packet is scheduled only
once; \labelcref{eq:all-capacities} ensures capacities in each subframe and backhaul link are not exceeded; \labelcref{eq:sufficient} ensures
sufficient scheduled blocks are allocated for each $(i,r)$ transmission;
and \labelcref{eq:used-once} ensures that each scheduled block is used at most
once in the subframe of each BS.
\else
The
scheduling decisions are represented using indicator variables $z_{i} \in
\{0,1\}$, $y_{i} \in \{0,1\}$, and
$x_{is}\in \{0,1\}$, where $z_{i}$ indicates if packet $i$ is transmitted wirelessly,
$y_{i}$ indicates if packet $i$ is forwarded over the backhaul
and
$x_{is}$ indicates if scheduled block $s$ is used by packet $i$.
The scheduler needs to solve the
following integer programming problem (with $\pmb{z} =
(z_{i})_{i\in\mathcal{I}}$, $\pmb{y} = (y_{i})_{i\in\mathcal{I}}$, and $\pmb{x} = (x_{is})_{i\in\mathcal{I},s \in \mathcal{S}}$).

\noindent{\bf OFDMA Joint Scheduling (OJS) Problem}:
\begin{alignat}{2}
\underset{\pmb{x},\pmb{y},\pmb{z}}{\text{max}}&
&& \sum_{i=1}^I z_i u(i,1) + y_i u(i,0) =: U(\pmb{z},\pmb{y}) \notag
\\
\text{s.t.}&
&& \ z_{i} + y_i \leq 1, \quad \forall\  i\in
\mathcal{I}, \label{eq:one-conf} \\
&
&&   \sum_{\{i: a\in h(i)\}} z_i \leq S, \forall a\in\mathcal{B};
\sum_{\{i:h(i)=l\}} y_i \leq K, \
\forall l\in\mathcal{C},  \label{eq:all-capacities} \\
&
&&  \sum_{s=1}^S x_{is}=z_{i},  \quad \forall
  i \in \mathcal{I};\ y_i=0, \quad \forall i \text{ s.t. } \beta(i)=1,  \label{eq:sufficient} \\
&
&&   \sum_{\{i:\ b\in h(i)\}} x_{is} \leq 1, \quad \forall b\in
  \mathcal{B}\ \forall s\in \mathcal{S}, \label{eq:used-once}\\
&
&&   z_{i} \in \{0,1\},\ y_i\in \{0,1\}, \quad \forall i \in \mathcal{I}, \label{eq:integer_z}\\
&
&&   x_{is} \in \{0,1\}, \quad \forall i \in \mathcal{I}\ \forall s \in \mathcal{S}. \label{eq:integer_x}
\end{alignat}
Constraint \eqref{eq:one-conf} ensures a packet is scheduled at most
once and resides in a single queue; \eqref{eq:all-capacities} ensures capacities in each subframe and backhaul link are not exceeded; \eqref{eq:sufficient} ensures
a scheduled block is allocated for each wireless transmission and packets in
$\widehat{Q}_n$ cannot be forwarded over the backhaul;
and \eqref{eq:used-once} ensures that each scheduled block is used at most
once in the subframe of each BS.

\begin{figure}
  \centering
  \subfigure[a wireless network with 3 users and 7 packets (denoted 1..7)]{\label{fig:example1}\includegraphics[width=0.3\columnwidth]{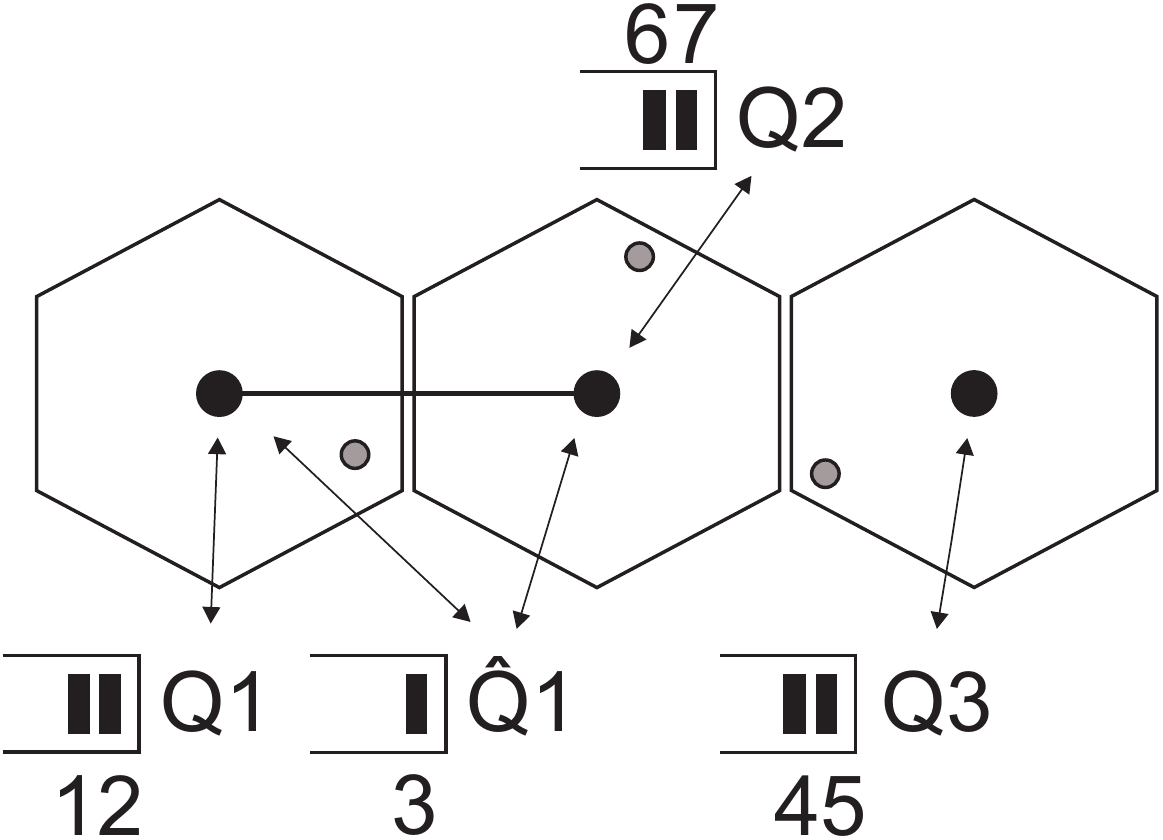}}
  \hspace{1cm}
   \subfigure[a possible allocation of the scheduled blocks]{\label{fig:example2}\includegraphics[width=0.2\columnwidth]{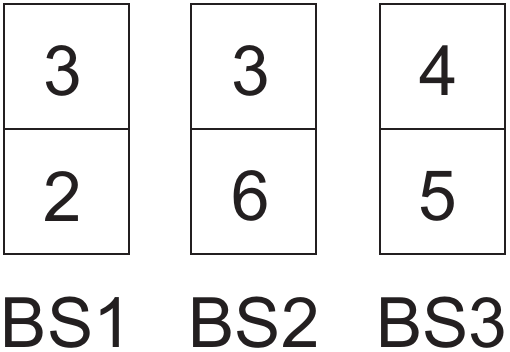}}
   \hspace{1cm}
   \subfigure[the resulting scheduled block graph]{\label{fig:example3}\includegraphics[width=0.3\columnwidth]{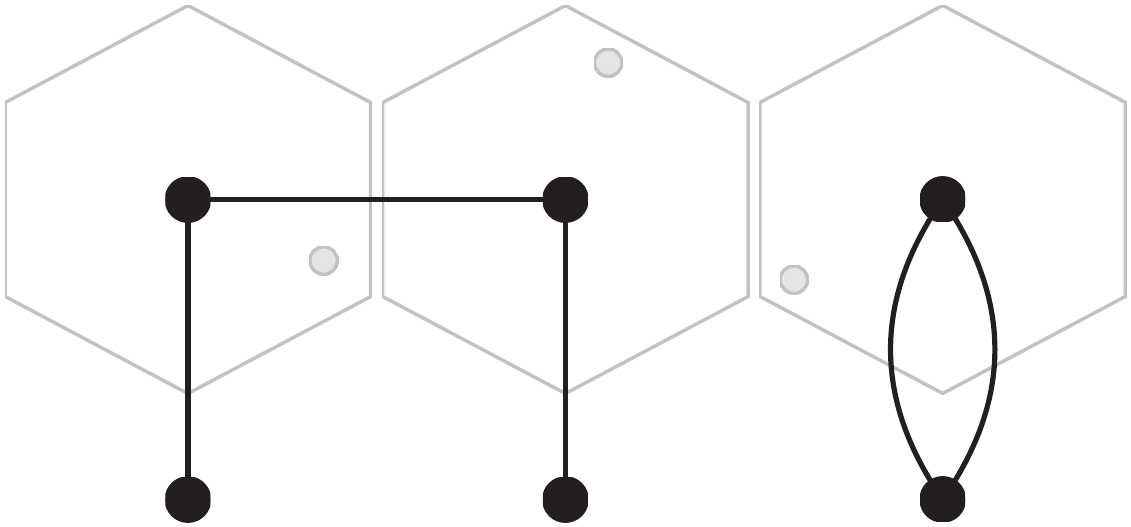}}
   \caption{An example of a scheduled block allocation and the resulting scheduled block graph.}
\end{figure}

To illustrate this problem, consider the network displayed in
Fig.~\ref{fig:example1}. This network comprises 3 BSs, 3 users and 7 packets,
numbered 1,$\dots$,7. BS1 and BS2 are connected with a backhaul, so they can
joint-transmit packets for user 1. Let $S=2$, and assume that we want to
allocate the scheduled block to achieve the following transmissions: (i)
packet 3 should be joint-transmitted; (ii) packets 2,4,5 and 6
single-transmitted; and (iii) packet 1 forwarded over the
  backhaul from $Q_1$ to $\widehat{Q}_1$. This schedule can be obtained with the assignment displayed in Fig.~\ref{fig:example2}, and the following solution to the OJS problem:
$$
\pmb{z} = (0,1,1,1,1,1,0), \quad \pmb{y} = (1,0,0,0,0,0,0), \quad \pmb{x} = \left(
\begin{array}{lllllll}
0   &   0   &   1   &   1   &   0   &   0   &   0\\
0   &   1   &   0   &   0   &   1   &   1   &   0
\end{array}
\right).
$$

\fi


\label{sec:hardness_results}
\ifTechRep
We now study the complexity of the OJS problem. We first show that an FPTAS is unlikely to
exist for this problem. 
\ifLongversion
\begin{proposition} \label{pro:NPhard-Knapsack} The OJS Problem with $B=1$ is
  NP-hard.
\end{proposition}

As a consequence of Proposition~\ref{pro:NPhard-Knapsack} we know that  OJS is NP-hard in general. We now proceed to show that an
efficient approximation scheme is unlikely to exist.
\fi

\begin{proposition} \label{pro:noFPTAS}
There is no FPTAS for OJS with $B\geq 2$ unless P=NP.
\end{proposition}



Next, we show that the problem is NP-hard even for a restricted set of instances.

\begin{proposition} \label{pro:NPhard-coloring}
  OJS is strongly NP-hard even for instances in which all of the following hold:

  \noindent (a) $\mathcal{R}=\{0, 1\}$;

  \noindent (b) $\Gamma(i,r)=1,\ \forall i\in\mathcal{I},\ \forall c\in\mathcal{R}$;

  \noindent (c) $u(i,0)=0$, $u(i,1)=1$, $\forall i\in\mathcal{I}$;

  \noindent (d) $L_n\leq 1$, $\widehat{L}_n\leq 1$, $\forall n\in \mathcal{N}$.
\end{proposition}

The proof of Proposition~\ref{pro:NPhard-coloring} uses a reduction from the
well-known problem of minimum edge coloring~\cite{H81}. The reduction
demonstrates that even for cases with sufficient bandwidth to accommodate all
packet transmissions in the BSs subframes, obtaining a feasible schedule where
joint-transmissions use an identical set of scheduled blocks is equivalent to minimum edge coloring~\cite{H81}.  In
Section~\ref{sec:algorithms}, we use algorithms for minimum edge coloring when
developing algorithms for OJS.



\else

We now describe the complexity of the OJS problem.
\begin{proposition} \label{pro:NPhard-coloring}
  OJS is strongly NP-hard even for instances in which all of the following hold:

  \noindent (a) $u(i,0)=1$, $u(i,1)=1$, $\forall i\in\mathcal{I}$;

  \noindent (b) $L_n\leq 1$, $\widehat{L}_n\leq 1$, $\forall n\in \mathcal{N}$.
\end{proposition}

The proof of Proposition~\ref{pro:NPhard-coloring} uses a reduction from the
well-known problem of minimum edge coloring~\cite{H81}. The reduction
demonstrates that even for cases with sufficient bandwidth to accommodate all
packet transmissions in the BSs subframes, obtaining a feasible schedule where
joint-transmissions use an identical set of scheduled blocks is equivalent to
the well-known problem of minimum edge coloring~\cite{H81}.  In
Section~\ref{sec:algorithms}, we use algorithms for minimum edge coloring when
developing algorithms for OJS.

\fi


\section{OJS Problem -- Algorithms}
\label{sec:algorithms}

\ifTechRep
\begin{table*}
\scriptsize
\begin{minipage}[t]{0.62\linewidth}
\centering
\caption{The performance and input required for different algorithms within the decomposition
  framework
  \DF{$A_{\text{JTK}}$}{$A_{\text{JTC}}$}
  \label{table:algorithms_summary}}
\begin{tabular}{|l|l|l|l|l|} \hline 
{\bf $A_{\text{JTK}}$} & {\bf $A_{\text{JTC}}$} & {\bf Ratio} & {\bf Running time} & {\bf Input $G_J$}
\\
\hline
\JTKMMK & \JTCBIP & $\alpha$ & $O(T_{\text{MMK}}(I,R,D,S))$ & bipartite  \\ \hline
\JTKMAT & \JTCBIP & $\frac {2  \alpha}
  {3  \Delta(G_J)}$ & $O(C \cdot T_{\text{MMK}}(I,R,3,S)$ & any \\ \hline
\JTKSTA & \JTCBIP & $\frac{\alpha}{\Delta(G_J)}$ & $O(B^2
T_{\text{MMK}}(I,R,2\Delta(G_J)+1,S))$ & any \\ \hline
\JTKPSP & \JTCPSP &$\alpha$ & $O(2^{B}+T_{\text{MMK}}(I,R,2^{B},S)$ &
planar ser.paral.\\
\hline
\end{tabular}
\end{minipage}
\hfill
\begin{minipage}[t]{0.35\linewidth}
\centering
\caption{Algorithms for MMK \label{table:MMK}}
\begin{tabular}{|l|l|l|} \hline 
{\bf $A_{\text{MMK}}$} & {\bf Ratio} & {\bf $T_{\text{MMK}}(I,R,D,S)$} \\
\hline
DP~\cite{Book:Kellerer04} & Optimal & $O(S^D I R D)$ \\
\hline
PTAS~\cite{PR12} & $1/(1+\epsilon)$ & $O((I R)^{(D/\epsilon)})$ \\ \hline
Greedy~\cite{Book:Kellerer04} & $\infty$ & $O(I R \log(IR))$ \\ \hline
\end{tabular}
\end{minipage}
\normalsize
\end{table*}

\else

\begin{table}
\scriptsize
\centering
\caption{The performance and input required for different algorithms $A_{\text{OJS}} = $
  \DF{$A_{\text{JTK}}$}{$A_{\text{JTC}}$}
  \label{table:algorithms_summary}}
\begin{tabular}{|l|l|l|l|l|} \hline 
{\bf $A_{\text{JTK}}$} & {\bf $A_{\text{JTC}}$} & {\bf Ratio} & {\bf Running time} & {\bf Input $G_J$}
\\
\hline
\JTKMMK & \JTCBIP & $\alpha$ & $O(T_{\text{MMK}}(I,B,C,S))$ & bipartite  \\ \hline
\JTKMAT & \JTCBIP & $\frac {2  \alpha}
  {3  \Delta(G_J)}$ & $O(C  T_{\text{MMK}}(I,2,1,S))$ & any \\ \hline
\multirow{2}{*}{\JTKSTA} & \multirow{2}{*}{\JTCBIP} & \multirow{2}{*}{$\frac{\alpha}{\Delta(G_J)}$} & $O(
T_{\text{MMK}}(I,\Delta(G_J)+$ & \multirow{2}{*}{any} \\
&&& $1,\Delta(G_J),S) B^2)$ &  \\ \hline
\multirow{2}{*}{\JTKPSP} & \multirow{2}{*}{\JTCPSP} &
\multirow{2}{*}{$\alpha$} & $O(2^{B}+T_{\text{MMK}}(I,$ & planar
  \\
&&& $B+2^B,C,S))$ & ser.\ paral.\ \\
\hline
\end{tabular}
\normalsize
\vspace*{-0.4cm}
\end{table}

\addtolength{\tabcolsep}{.35em}

\fi

In this section we develop algorithms to solve the OJS problem.
First, we describe a framework for solving
the OJS problem by decomposing it into problems related to
knapsack and coloring, see Section~\ref{sec:decomp}. We then use this
decomposition framework to develop efficient algorithms for OJS. In particular, in Section~\ref{sec:alg-bip} we develop algorithms for instances where the joint transmission graph $G_J$ (consisting of the BSs and the backhaul) is bipartite, in Section~\ref{sec:alg-sp} we develop algorithms for instances where the joint transmission graph is planar and series-parallel, and in Section~\ref{sec:alg-general} we develop algorithms for general joint transmission graphs. Note that joint transmission graphs encountered in practice need not always be bipartite or planar and series-parallel. However, if this is the case, by using an algorithm that exploits the structure of the graph, we can guarantee lower computational complexity and more accurate results. The algorithms for the general case in Section~\ref{sec:alg-general} are based on those for bipartite graphs in Section~\ref{sec:alg-bip}.

We denote the approximation ratio of a
given algorithm by $\alpha$ ($0<\alpha\leq 1$). If the algorithm is
optimal, we have $\alpha=1$.

\subsection{Decomposition Framework}\label{sec:decomp}

\ifTechRep
As shown in Propositions~\ref{pro:noFPTAS} and~\ref{pro:NPhard-coloring},
\else
From Proposition \ref{pro:NPhard-coloring} we see that,
\fi
unless P=NP, an efficient optimal algorithm for general instances of OJS does not exist. In order to
develop efficient approximations for the general case and optimal solutions for a subset of instances,
we present two additional scheduling problems and explore their relation to OJS. These two problems
are obtained by partitioning OJS into two parts,
exploiting the fact that $\pmb{x}$ only appears in
\eqref{eq:sufficient}, \eqref{eq:used-once}, and \eqref{eq:integer_x}.  

\ifTechRep
\noindent \textbf{Joint Transmission Knapsack (JTK) Problem:}
\begin{alignat*}{2}
\underset{\pmb{z}}{\text{max}} & & \:\: & U(\pmb{z})\\
\text{ s.t.} &&& \eqref{eq:one-conf},
\eqref{eq:all-capacities}, \eqref{eq:integer_z},\  \exists
\pmb{x}:\ \eqref{eq:sufficient}, \eqref{eq:used-once}, \eqref{eq:integer_x} \text{ hold}
\end{alignat*}
\noindent \textbf{Joint Transmission Coloring (JTC) Problem:}
\begin{alignat*}{2}
\text{given } \pmb{z} \text{, find}& &\:\: & \pmb{x} \text{
  s.t. \eqref{eq:sufficient}, \eqref{eq:used-once}, \eqref{eq:integer_x}
hold}
\end{alignat*}
\else
\noindent \textbf{Joint Transmission Knapsack (JTK) Problem:}
\begin{alignat*}{2}
\underset{\pmb{z}, \pmb{y}}{\text{max}} & & \:\: & U(\pmb{z},\pmb{y})\\
\text{ s.t.} &&& \eqref{eq:one-conf},
\eqref{eq:all-capacities}, \eqref{eq:integer_z},\  \exists
\pmb{x}:\ \eqref{eq:sufficient}, \eqref{eq:used-once}, \eqref{eq:integer_x} \text{ hold}
\end{alignat*}
\noindent \textbf{Joint Transmission Coloring (JTC) Problem:}
\begin{alignat*}{2}
\text{given } \pmb{z},\pmb{y} \text{, find}& &\:\: & \pmb{x} \text{
  s.t. \eqref{eq:sufficient}, \eqref{eq:used-once}, \eqref{eq:integer_x}
hold}
\end{alignat*}
\fi



Note that the JTK problem resembles
an assignment problem rather than a knapsack problem. We remark that this is due to`                                                                                        the assumptions that all packets have the same length and use a single MCS.
\ifTechRepFinal
 We show in Appendix~\ref{sec:extension} that when relaxing these assumptions, JTK indeed generalizes to a knapsack-like problem.
\else
 We show in~\cite{GBVZ15} that when relaxing these assumptions, JTK indeed generalizes to a knapsack-like problem.
\fi

We use $A_{\text{JTK}}$ and $A_{\text{JTC}}$ to denote an algorithm for
JTK and JTC, respectively.
A specific algorithm for problem P is
denoted P-D where D identifies the algorithm. For instance, we write $A_{\text{JTK}} = \text{JTK-GREEDY}$ if we solve the JTK problem using a greedy algorithm. An instance of problem P consists of specific values for all
variables in its constraints, except for
\ifTechRep
$\pmb{x}$ and $\pmb{z}$.
\else
$\pmb{x}$, $\pmb{y}$, and $\pmb{z}$ (JTK) and $\pmb{x}$ (JTC).
\fi

\begin{figure}
\centering
\ifTechRep
\includegraphics[width=0.6\columnwidth]{fig/framework_V3.pdf}
\else
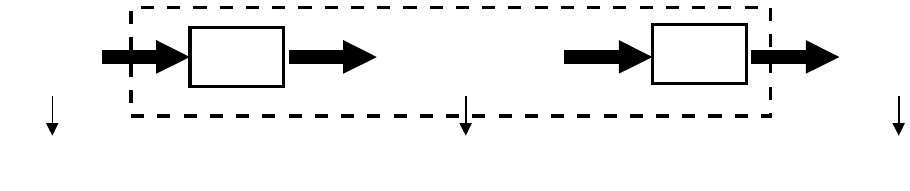
\fi
\caption{The decomposition framework leading to algorithm $A_{\text{OJS}} = $\DF{$A_{\text{JTK}}$}{$A_{\text{JTC}}$}.} \label{fig:framework}
\vspace*{-0.5 cm}
\end{figure}

The JTK problem differs from OJS in that it does
not attempt to find $\pmb{x}$ but guarantees that such $\pmb{x}$
exists for its solution
\ifTechRep
$\pmb{z}$.
\else
$\pmb{z},\pmb{y}$.
\fi
The JTC problem then finds $\pmb{x}$ given $\pmb{z}$ and $\pmb{y}$, which is later shown to correspond to a coloring problem. It is 
ensured that if a solution exists for JTK, the corresponding JTC problem instance can also be solved. Thus we can decompose OJS by first solving JTK to obtain
\ifTechRep
$\pmb{z}$
\else
$\pmb{z},\pmb{y}$
\fi
and then solving JTC to obtain
$\pmb{x}$, see Fig.~\ref{fig:framework} (some details
in the figure are explained later).
In Sections \ref{sec:alg-bip}-\ref{sec:alg-general},
we identify instances where the existence of a feasible $\pmb{x}$ is guaranteed without the
need to find $\pmb{x}$, and use this to develop efficient algorithms for OJS.

 If algorithms $A_{\text{JTK}}$ and
$A_{\text{JTC}}$ are used in this decomposition to solve JTK and JTC,
respectively, we denote the corresponding OJS algorithms as
$A_{\text{OJS}} = $\DF{$A_{\text{JTK}}$}{$A_{\text{JTC}}$}.
As we shall demonstrate in this section, making this decomposition allows us to find efficient algorithms for solving OJS. Throughout
the paper only optimal $A_{\text{JTC}}$ algorithms are considered. The following lemma immediately follows.

\begin{lemma} \label{lem:basic-framework}
  If $A_{\text{JTK}}$ is an $\alpha$-approximation algorithm for the JTK
  problem and $A_{\text{JTC}}$ an optimal algorithm for the JTC problem, the algorithm $A_{\text{OJS}} = $\DF{$A_{\text{JTK}}$}{$A_{\text{JTC}}$} is an
  $\alpha$-approximation for the OJS problem.
\end{lemma}

We introduce the following definitions
that will be used to solve JTC.
Recall from Section~\ref{sec:model} that $G_J=(\mathcal{B},\mathcal{C})$ denotes the joint transmission
graph that describes what BSs are connected by a backhaul link.

\ifTechRep
\begin{definition} \label{def:sb-graph}
The \emph{scheduled blocks graph}
  $G_{\text{SB}} = G_{\text{SB}}(G_J,\pmb{z}) =(V_{\text{SB}},E_{\text{SB}})$ of a JTC instance is defined
  by $V_{\text{SB}}=V_J\cup\{B+1,\ldots, 2B\}$ and the set of edges
  $E_{\text{SB}}$ constructed from the JTC instance as follows. Every pair
  $(i,r)$ such that $r\ge 1$ and $z_{ir}=1$ contributes $\Gamma(r,c)$ edges to
  $E_{\text{SB}}$: if $\beta(i)=1$ every edge is between the vertices in
  $h(i,r)$; if $\beta(i)=0$ then $h(i,r)=\{b\}$ and the edges are between the
  vertices $b$ and $B+b$.
\end{definition}

As an example for $G_J$ and $G_{SB}$ consider the network depicted in
Fig.~\ref{fig:basic_example} and assume all backhaul capacities are
positive. In this case, $G_J$ is a cycle with 3 vertices. Consider a
$\pmb{z}'$ input to JTC for which $z'_{i1}=1$ for $\{i_1, i_2, i_3\}$ (these
items are shown in Fig. \ref{fig:basic_example}),
$\Gamma(i_2,1)=\Gamma(i_3,1)=1$, and $\Gamma(i_1,1)=2$. Fig.~\ref{fig:GSB}
shows the graph $G_{\text{SB}}$ corresponding to $\pmb{z}'$.
\else
\begin{definition} \label{def:sb-graph}
The \emph{scheduled blocks graph}
  $G_{\text{SB}} = G_{\text{SB}}(G_J,\pmb{z}) =(V_{\text{SB}},E_{\text{SB}})$ of a JTC instance is defined
  by
  $$V_{\text{SB}}=V_J\cup\{B+1,\ldots, 2B\}$$
  and
  $$
  E_{\text{SB}} =\Big\{\{\text{BS}(n(i)),\text{BS}(n(i))+B\} \mid z_i = 1 {\rm ~and~} \beta(i) = 0 \Big\} \cup \Big\{\{\text{BS}(n(i)),\widehat{\text{BS}}(n(i))\} \mid z_i = 1 {\rm ~and~} \beta(i) = 1 \Big\}.
  $$
\end{definition}
To interpret the definition of $G_{\text{SB}}$, recall that $z_i$ denotes whether packet~$i$ does a wireless transmission ($z_i = 1$) or not ($z_i = 0$). So each edge in the scheduled block graph represents a wireless transmission, connecting the BSs that are engaged in the transmission. To avoid self-loops, we introduce dummy vertices $B+1,\dots,2B$ in case these BSs are involved in a single-transmission. Note that while $G_J$ is a simple graph, $G_{\text{SB}}$ need not be. As an example for $G_J$ and $G_{SB}$ consider the network depicted in
Fig.~\ref{fig:example1}. Using the solution described in the example at the end of Section~\ref{sec:scheduling_problem}, the resulting scheduled block graph is depicted in Fig.~\ref{fig:example3}
\fi

We now show that, using the scheduled block graph, JTC can be rewritten as an instance of the well-known
edge-coloring (EC) problem~\cite{COS01}, and solved accordingly. The input to the EC problem is a
graph $G=(V,E)$ and the output is a coloring on the edges that uses a minimum
number of colors.

\begin{lemma} \label{lem:JTC-coloring} JTC is equivalent to finding an edge
  coloring using at most $S$ colors on
  $G_{\text{SB}}=(V_{\text{SB}},E_{\text{SB}})$.
\end{lemma}

As a consequence, JTC can be solved optimally
  by invoking an optimal algorithm $A_{\text{EC}}$ for the EC problem on
  $G_{\text{SB}}$. Note that the EC problem is NP-hard.


\ifLongversion
Currently, the best known $A_{\text{EC}}$ for general graphs are
from~\cite{BHK09}. One algorithm from \cite{BHK09} finds an edge coloring in
$O(2^{|E|} |E|^{O(1)})$ using exponential space. Another algorithm proposed in
\cite{BHK09} runs in $O(2.2461^{|E|})$ time using polynomial space. In
sections \labelcref{sec:alg-bip,sec:alg-general,sec:alg-sp} we discuss
cases where more efficient algorithms can be used.


The OJS Problem can be solved by iterating over all possible assignments of
$\pmb{z}$ such that \labelcref{eq:one-conf,eq:all-capacities} in the OJS
problem hold, solving JTC for each such assignment, and eventually returning
the assignment that maximizes the total utility.  \ifLongversion This idea is
presented in Algorithm \ref{alg:opt}.
\fi

\algrenewcommand\algorithmicthen{}
\algrenewcommand\algorithmicindent{1.5em}

\begin{varalgorithm}{OJS-OPT}
\caption{An optimal algorithm}
\label{alg:opt}
\begin{algorithmic}[1]
  \State{max$_z\gets nil$; max$_x\gets nil$; maxval$\gets -1$}
\Loop { over each assignment $z_{ir}$} \label{ln:all-z}
  \If{\parbox[t]{\dimexpr\linewidth-\algorithmicindent}{{\bf[} \labelcref{eq:one-conf,eq:all-capacities}
      hold and $U(f,z_{ir}) >$maxval{\bf]}}}
  \State{Solve JTC} \label{ln:run-sbap}
  \If{\parbox[t]{\dimexpr\linewidth-\algorithmicindent-\algorithmicindent}{{\bf[}
    feasible $x_{irs}$ are found{\bf]}}}
  \State {\parbox[t]{\dimexpr\linewidth-\algorithmicindent-\algorithmicindent-\algorithmicindent}{max$_z\gets z_{ir}$; max$_x\gets x_{irs}$;
    maxval$\gets U(f,z_{ir})$}}
  \EndIf
  \EndIf
\EndLoop
\State{Return max$_z$, max$_x$}
\end{algorithmic}
\end{varalgorithm}

The loop in line~\ref{ln:all-z} of Algorithm~\ref{alg:opt} has
$2^{|\mathcal{I}|  |\mathcal{C}|}$ iteration. Each iteration is dominated
by the running time of line~\ref{ln:run-sbap}. Therefore, the running time of
Algorithm~\ref{alg:opt} is $O(2^{|\mathcal{I}|  |\mathcal{C}|}
T_{\text{JTC}})$.
\fi

Table~\ref{table:algorithms_summary} summarizes the different options
we will describe in sections \ref{sec:alg-bip}-\ref{sec:alg-general} for
solving OJS using the decomposition framework.

\subsection{Algorithms for Bipartite Network Graphs} \label{sec:alg-bip}

We now develop algorithms for OJS instances in which $G_J$ is bipartite.  The results in this section can be used to solve such networks, and will provide the building blocks for the algorithms for general joint transmission graphs in Section \ref{sec:alg-general}. We
start by describing an algorithm for JTK and an algorithm for JTC, and show
how using them in the decomposition framework will result in an approximation
algorithm for OJS. We require the following two lemmas. Denote by $\Delta(G)$ the
  maximum vertex degree of $G$

\begin{lemma} \label{lem:bip}
  If $G_J$ is bipartite, then $G_{\text{SB}}(G_J,\pmb{z})$ is bipartite for every $\pmb{z}$.
\end{lemma}

\begin{lemma} \label{lem:bip-exists-x}
  If $G_{\text{SB}}$ is bipartite and $\Delta(G_{\text{SB}})\leq S$, then
  $\exists \pmb{x}$ such that
  $\eqref{eq:sufficient}, \eqref{eq:used-once}, \eqref{eq:integer_x} \text{ hold.}$
\end{lemma}

We now describe an algorithm for JTK based on the well-known Multidimensional
Multiple-choice Knapsack (MMK) Problem~\cite{Book:Kellerer04}. Observing the
formulation of MMK in~\cite{Book:Kellerer04}, the input to MMK is a subset of
the input to OJS. We define the algorithm $A_{\text{JTK}} = $\JTKMMK as simply running some
algorithm $A_{\text{MMK}}$ for solving MMK with a running time of
$T_{\text{MMK}}$ (for different $A_{\text{MMK}}$ algorithms see
Table~\ref{table:MMK}) and show that it solves JTK for bipartite networks.

\begin{lemma} \label{lem:bip-MMK}
  If $G_J$ is bipartite and algorithm $A_{\text{MMK}}$ is an $\alpha$-approximation
  algorithm for MMK,
  $A_{\text{JTK}} = A_{\text{MMK}}$ is an $\alpha$-approximation algorithm for JTK.
\end{lemma}

Next, we describe an algorithm for JTC when the network graph is bipartite, by exploiting the connection to graph-coloring problems from Lemma~\ref{lem:JTC-coloring}. Let
\JTCBIP be the edge coloring algorithm from~\cite{COS01}. Using
Lemma \ref{lem:bip}, $G_{\text{SB}}$ is bipartite and since also
$\Delta(G_{\text{SB}})\leq S$ it follows from~\cite{COS01} that \JTCBIP finds
an edge coloring using at most $S$ colors. Using Lemma \ref{lem:JTC-coloring}
we conclude that \JTCBIP solves JTC.
The running time of \JTCBIP is $O(|E_{\text{SB}}| \log \Delta(G_{\text{SB}}))= O(B S \log S)$.
The following theorem is the main result
of this section.

\begin{theorem} \label{th:framework}
  For bipartite networks, if $A_{\text{JTK}} = $\JTKMMK is an $\alpha$-approximation for JTK, then
$A_{\text{OJS}} = $\DF{\JTKMMK}{\JTCBIP} is an $\alpha$-approximation for OJS.
\end{theorem}

\ifTechRep
\else
\addtolength{\tabcolsep}{-.2em}
\begin{table}
\scriptsize
\centering
\caption{Algorithms for MMK \label{table:MMK}}
\begin{tabular}{|c|c|c|} \hline 
{\bf $A_{\text{MMK}}$} & {\bf Ratio} & {\bf $T_{\text{MMK}}(I,B,C,S)$} \\
\hline
DP~\cite{Book:Kellerer04} & Optimal & $O(S^{(B+C)} I (B+C))$ \\
\hline
PTAS~\cite{PR12} & $1/(1+\epsilon)$ & $O(I^{((B+C)/\epsilon)})$ \\ \hline
Greedy~\cite{Book:Kellerer04} & $\infty$ & $O(I \log(I))$ \\ \hline
\end{tabular}
\vspace*{-0.4 cm}
\normalsize
\end{table}
\addtolength{\tabcolsep}{.2em}
\fi

\subsection{Algorithm for Planar Series-Parallel Graphs} \label{sec:alg-sp}

We now develop an algorithm for OJS instances in which $G_J$ is planar and
series-parallel. We describe algorithms for JTK and JTC,
and use them in the decomposition framework to devise an
approximation algorithm for OJS. In this section we use similar ideas to those
in Section~\ref{sec:alg-bip}.  We need the following lemma whose proof is
similar to that of Lemma~\ref{lem:bip}:

\begin{lemma}\label{lem:sp}
  If $G_J$ is planar and series-parallel, $G_{\text{SB}}$
  is planar and series-parallel for every $\pmb{z}$.
\end{lemma}

We first describe Algorithm \JTKPSP that solves JTK when $G_J$ is planar and
series-parallel.  The algorithm uses $A_{\text{MMK}}$ to solve an MMK instance
defined as follows. The number of dimensions is
$D'=D+|\mathcal{B}_{\text{odd}}|$ where
$\mathcal{B}_{\text{odd}}=\{\mathcal{B}'\subseteq\mathcal{B}:\
|\mathcal{B}'|\text{ is odd and} \geq 3\}$. The capacity for each new
dimension associated with a set $\mathcal{B}'\in\mathcal{B}_{\text{odd}}$ is
$S(|\mathcal{B}'|-1)/2$.  The weight in each new dimension for each $(i,r)$
such that $r>0$ and $h(i,r)\subseteq \mathcal{B}'$ is set to $\Gamma(i,r)$;
for all other cases it is set to zero. The algorithm concludes by scheduling
packets for transmission according to the configurations selected in the
solution returned by $A_{\text{MMK}}$.

We note that $|\mathcal{B}_{\text{odd}}|=O(2^{B})$ and therefore in
general Algorithm \JTKPSP may be impractical due to a very large running
time. This algorithm is therefore more appropriate for small $B$. We now
show that in some instances the running time can be improved.  Note that if a
set $\mathcal{B}'\in \mathcal{B}_{\text{odd}}$ has no more than
$|\mathcal{B}'|-1$ edges in $\mathcal{C}$ that connects two nodes in $\mathcal{B}'$,
it can be removed from $\mathcal{B}_{\text{odd}}$; if the number of such sets is
large this can significantly decrease the running time.

For networks that are planar and series-parallel, JTC can be solved using the
edge-coloring algorithm from \cite{ZNSN92}. We call
this algorithm \JTCPSP, and note that its running time is $O(B
\Delta(G_{\text{SB}}))=O(B \cdot S)$.
The following theorem applies the decomposition framework to planar and
series-parallel networks.

\begin{theorem} \label{th:psp-framework}
  For planar and series-parallel networks, if \JTKMMK is an
  $\alpha$-approximation for MMK then \DF{\JTKPSP}{\JTCPSP} is an
  $\alpha$-approximation for OJS.


\end{theorem}

\subsection{Algorithms for General Graphs} \label{sec:alg-general}

We now develop algorithms for general OJS instances, without imposing any conditions on $G_J$. We start with describing two
approximation algorithms for JTK. For each approximation algorithm we show how
using it in the decomposition framework will result in an approximation
algorithm for OJS.

First, we describe Algorithm \JTKMAT which is based on computing a
matching. For each $\{a,b\}\in \mathcal{C}$, the algorithm solves
an instance of JTK defined by a network that has only two BSs $a$,$b$ and the
backhaul link with capacity
\ifTechRep
$l_{ab}$. Only pairs $(i,r)$ which are relevant in
\else
$K$. Only packets that can be scheduled in
\fi
such network are considered, and $A_{\text{MMK}}$ with
\ifTechRep
$D=3$ is used to solve this limited instance.
\else
$B=2$ and $C=1$ is used to solve this limited instance.
\fi
Each edge in $\mathcal{C}$ is assigned a weight equal to the
total utility obtained when solving its limited JTK instance. Then, maximum
weighted matching is found and the union of all solutions for edges in the
matching is returned. This solution is feasible for the general JTK problem.

\ifTechRep
\begin{varalgorithm}{\JTKMAT}
\caption{Based on matching}
\label{alg:JTK-MAT}
\begin{algorithmic}[1]
\For{$b\in \mathcal{B}$}
\State\parbox[t]{\dimexpr\linewidth-\algorithmicindent} {If $b$ has no
  backhaul link, run $A_{\text{MMK}}$ to solve a JTK instance with
$\mathcal{B}'=\{b\}$, $\mathcal{I}'=\{i\in\mathcal{I}:\ h(i)=\mathcal{B}'\}$,
$\mathcal{C}'=\{\}$ and set $z_{ir}$ as determined by $A_{\text{MMK}}$.} \label{ln:leaf-node}
\EndFor
  \For{$e=\{a,b\} \in \mathcal{C}$}
  \State \parbox[t]{\dimexpr\linewidth-\algorithmicindent} {Run
    $A_{\text{MMK}}$ to solve a JTK instance with $\mathcal{B}'=\{a,b\}$, $\mathcal{I}'=
    \{i\in\mathcal{I}:\ \exists r\in\mathcal{R},\ h(i,r)\subseteq\mathcal{B}'\}$} \label{ln:MMK}
  \State \parbox[t]{\dimexpr\linewidth-\algorithmicindent}{Assign
    $U(\pmb{z})$ found by $A_{\text{MMK}}$ as a weight for $e$ \label{ln:edge-utility}}
\EndFor
\State{Compute maximum weight matching on $G_J$ and store the result in the edge set
  $\mathcal{E}$ \label{ln:max-matching}}
\State{Set $z_{ir}=1$ only if $\exists e\in\mathcal{E}$
  such that $z_{ir}=1$ in the solution returned in line~\ref{ln:MMK} for
  edge $e$} \label{ln:ret-mat-sol}
\end{algorithmic}
\end{varalgorithm}

\else
\begin{varalgorithm}{\JTKMAT}
\caption{Based on matching}
\label{alg:JTK-MAT}
\begin{algorithmic}[1]
\For{$b\in \mathcal{B}$}
\State\parbox[t]{\dimexpr\linewidth-\algorithmicindent} {If $b$ has no
  backhaul link, run $A_{\text{MMK}}$ to solve a JTK instance with
$\mathcal{B}'=\{b\}$, $\mathcal{I}'=\{i\in\mathcal{I}:\ h(i)=\mathcal{B}'\}$,
$\mathcal{C}'=\{\}$ and set $z_{i}$ and $y_i$ as determined by $A_{\text{MMK}}$.} \label{ln:leaf-node}
\EndFor
  \For{$e=\{a,b\} \in \mathcal{C}$}
  \State {Run
    $A_{\text{MMK}}$ to solve a JTK instance with $\mathcal{B}'=\{a,b\}$, $\mathcal{I}'=
    \{i\in\mathcal{I}:\ h(i)\subseteq\mathcal{B}'\}$, $\mathcal{C}'=\{e\}$} \label{ln:MMK}

  \State \parbox[t]{\dimexpr\linewidth-\algorithmicindent}{Assign
    $U(\pmb{z},\pmb{y})$ found by $A_{\text{MMK}}$ as a weight for $e$ \label{ln:edge-utility}}
\EndFor
\State{Compute maximum weight matching on $G_J$ and store the result in the edge set
  $\mathcal{E}$ \label{ln:max-matching}}
\State{Set $z_{i}=1$ ($y_i=1$) only if $\exists e\in\mathcal{E}$
  such that $z_{i}=1$ ($y_i=1$) in the solution returned in line~\ref{ln:MMK} for
  edge $e$} \label{ln:ret-mat-sol}
\end{algorithmic}
\end{varalgorithm}
\fi

\begin{theorem} \label{th:JTK-MAT-approx}
  If algorithm $A_{MMK}$ is an $\alpha$-approximation algorithm, then
$A_{OJS} = $\DF{\JTKMAT($A_{MMK}$)}{\JTCBIP}
is a $(2  \alpha)/
  (3  \Delta(G_J))$-approximation algorithm for OJS.
\end{theorem}

The maximum weight matching algorithm from~\cite{GMG86} that takes $O(|E|
|V|\log |V|)$ time can be used in line~\ref{ln:max-matching}; in this case the
running time of Algorithm \JTKMAT is dominated by $A_{\text{MMK}}$ in
line~\ref{ln:MMK}. Therefore, the running time of Algorithm \JTKMAT is
\ifTechRep
$O(C\cdot T_{\text{MMK}}(I,R,3,S))$,
where $T_{\text{MMK}}(I, R, D, S)$ is the
running time of $A_{\text{MMK}}$.
\else
$O(C T_{\text{MMK}}(I,2,1,S))$,
where $T_{\text{MMK}}$ is the
running time of $A_{\text{MMK}}$ (see Table~\ref{table:MMK}).
\fi

We now describe Algorithm \JTKSTA which iterates over star subgraphs.
It is similar to Algorithm \JTKMAT, but iterates over the vertices $b\in \mathcal{B}$
instead of the edges $e \in \mathcal{C}$.
The approximation ratio of \JTKSTA is better than that of \JTKMAT, but its
running time is worse.
For each
vertex $b$, a JTK instance is constructed using only $b$ and its
neighbors in $G_J$. The solution of this instance,
\ifTechRep
$\pmb{z}$,
\else
$\pmb{z},\pmb{y}$,
\fi
is assigned to $b$. Next,
the algorithm finds the vertex $b_{\max}$ associated with maximum total
utility. If
\ifTechRep
$\pmb{z}_{\max}$
\else
$\pmb{z}_{\max}, \pmb{y}_{\max}$
\fi
 is the solution associated with $b_{\max}$, the
algorithm schedules the packets indicated by
\ifTechRep
$\pmb{z}_{\max}$.
\else
$\pmb{z}_{\max},\pmb{y}_{\max}$.
\fi
 The vertex
$b_{\max}$ and its neighbors are removed from $\mathcal{B}$. This process is repeated
until $\mathcal{B}$ is empty.




Note that after the first vertex is removed from $\mathcal{B}$,
in order to update the
weights it is sufficient to consider 2-hop neighbors of $b_{\max}$ in
line~\ref{ln:update_weight},
since weights of other vertices remain unchanged.
The running time of \JTKSTA is
\ifTechRep
$O(B^2\cdot T_{\text{MMK}}(I,R,2\Delta(G_J)+1,S) )$.
\else
$O(
T_{\text{MMK}}(I,\Delta(G_J)+1,\Delta(G_J),S) B^2)$.
\fi

The following theorem proves that OJS can be solved approximately when \JTKSTA
is used in the decomposition framework.
\begin{theorem} \label{th:JTK-STA-approx}
  If algorithm $A_{\text{MMK}}$ is an $\alpha$-approximation algorithm, then,
$A_{OJS} = $\DF{\JTKSTA($A_{MMK}$)}{\JTCBIP}
is an $(\alpha/\Delta(G_J))$-approximation algorithm for OJS.
\end{theorem}

\ifTechRep
\begin{varalgorithm}{\JTKSTA}
\caption{Based on star subgraphs}
\label{alg:JTK-STA}
\begin{algorithmic}[1]
\Function{SOL-STAR}{$b$, $\mathcal{B}'$, $\mathcal{C}'$, $\mathcal{I}'$}
  \State \parbox[t]{\dimexpr\linewidth-\algorithmicindent}{Run
    $A_{\text{MMK}}$ to solve a JTK instance defined with
    $\widetilde{\mathcal{B}'}=\{b\}\cup\{a:\ \{a,b\}\in \mathcal{C}'\}$, $\widetilde{\mathcal{C}'}=\{
    \{a,b\}:\ \{a,b\}\in \mathcal{C}',\ a\in \widetilde{\mathcal{B}'},\ b\in \widetilde{\mathcal{B}'}\}$,
$\widetilde{\mathcal{I}'}=
    \{i\in\mathcal{I}':\ h(i)\subseteq\widetilde{\mathcal{B}'}\}$} \label{ln:STA-DMCKP}
  \Return $\pmb{z}$ as determined by $A_{\text{MMK}}$
\EndFunction

  \For{$b \in \mathcal{B}$}
  \State {$\pmb{Z}[b] \gets$ SOL-STAR($b$, $\mathcal{B}$,
    $\mathcal{C}$, $\mathcal{I}$)}
  \State \parbox[t]{\dimexpr\linewidth-\algorithmicindent}{Assign
    $U(\pmb{Z}[b])$ as a weight for $b$ in
    $\mathcal{B}$ \label{ln:vertex-utility}}
\EndFor
\State{Initialize $\mathcal{B}''\gets \mathcal{B}$; $\mathcal{C}''\gets
  \mathcal{C}$; $\mathcal{I}''\gets
\mathcal{I}$}
\Repeat \label{ln:repeat-loop}
\State \parbox[t]{\dimexpr\linewidth-\algorithmicindent}{Find the vertex $b_{\max}$ in $\mathcal{B}''$ with maximum weight}

\State {$\widetilde{\mathcal{I}}\gets
  \{i\in\mathcal{I}'': \ \exists r,\  \pmb{Z}[b_{\max}]_{ir}=1\}$} \label{ln:find-tilde-I}
\State{for all $i\in \tilde{\mathcal{I}}$, $z''_{ir}\gets \pmb{Z}[b_{\max}]_{ir}$} \label{ln:set-z}
\State\parbox[t]{\dimexpr\linewidth-\algorithmicindent}{$\mathcal{J}\gets \{
  a\in \mathcal{B}'':\exists b\in \mathcal{B}'',\{b_{\max},b\}\in \mathcal{C}'',\{a,b\}\in \mathcal{C}''\}$}
\State{$\mathcal{I}'' \gets \mathcal{I}''\setminus \tilde{\mathcal{I}}$;
  $\mathcal{B}'' \gets \mathcal{B}''\setminus \{a:\ \{a,b_{\max}\}\in
  \mathcal{C}''\}$} \label{ln:update-I''} \label{ln:remove-neighbors}
\State{Remove from $\mathcal{C}''$ edges with an endpoint not in $\mathcal{B}''$}
\For{$b\in \mathcal{J}$} \label{ln:update_weight}
\State{$\pmb{Z}[b] \gets$ SOL-STAR($b$, $\mathcal{B}''$,
  $\mathcal{C}''$, $\mathcal{I}''$)}
\State{Update $U(\pmb{Z}[b])$ as a weight for $b$ in
  $\mathcal{B}''$}
\EndFor

\Until{$\mathcal{B}''$ is empty \label{ln:empty-vtag}}
\State{\text{{\bf return} }$\pmb{z}''$} \label{ln:return-z''}
\end{algorithmic}
\end{varalgorithm}

\else
\begin{varalgorithm}{\JTKSTA}
\caption{Based on star subgraphs}
\label{alg:JTK-STA}
\begin{algorithmic}[1]
\Function{SOL-STAR}{$b$, $\mathcal{B}'$, $\mathcal{C}'$, $\mathcal{I}'$}
  \State \parbox[t]{\dimexpr\linewidth-\algorithmicindent}{Run
    $A_{\text{MMK}}$ to solve a JTK instance defined with
    $\widetilde{\mathcal{B}'}=\{b\}\cup\{a:\ \{a,b\}\in \mathcal{C}'\}$, $\widetilde{\mathcal{C}'}=\{
    \{a,b\}:\ \{a,b\}\in \mathcal{C}',\ a\in \widetilde{\mathcal{B}'},\ b\in \widetilde{\mathcal{B}'}\}$,
$\widetilde{\mathcal{I}'}=
    \{i\in\mathcal{I}':\ h(i)\subseteq\widetilde{\mathcal{B}'}\}$} \label{ln:STA-DMCKP}
  \Return $\pmb{z},\pmb{y}$ as determined by $A_{\text{MMK}}$
\EndFunction

  \For{$b \in \mathcal{B}$}
  \State {$\pmb{Z}[b],\pmb{Y}[b] \gets$ SOL-STAR($b$, $\mathcal{B}$,
    $\mathcal{C}$, $\mathcal{I}$)}
  \State \parbox[t]{\dimexpr\linewidth-\algorithmicindent}{Assign
    $U(\pmb{Z}[b],\pmb{Y}[b])$ as a weight for $b$ in
    $\mathcal{B}$ \label{ln:vertex-utility}}
\EndFor
\State{Initialize $\mathcal{B}''\gets \mathcal{B}$; $\mathcal{C}''\gets
  \mathcal{C}$; $\mathcal{I}''\gets
\mathcal{I}$}
\Repeat \label{ln:repeat-loop}
\State \parbox[t]{\dimexpr\linewidth-\algorithmicindent}{Find the vertex $b_{\max}$ in $\mathcal{B}''$ with maximum weight}

\State {$\widetilde{\mathcal{I}}\gets
  \{i\in\mathcal{I}'': \ \pmb{Z}[b_{\max}]_i+\pmb{Y}[b_{\max}]_i=1\}$} \label{ln:find-tilde-I}
\State{for all $i\in \tilde{\mathcal{I}}$, $z''_{i}\gets \pmb{Z}[b_{\max}]_{i}$, $y''_i \gets \pmb{y}[b_{\max}]_i$} \label{ln:set-z}
\State\parbox[t]{\dimexpr\linewidth-\algorithmicindent}{$\mathcal{J}\gets \{
  a\in \mathcal{B}'':\exists b\in \mathcal{B}'',\{b_{\max},b\}\in \mathcal{C}'',\{a,b\}\in \mathcal{C}''\}$}
\State{$\mathcal{I}'' \gets \mathcal{I}''\setminus \tilde{\mathcal{I}}$;
  $\mathcal{B}'' \gets \mathcal{B}''\setminus \{a:\ \{a,b_{\max}\}\in
  \mathcal{C}''\}$} \label{ln:update-I''} \label{ln:remove-neighbors}
\State{Remove from $\mathcal{C}''$ edges with an endpoint not in $\mathcal{B}''$}
\For{$b\in \mathcal{J}$} \label{ln:update_weight}
\State{$\pmb{Z}[b],\pmb{Y}[b] \gets$ SOL-STAR($b$, $\mathcal{B}''$,
  $\mathcal{C}''$, $\mathcal{I}''$)}
\State{Update $U(\pmb{Z}[b],\pmb{Y}[b])$ as a weight for $b$ in
  $\mathcal{B}''$}
\EndFor

\Until{$\mathcal{B}''$ is empty \label{ln:empty-vtag}}
\State{\text{{\bf return} }$\pmb{z}''$} \label{ln:return-z''}
\end{algorithmic}
\end{varalgorithm}

\fi








\section{Queueing Dynamics}
\label{sec:queueing_intro}
The OJS problem discussed in Sections~\ref{sec:hardness_results} and~\ref{sec:algorithms} 
schedules
transmissions within a single subframe. We now expand the
scope to multiple subframes, where packets arrive and depart over time, and study the
evolution of the users' queues.  Our objective is to identify a scheduling
policy that is maximum stable (or throughput optimal). Under
such a policy, the queue-length process is positive recurrent for any arrival
for which a stabilizing policy exists, see,
e.g.,~\cite{MAW96,TE92,andrews2004scheduling,ESP05}.
We prove that by using a specific utility function and an algorithm for
solving the OJS problem (we refer to this combination as a scheduling policy), we obtain a
MaxWeight-like scheduling policy (see, e.g., \cite{TE92}), which is throughput
optimal.

\ifTechRep to simplify the notations in this section, we define the indicator
variables $\pmb{y}$ where $y_i=1$ indicates packet $i$ is forwarded over the
backhaul ($z_{i0}=1$) and $y_i=0$ otherwise ($z_{i0}=0$). Also, we assume that
a single MCS is available, namely, $M=1$ and $R=2$. Therefore, the notation is
further simplified by using $z_i$ instead of $z_{ir}$ to indicate packet $i$
is wirelessly transmitted.  \fi

Let $\bfL(t) = (L_1(t),\widehat{L}_1(t),\dots,L_n(t),\widehat{L}_n(t))$ denote the queue length process at time $t$, so $\{\bfL(t)\}_{t\ge0}$ is the stochastic process that tracks the queue-length evolution over time. We solve the OJS problem in each subframe, given a certain utility function.
Let us denote by $\pmb{y}(t) = (y_{i}(t))_{i\in\mathcal{I}}$ and $\pmb{z}(t) = (z_{i}(t))_{i\in\mathcal{I}}$ the solution of OJS in slot $t$. Here $(\pmb{y}(t),\pmb{z}(t))$ can represent both an exact solution or an approximation.

Recall that $W_n(t)$ is the number of packets generated for user~$n$ at time~$t$. Let $\lambda_n
= \expect{W_n(0)}$, and define $\bflambda = (\lambda_1,\dots,\lambda_N)$ to be the arrival rates. We
denote by $\mu_n^{(1)}(t;\pmb{z}(t))$, $\mu^{(2)}_n(t;\pmb{z}(t))$, and
$\mu_n^{(3)}(t;\pmb{y}(t))$ the number of packets transmitted towards user $n$ in
subframe $t$ using single and joint transmission, and the number of packets forwarded across the backhaul, respectively, given solution $(\pmb{y}(t),\pmb{z}(t))$ of OJS.

Denote $\mathcal{I}_n(t)$ the set of packets in $Q_n$ at time $t$ (so $|\mathcal{I}_n(t)| = L_n(t)$), and $\widehat{\mathcal{I}}_n(t)$ the set of packets in $\widehat{Q}_n$. The $\mu^{(j)}_n$, $j=1,2,3$ can be written as
\begin{equation*}
\mu^{(1)}_n(t;\pmb{z}) = \sum_{i \in \mathcal{I}_n(t)} z_{i}(t) Y_{it},\quad
\mu^{(2)}_n(t;\pmb{z}) = \sum_{i \in \widehat{\mathcal{I}}_n(t)} z_{i}(t) Y_{it}, \quad
\mu^{(3)}_n(t;\pmb{y}) = \sum_{i \in \mathcal{I}_n} y_{i}(t),
\end{equation*}
where the $Y_{it}\sim {\rm Ber}(p(i))$
are mutually independent Bernoulli distributed random variables that
represent whether packet transmissions are successful. For notational
convenience, we write $\mu^{(j)}_n(t)$ to represent the transmission rates at time $t$.

It is readily seen that the joint queue-length process $\{\bfL(t)\}_{t\ge0}$ is Markovian.

We now analyze the traffic intensity that can be sustained by the queueing system described
by~\eqref{eqn:evolution_1} and~\eqref{eqn:evolution_2}.  The
stability region of a particular policy is defined as the set of all arrival
rates such that the $\{\bfL(t)\}_{t\ge0}$ is positive recurrent is called the \emph{stability region} of this particular
policy. The \emph{capacity region} of a network is defined as the union of the
stability region over all policies.
If the stability region of an algorithm OJS-ALG and utility function $u$ is equal to the capacity region, we say that policy (OJS-ALG,$u$) is throughput-optimal.

In order to investigate the network capacity region in more detail, we first
introduce some definitions.
We denote by
$f_n^{(1)}$ the rate (long-term average traffic flow) of single-transmission
packets for user $n$, by $f_n^{(2)}$ the rate of joint-transmission traffic for
user $n$, and by $f_n^{(3)}$ the rate of user-$n$ traffic sent across
the backhaul. Define the vector $\bff =
(f_1^{(1)},f_1^{(2)},f_1^{(3)},\dots,f_N^{(1)},f_N^{(2)},f_N^{(3)})$. Then, for a given arrival rate vector $\bflambda$, the set of all $\bflambda$-admissible traffic flows can be defined as
\begin{equation}\label{flow_conditions}
F_{\bflambda} = \Big\{\bff \in \mathbb{R}_+^{3 N}\ \Big\rvert \ \lambda_n = f_n^{(1)} + f_n^{(3)},\ f_n^{(3)} = f_n^{(2)},\ n\in\mathcal{N} \Big\}.
\end{equation}
That is, a flow is $\bflambda$-admissible if for all queues $Q_n$,
$\widehat{Q}_n$, $n=1\dots,N$, the traffic arrival rate is equal to the
departure rate.


We now introduce 
the set of all arrival rate vectors such that at
least one $\bflambda$-admissible flow can be realized:
\begin{equation*}
\Lambda = \Big\{ \bflambda \in \mathbb{R}_+^{3N}\mid \exists \bff \in F_{\bflambda}\ \exists \bfr \in {\rm conv}(R)
 f_n^{(j)} < r_n^{(j)} {\rm ~if~} f_n^{(j)} > 0,\ n \in\mathcal{N},\ j = 1,2,3  \Big\},
\end{equation*}
where ${\rm conv}(R)$ denotes the convex hull of $R$, the set of all  rates
across the various links that can be achieved in saturation.


\begin{figure*}
\begin{minipage}[]{0.58\linewidth}
\subfigure[$3$-cluster]{\label{fig:3cluster}\includegraphics[scale=0.18,bb=0 0 400 400]{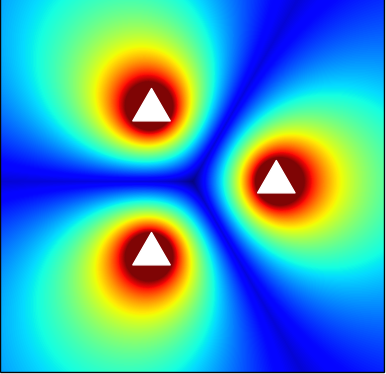}\includegraphics[scale=0.18,bb=0 0 100 0]{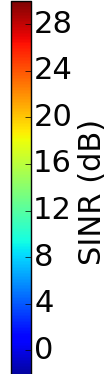}}
\subfigure[Cycle topology]{\label{fig:7cluster_cycle}\includegraphics[width=0.32\columnwidth,bb=0 0 500 500]{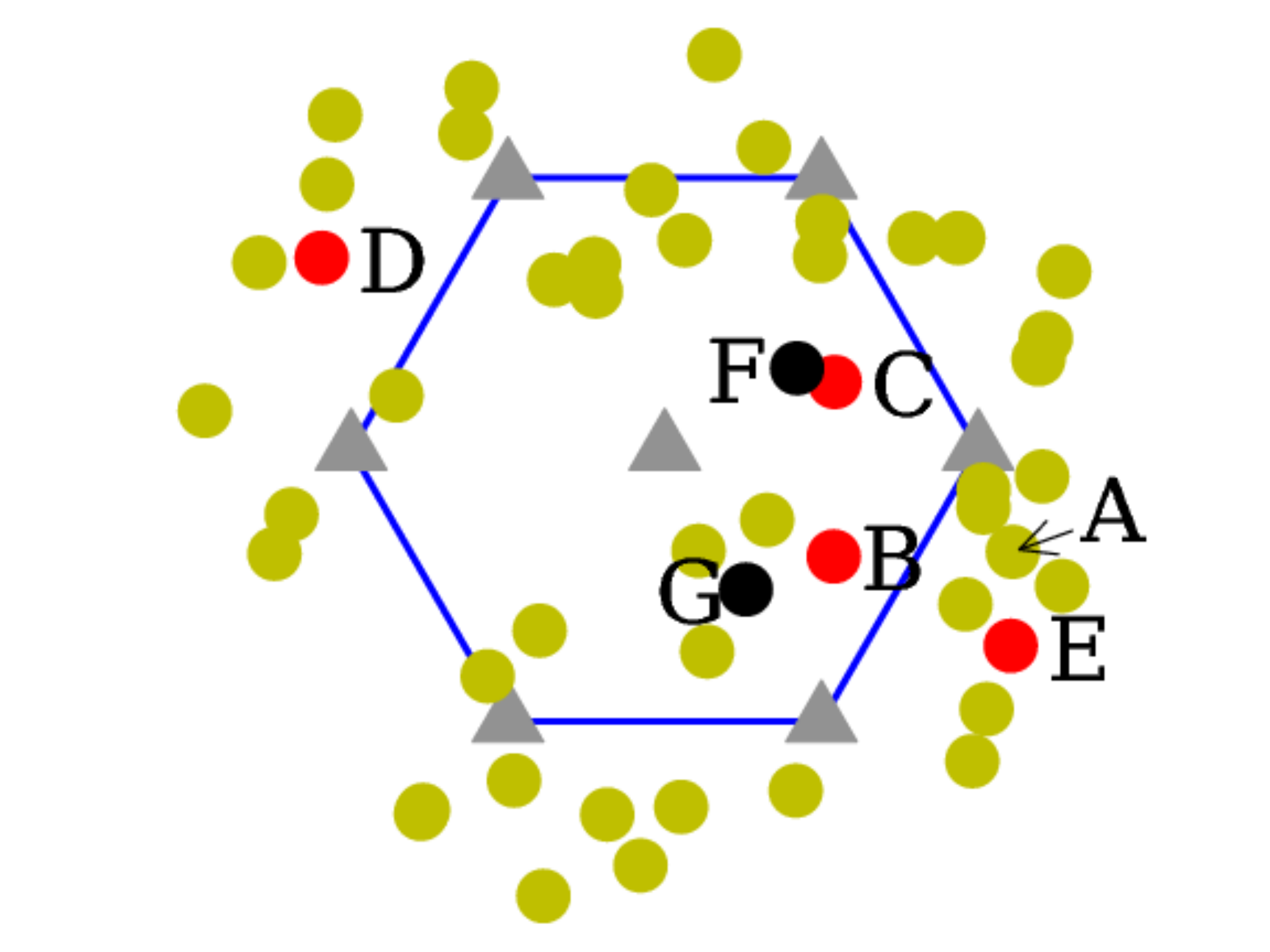}}
\subfigure[Star topology]{\label{fig:7cluster_star}\includegraphics[width=0.32\columnwidth,bb=0 0 500 500]{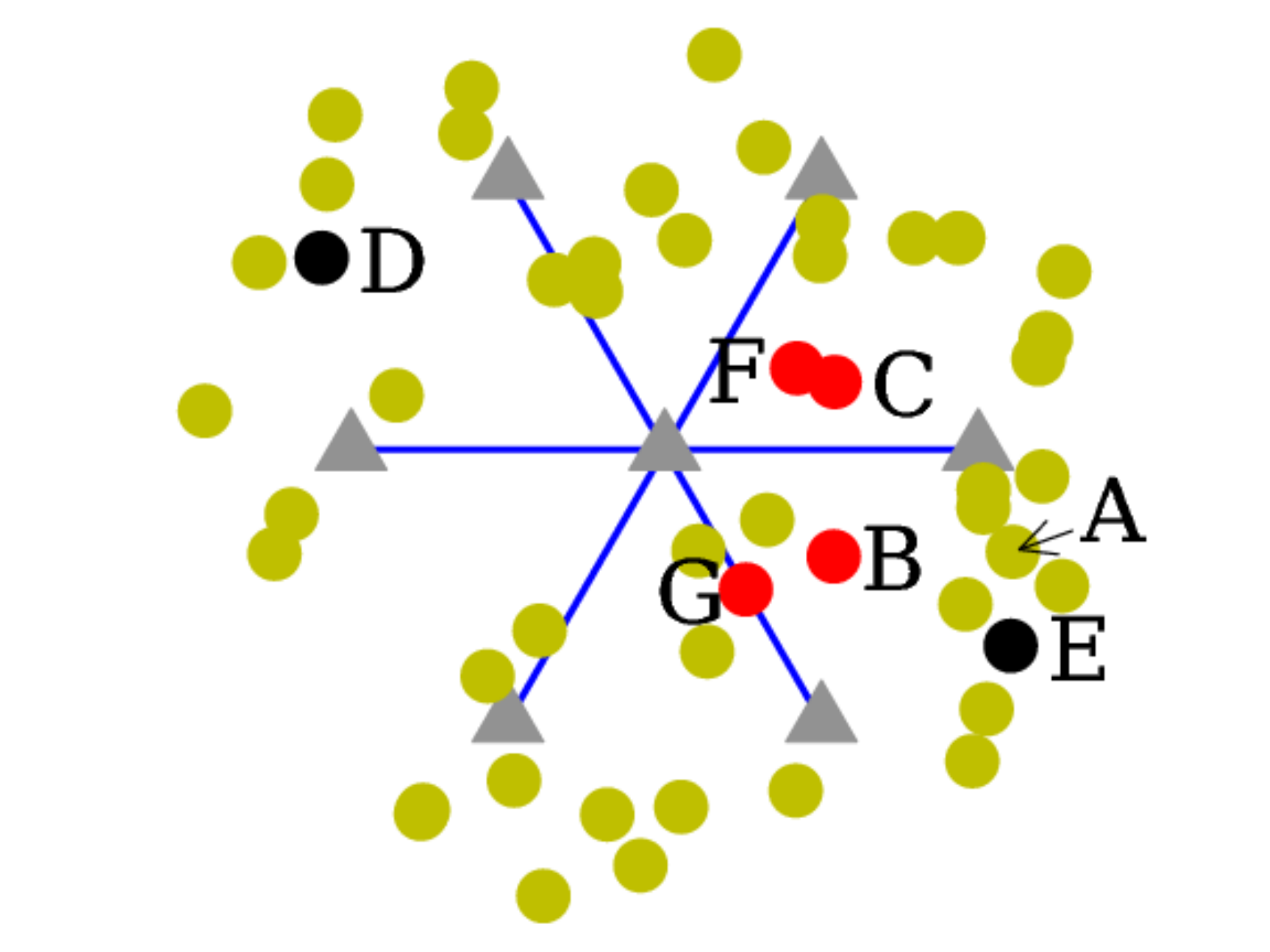}}
\caption{Simulated network topologies. \subref{fig:3cluster} SINR for the cluster of
3 BSs. The dark blue area denotes the location of inter-cell users.
\subref{fig:7cluster_cycle}-\subref{fig:7cluster_star} Two 7 BS topologies.
The color of the nodes illustrate the throughput gains of users for a sample
run. Red users benefit from JT and the throughput
of yellow and black users remains roughly the same. Black users have
throughput of 0 both with and without JT.
}
\label{fig:user_heatmap}
\end{minipage}\hspace*{.03\linewidth}\begin{minipage}[]{0.39\linewidth}
\centering
\subfigure[\vspace*{-20pt}]{\includegraphics[scale=0.25,bb=0 0 350 350]{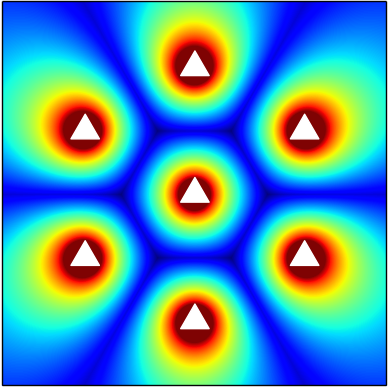}}
\subfigure[\vspace*{-20pt}\quad\quad\quad]{\includegraphics[scale=0.25,bb=0 0 350 350]{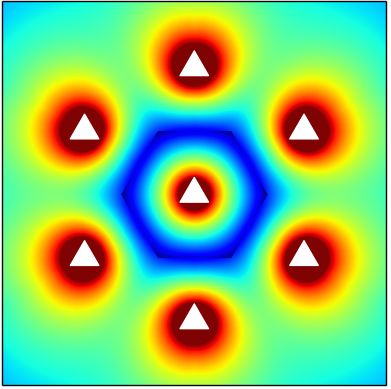}\includegraphics[scale=0.188,bb=0 0 10 0]{fig/hist_3bs_single_colorbar.png}}
\caption{SINR for the cycle topology for (a) single-transmissions and (b) joint-transmissions.}
\vspace*{15pt}
\label{fig:SINR_cycle_heatmap}
\end{minipage}
\vspace*{-10pt}
\end{figure*}




We now show that any $\bflambda \in \Lambda$ can be stabilized, and that any $\bflambda$ outside of the closure of
$\Lambda$ cannot. For stabilizing $\bflambda \in \Lambda$ we use the policy $($OJS-OPT,$u_Q)$,
where OJS-OPT represents any algorithm that solves OJS exactly,
and $u_Q$ is the queue-length based utility function from~\eqref{eqn:queue_based_utility}. The following theorem then implies that
$\Lambda$ is indeed the capacity region, and that    $($OJS-OPT,$u_Q)$ is
throughput-optimal. The proof relies on a standard drift argument using a quadratic Lyapunov function~\cite{TE92}.

\begin{theorem}\label{thm:stability}
Let $\bflambda \in \Lambda$, then the queue-length process is stable under policy $($OJS-OPT,$u_Q)$. If $\bflambda \not\in \bar{\Lambda}$, then there exists no policy that stabilizes the network.
\end{theorem}

The reason for choosing the queue-length based utility
function~\eqref{eqn:queue_based_utility} becomes
clear when substituting it into the objective function of OJS.
\begin{align}
 U(\pmb{z})
=   \sum_{n=1}^N \Big(L_n(t) \expect{\mu_n^{(1)}(t)} + \widehat{L}_n(t)
\expect{\mu_n^{(3)}(t)} + \big(L_n(t)-\widehat{L}_n(t)\big) \mu_n^{(2)}(t)\Big). \label{eqn:MWS}
\end{align}
This yields the objective function of the celebrated
MaxWeight scheduling algorithm.
This algorithm, first introduced
in~\cite{TE92}, has been shown to be throughput-optimal in a wide range of
settings, see, e.g.,~\cite{TE92,MAW96}. 
Note that although our objective function of maximizing the queue-weighted
throughput is similar to that used in traditional work on MaxWeight
scheduling, the constraints of the OJS problem are markedly different. Specifically, the MaxWeight scheduling literature is typically concerned with maximum (weighted) set problems, which are
fundamentally different from the OJS problem.

Since the maximization for utility
function~\eqref{eqn:queue_based_utility} is a specific instance of the OJS
problem, it follows from Proposition~\ref{pro:NPhard-coloring} that solving this problem is
NP-hard. Thus, using an optimal algorithm OJS-OPT (as in Theorem~\ref{thm:stability}) is typically
not feasible in practice for general graphs.
In Section~\ref{sec:simulation} we investigate the performance of
a wider set of (suboptimal) algorithms via simulation. Theoretical results
on the capacity loss for general algorithms will be the subject of future
work.



\section{Numerical Results}
\label{sec:simulation}
\newcommand{\attn}[1]{{\color{red}#1}}
We conducted a simulation study to evaluate the performance of the various
algorithms introduced in Section~\ref{sec:algorithms}. Throughout this section
we consider the case where a packet can be transmitted
using one of several Modulation and Coding Schemes (MCSs); Details on
extending the
algorithms to support several MCSs are given in
\ifTechRepFinal
Appendix~\ref{sec:extension}.
\else
~\cite{GBVZ15}.
\fi
The simulation results
provide insights on the network-level benefits and tradeoffs of JT
under various network scenarios.


\subsection{Simulation Setup}

{\bf OJS algorithm.} We implemented the four algorithms presented in
Table~\ref{table:algorithms_summary}.  The majority of the algorithms and the
queueing dynamics are implemented in Python, while the $A_{\text{MMK}}$
procedure is written in C. 
We did not implement any JTC algorithms, since it was proved in
Section~\ref{sec:algorithms} that JTK guarantees that a feasible solution always
exists. Furthermore, as we show in Section~\ref{sec:perf_approx_algo}, the greedy algorithm shown in
Table~\ref{table:MMK} performs well in most considered scenarios.


\ifLongversion
Below, we describe the
various network parameters used, and explain our evaluation methodology
for a single subframe (Section~\ref{sec:algorithms}) and for multiple
subframes (Section~\ref{sec:queueing_intro}).
\fi

{\bf Network setup.}
We consider three network topologies.  In
Section~\ref{sec:impact_backhaul_capacity}, we analyze a network of 3 BSs,
with backhaul links between each pair (Fig.~\ref{fig:3cluster}). In
Section~\ref{sec:impact_topology}, we
look at two different backhaul topologies for a 7 BS network,
shown in Fig.~\ref{fig:7cluster_cycle}-\ref{fig:7cluster_star}. The backhaul
capacity (BC) is the same for all links.

We use a fixed packet size of
\unit[73]{bytes}, and the backhaul capacities are normalized to units of
\unit{packets/subframe}. The distance between neighboring BSs is
\unit[700]{m}. The height of each BS's antenna is \unit[20]{m}. The BSs'
transmit power is \unit[39]{dBm} and \unit[30]{dBm} for the 3 and 7 BSs
network, respectively. Lower transmission power is used for the larger network,
since more BSs transmit interfering signals.

We simulate $N=20$ users for the 3-cluster, and $N=50$ for the 7-cluster
topologies.  The users are placed uniformly within a circle that contains the
entire simulation area, and with radius \unit[1050]{m}.

{\bf Wireless model.}
We set $S = 50$ scheduled blocks, corresponding to a 10MHz LTE
system.
Once the location for a user is determined, the received power level
from each BS is computed based on the Hata propagation
model~\cite{hatay1980empirical} which was shown suitable for LTE in urban areas~\cite{NAOA11}.
The power levels from the different BSs are
used to compute the SINR for single and joint transmission to the user.
The SINR values for single and joint transmissions for the cycle topology are plotted in
Fig.~\ref{fig:SINR_cycle_heatmap}.
Given the SINR values, the success probability for single and joint transmission
is computed for each MCS (QPSK-1/2, QAM64-1/2, and QAM64-3/4) using data taken
from~\cite{CCJKKRRSS07}.

{\bf Queueing Dynamics.} The queueing dynamics are implemented as in
Section~\ref{sec:queueing_intro}. Unless otherwise noted, packet arrivals for
users follow a binomial distribution with $n=3$ and $p=0.5$.
In every subframe, an algorithm
for solving OJS is executed with the utility
function~\eqref{eqn:queue_based_utility}.
Throughout the simulation, we track the \emph{normalized throughput} of a user,
defined as the fraction of arrived packets that have been
successfully transmitted.
The average normalized throughput is computed over all users. The simulation
duration is 1,000 subframes and each data point is obtained by averaging
over 1,000 runs.

We also distinguish the performance of \emph{inter}-cell users from that of
the \emph{intra}-cell users in order to evaluate the benefits of CoMP JT for
both types. In this section, we refer to inter-cell users as those whose
power levels from two BSs is above a threshold. This threshold is determined numerically by
observing the physical location of these users (illustrated by the darker
regions of Fig. \ref{fig:3cluster}).


\subsection{Simulation Results}
\label{sub:simulation_results}
\ifLongversion
We first evaluate the performance of the different algorithms for a single
subframe, and then simulate the queueing system to reach conclusions regarding the
benefits of CoMP.
\fi

\subsubsection{Performance of the Approximation Algorithms}\label{sec:perf_approx_algo}
In practice, approximation algorithms may perform significantly better than
their guaranteed approximation ratios.
Hence, we carry out a single-subframe evaluation with the goal of
isolating the performance of the algorithms from the long-term
effects of the queueing dynamics.
We consider the OJS formulation with the throughput utility
function~\eqref{eqn:throughput_based_utility}. We consider all algorithms, for
two different topologies with 3 BSs: complete graph and a bipartite graph.

\begin{figure}
\centering
\includegraphics[width=0.5\columnwidth]{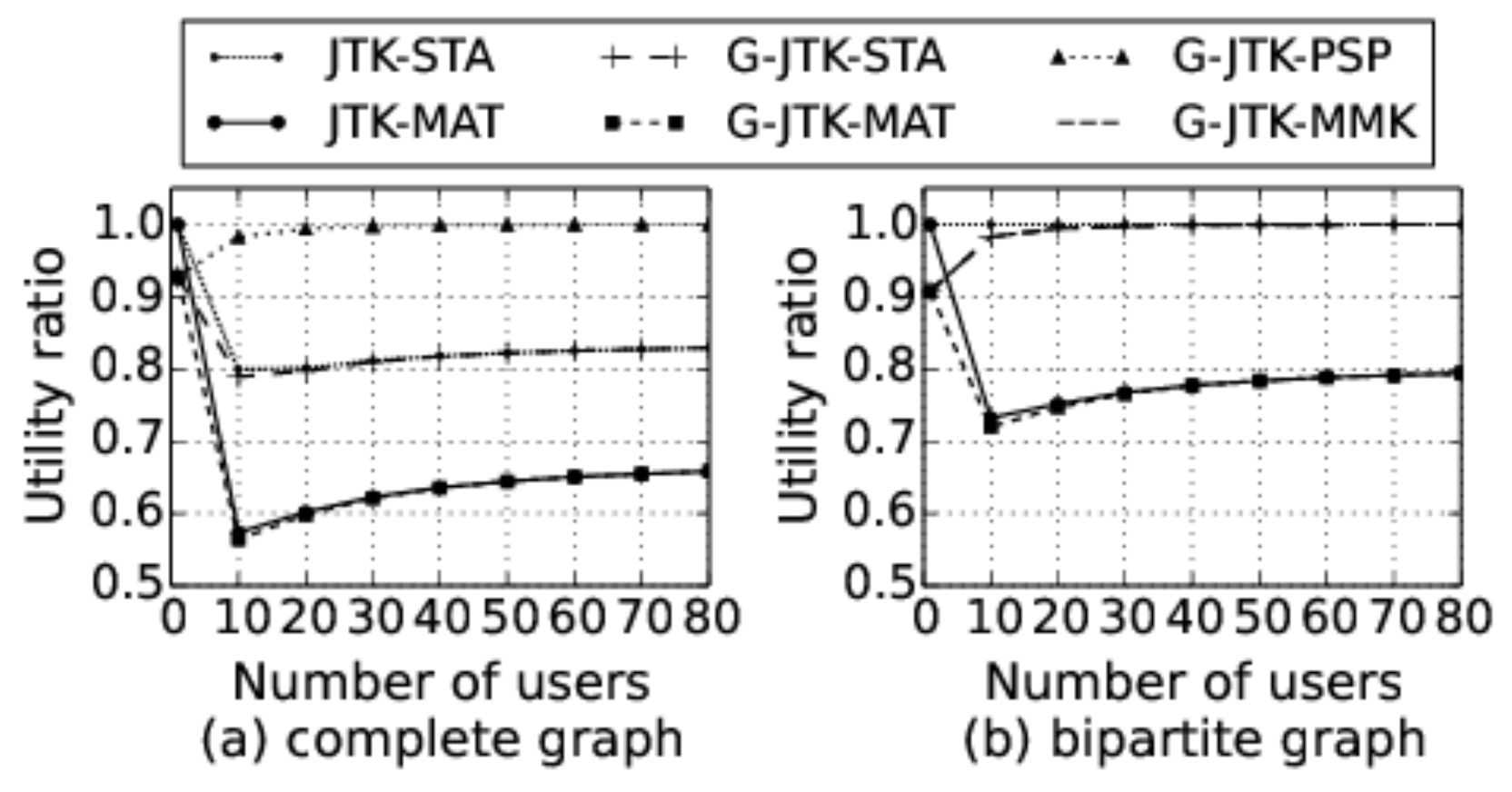}
\caption{The ratio between the optimal utility for the OJS problem for a
  single subframe, and the utility obtained by the different algorithms
under the two topologies.}
\label{fig:online_users}
\end{figure}

When an optimal $A_{\mathrm{MMK}}$ subroutine is used, \JTKMMK and \SPJCON are
optimal under the complete graph and bipartite topologies, respectively,
by Theorems~\ref{th:framework} and \ref{th:psp-framework}.
These two algorithms, are therefore, used as benchmarks.
\JTKMAT and \JTKSTA operate on general networks, but only consider a subset of
the backhaul links. Therefore, they achieve a fraction of the optimal utility,
denoted \emph{utility ratio} in Fig.~\ref{fig:online_users}.  We vary the
number of users from 1 to 80, reflecting the range of users that can be
expected in small cell deployments.
In each run, a set of items $\mathcal{I}$ is sampled randomly.
To obtain a single point, 10,000 iterations are averaged.

We first use the optimal DP algorithm for
$A_{\text{MMK}}$, for which $\alpha=1$ (Table~\ref{table:MMK}).
Since the maximum vertex degree is $\Delta(G_J)=2$ for both topologies under
consideration, \JTKMAT and \JTKSTA are $1/3$- and $1/2$-approximations,
respectively (Table~\ref{table:algorithms_summary}).
For the complete graph topology, \JTKMAT achieves a
utility ratio of $0.6$ at its worst, while \JTKSTA does better with
a ratio of $0.8$ (Fig.~\ref{fig:online_users}(a)).
Similar insights hold for the bipartite topology in Fig.~\ref{fig:online_users}(b).

%

In addition to using the optimal $A_{\mathrm{MMK}}$, we ran the same
simulations for the case when a greedy algorithm is used for
$A_{\mathrm{MMK}}$ (Table~\ref{table:MMK}), denoted with the
prefix \mbox{``G-''} in Fig.~\ref{fig:online_users}. In this case the
approximation ratios no longer hold, as there are no
performance guarantees for the greedy algorithm. However,
we found that when $A_{\mathrm{MMK}}$ is solved greedily,
the algorithms are very close to optimal for more than 10 users.
Moreover, the running time of the greedy algorithm is
  significantly lower than the duration of a subframe.
Due to their improved running time, we only use the greedy version in
the following sections. For clarity, we omit the \mbox{``G-''} prefix.

\subsubsection{Impact of Backhaul Capacity}
\label{sec:impact_backhaul_capacity}
Backhaul links are typically expensive to deploy, and operators frequently
have to lease them.  Therefore, it is important for the operator
to strike a balance between improving performance and containing backhaul
costs.  To obtain a better understanding of the required backhaul capacity,
we evaluated its impact on the long-term throughput of the
the queueing system.

\begin{figure}
\centering
\includegraphics[height=100pt]{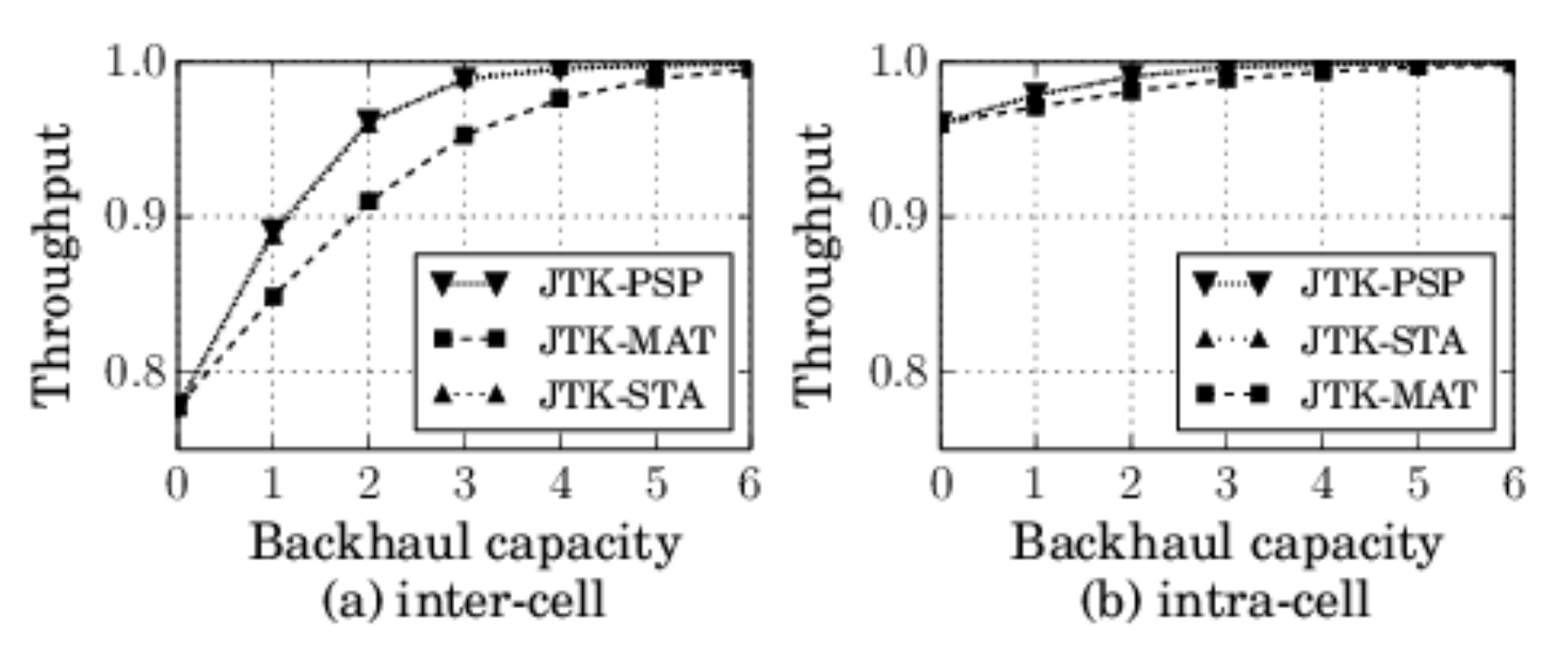}
\vspace*{-15pt}
\caption{The throughput of both intercell and intracell users for a range of
backhaul capacity levels.}
\label{fig:throughput_vs_backhaul_capacity}
\vspace*{-5pt}
\end{figure}

In Fig.~\ref{fig:throughput_vs_backhaul_capacity}, the user-averaged
normalized throughput is shown when the backhaul capacity between each pair
of BSs is scaled from $0$ to \unit[6]{packets/subframe}.
Inter-cell users in particular gain from JT
(Fig.~\ref{fig:throughput_vs_backhaul_capacity}(a)) when network-level behavior is
considered.
A 28\% throughput gain is observed for those users with
the addition of backhaul. Half of this gain is
achievable with 1 unit of backhaul capacity, while 2 units realizes 80\% of the
potential gains. However, intra-cell users gain 5\% throughput. As cell sizes become smaller, the
portion of inter-cell users increases and the overall gain from using CoMP
JT will be higher.


The achieved throughput depends on the used
algorithms.  The largest benefits are possible with the \JTKPSP and \JTKSTA
algorithms, which utilize 3 or 2 backhaul links in every subframe,
respectively. It is also observed that in clusters of this size, \JTKSTA
performs as well as the optimal \JTKPSP, despite its lower running
time. \JTKMAT requires more backhaul capacity to achieve the same
throughput. Overall, in this case \JTKSTA is the best choice while \JTKMAT can
be used to reduce computational resources at the cost of a larger investment
in infrastructure.


\subsubsection{Impact of Topologies}
\label{sec:impact_topology}
We now consider the \emph{star} and \emph{cycle} topologies with $7$ BSs, illustrated in
Fig.~\ref{fig:7cluster_cycle} and \ref{fig:7cluster_star}, respectively. Since both of these topologies are bipartite,
we use the optimal algorithm \JTKMMK in this section.

To study the impact of network topology on JT, we study the throughput gains
that are obtained when backhaul links with a 3 unit capacity are introduced in
the different topologies.
We invoke the algorithm for the same user placement and average
the results. The throughput gains of individual users under the cycle and star
topologies are illustrated in Fig.~\ref{fig:7cluster_cycle} and
\ref{fig:7cluster_star}, respectively.
Overall, we see that in each topology there are 4 users that observed
an increased throughput from JT while the throughput of the other users remained
very similar. While we do not list here the change in throughput for each
individual user, the trend is as follows. The red users have throughput of 0
without JT, since the SINR for a single transmission to these users leads to
a packet transmission success probability of zero. With JT, the throughput of
the red users is very close to 1 and the throughput of each yellow users is
reduced by at most 0.001.

In both topologies we see improvement in the throughput of 4 users alongside a
negligible decrease in throughput of some of other users. However, we observe
that the users that benefit from JT (i.e., red users) are different for the
different topologies. The throughput of users B,C,D, and E increased in the
cycle topology but for the star topology the throughput of users B,C,F, and G
increased.  Clearly, this is to be expected as the users who gain from JT are
those that reside between two BSs that are connected via a backhaul
link. Additionally, we notice that in each topology there are two users that
exhibit a throughput of 0 both with and without using JT. For the cycle
topology (Fig.~\ref{fig:7cluster_cycle}) the users are F and G, while for the
star topology (Fig.~\ref{fig:7cluster_star}) the users are D and E. To
conclude, the two different topologies result with different distribution of
the throughput among the users that are located between the BSs. Since the
users are located somewhat evenly in the simulation area, the number of red
users is similar for both topologies. However, for a different user placement a
specific topology may benefit a larger number of users.

\begin{figure}
\centering
\subfigure[a user's queue sizes]{\label{fig:user12_stability_region}\includegraphics[height=100pt]{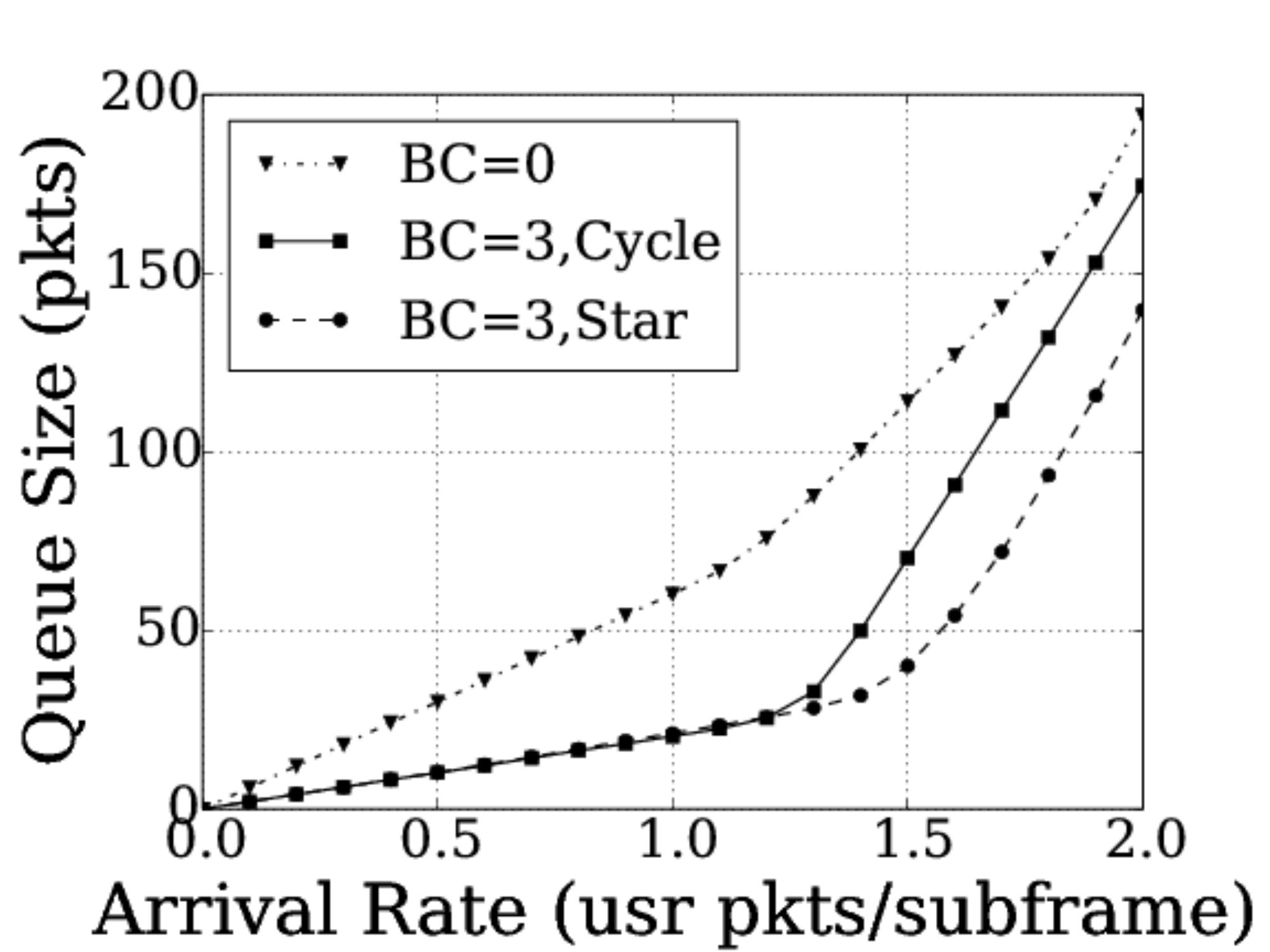}}
\subfigure[Throughput for cycle toplogy]{\label{fig:7bs_throughput_cycle}\includegraphics[height=100pt]{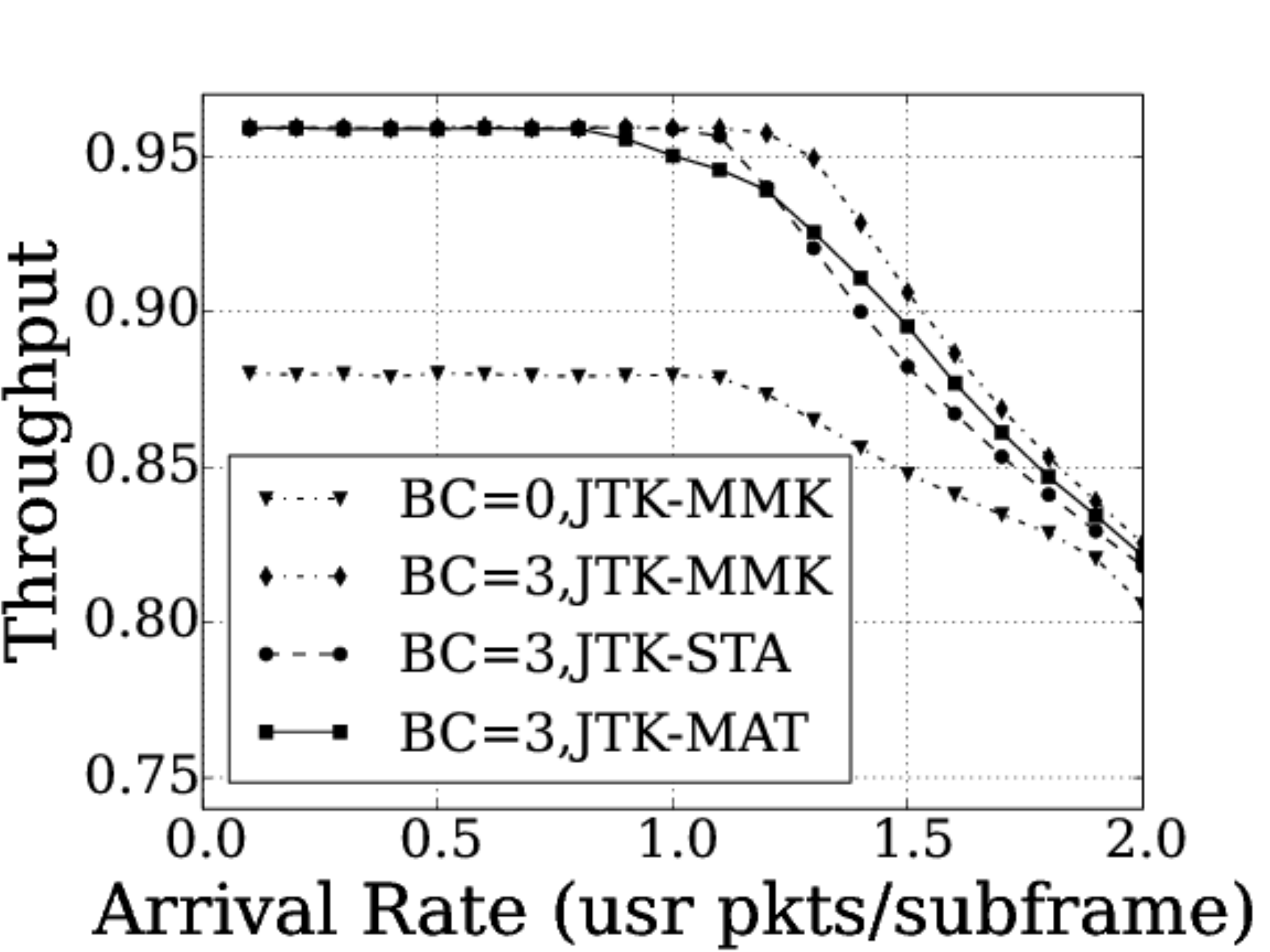}}
\subfigure[Throughput for star topology]{\label{fig:7bs_throughput_star}\includegraphics[height=100pt]{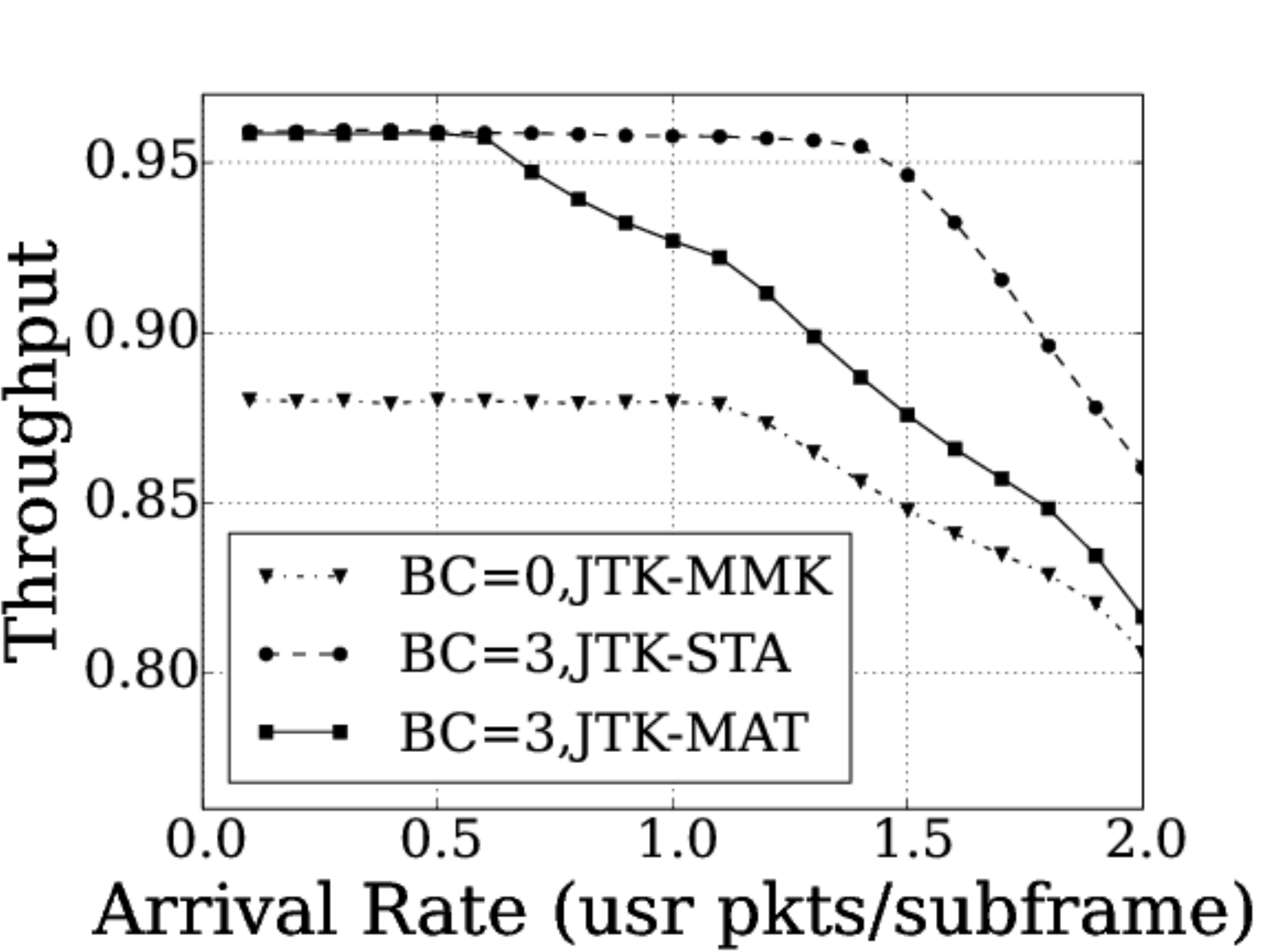}}
\caption{
 \subref{fig:user12_stability_region} The final sum of queue sizes
of user A from Fig.~\ref{fig:7cluster_cycle}, as the arrival rate is varied.
Larger backhaul capacities (BCs) keep the queues bounded for higher arrival
rates; \subref{fig:7bs_throughput_cycle}-\subref{fig:7bs_throughput_star}
Throughput obtained by the proposed scheduling algorithms vs. the arrival rate
for the star and cycle topologies.
}
\vspace*{-10pt}
\end{figure}

The impact of the topology on the stability region of user A from
Fig.~\ref{fig:7cluster_cycle}-\ref{fig:7cluster_star} is illustrated in
Fig.~\ref{fig:user12_stability_region}. The aggregate queue size at the end of
the simulation run, under different arrival rates and backhaul capacities, highlights
that this user's queues can be stabilized for higher arrival rates under the
star topology. This behavior is representative of other inter-cell users.




To further study the performance of our algorithms under the different
topologies we plot in Fig.~\ref{fig:7bs_throughput_cycle} and
Fig.~\ref{fig:7bs_throughput_star} the throughput as a function of the arrival
rate, for the cycle and star topologies, respectively. In both figures JT
improves the network throughput by $9\%$, even for low loads. This can be
explained by the observation that certain inter-cell users may never receive a
packet through single-transmission. For the cycle topology
(Fig.~\ref{fig:7bs_throughput_cycle}), the performance of \JTKSTA and \JTKMAT
is comparable, despite the different approximation ratio. This is because in
the cycle topology a matching may include 5 out of the 6 available backhaul
links, making \JTKMAT comparable to \JTKSTA. It turns out that for higher rate
values, \JTKMAT performs slightly better than \JTKSTA. Therefore, an operator
may choose to run both \JTKMAT and \JTKSTA and select the schedule that yields
the highest throughput in each subframe. Studying such algorithm is out of
scope and left for future research.  In the star topology
(Fig.~\ref{fig:7bs_throughput_star}), since \JTKSTA is an optimal algorithm,
we omit the curve for \JTKMMK from the figure. For lower arrival rate the
performance of \JTKMAT is very close to optimal; As the arrival rate increases
\JTKSTA is clearly favorable. This is expected since for the star topology the
matching computed by \JTKMAT will include at most one backhaul link.

\begin{figure}
\centering
\subfigure[Throughput for cycle toplogy]{\label{fig:7bs_throughput_cycle_bc}\includegraphics[height=100pt]{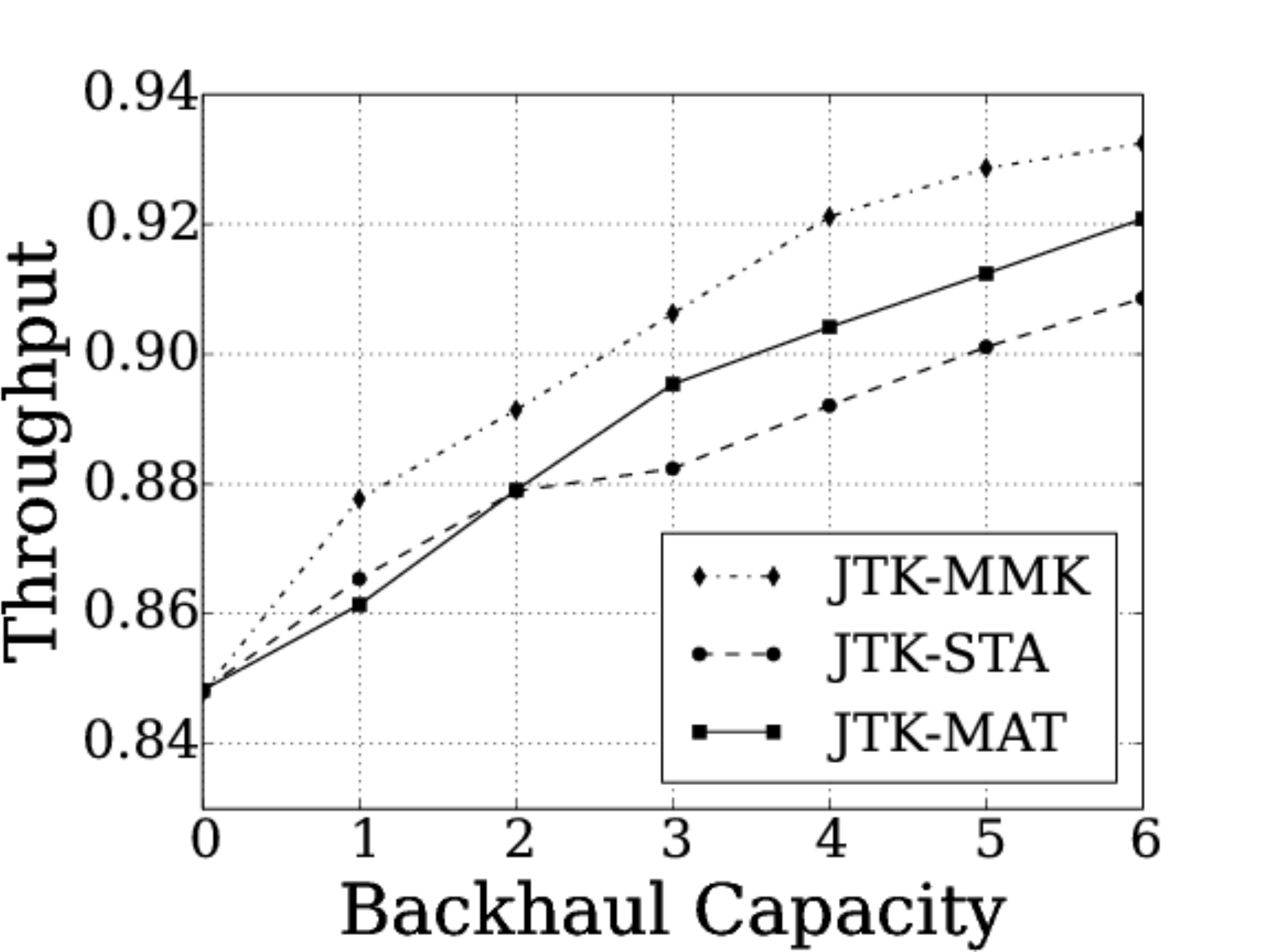}}
\subfigure[Throughput for star topology]{\label{fig:7bs_throughput_star_bc}\includegraphics[height=100pt]{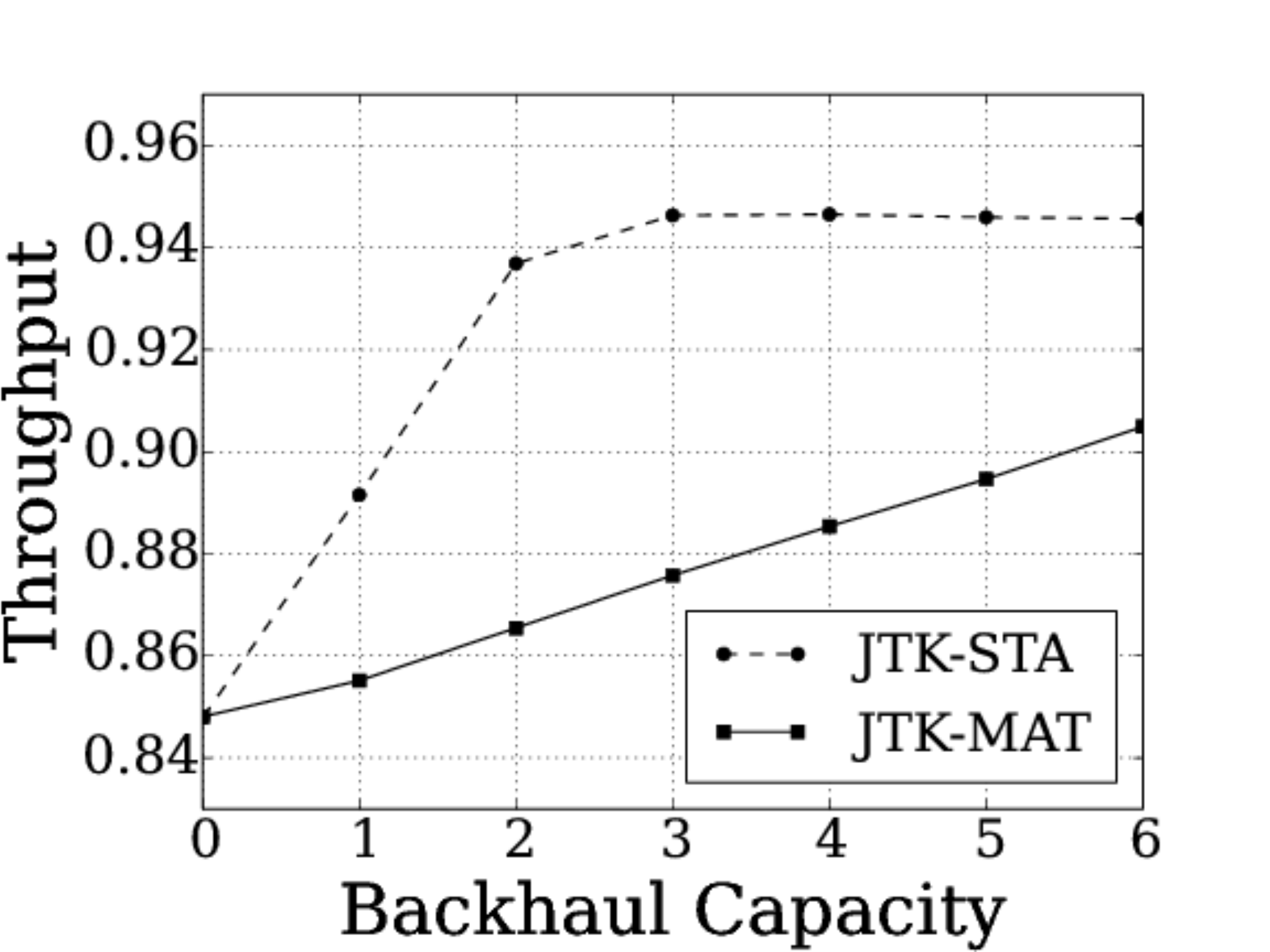}}
\caption{
Throughput obtained by the proposed scheduling algorithms vs. the backhaul capacity
for the star and cycle topologies, for an arrival rate of 1.5 packets per user
per subframe.
}
\vspace*{-10pt}
\end{figure}

The joint impact of backhaul capacity for star and cycle topologies is shown
in Fig.~\ref{fig:7bs_throughput_cycle_bc} and
Fig.~\ref{fig:7bs_throughput_star_bc},
respectively. Recall that \JTKSTA is optimal in
Fig.~\ref{fig:7bs_throughput_star_bc} and note that, compared to the optimal
\JTKMMK in Fig.~\ref{fig:7bs_throughput_cycle_bc}, bakchaul capacity of 3 or
more is beneficial only for the cycle topology. This is due to the fact that
users residing between the center BS and another BS exhibit greater
interference than users that reside between two non-center BSs. Therefore, in
the star topology, such users gain significant increase in their throughput by
just forwarding their packets. As we observed in
Fig.~\ref{fig:7cluster_star}, there are two such
users and since the arrival rate is $1.5$, a backhaul capacity of $3$ suffices
for the star topology. However, in the cycle topology, while mainly two users
gain a significant throughput increase, other users which can get their
packets using single transmission can still benefit from JT. Therefore,
increasing the backhaul capacity further improves the overall throughput.

Fig.~\ref{fig:7bs_throughput_cycle_bc} and
Fig.~\ref{fig:7bs_throughput_star_bc} also demonstrate the relationship
between the network topology and the performance of our algorithms. For the
cycle topology (Fig.~\ref{fig:7bs_throughput_cycle_bc}), a matching may
include 5 out of the 6 backhaul links, for this reason \JTKMAT is relatively
close to optimal and even surpasses \JTKSTA for most backhaul capacity
values. On the other hand, for star topology
(Fig.~\ref{fig:7bs_throughput_star_bc}), while \JTKSTA is guaranteed to be
optimal, the matching computed in \JTKMAT contains at most one backhaul link
and therefore the algorithm obtains low performance. Note that \JTKMAT with
backhaul capacity $6$ performs closely to \JTKSTA with backhaul capacity $1$
since it utilizes at most one out of the 6 backhaul links in every subframe.

\section{Conclusions}
\label{sec:conclusions}
In this paper, we considered a cellular network with Coordinated Multi-Point
(CoMP) Joint Transmission (JT) capabilities that
allow multiple BSs to transmit simultaneously to each user. We first formulated the OFDMA
Joint Scheduling (OJS) problem of determining a subframe schedule and deciding 
if to use JT.
By exploiting the characteristics of this problem, we developed efficient
scheduling algorithms for bipartite graphs and approximation algorithms for
general graphs.


We then considered a queueing model that evolves over time.
In this model, we proved that solving the OJS problem with a specific queue-based utility 
function (in every subframe) achieves maximum throughput in 
CoMP-enabled networks.

 Via extensive simulations we showed that the bulk of the gains from CoMP with
 JT can be achieved with low capacity backhaul links.
We showed that our algorithms distribute the network resources evenly,
increasing the inter-cell users' throughput at only a slight cost to the intra-cell users.


This paper is the first step towards a rigorous, \emph{network-level}
understanding of the impact of cross-layer scheduling algorithms on CoMP networks with JT.
 In future research, we will 
extend the model by allowing more than two BSs to jointly transmit, and by
allowing longer backhaul delays. Moreover, we will apply our techniques to CoMP related technologies such as network-MIMO, multi-cell MIMO, and MU-MIMO.
Finally, we will study the design considerations of the backhaul network and the impact of decentralization on the performance. 


%


\section*{acknowledgements}

This work was supported in part by NSF grant CNS-10-54856 and CIAN NSF ERC under grant EEC-0812072, and the People Programme (Marie Curie Actions) of the European Union’s Seventh Framework Programme (FP7/2007–2013) under REA grant agreement no.\ [PIIF-GA-2013-629740].11. The authors also gratefully acknowledge Marc Kurtz for his contributions to the development of the simulation code and performance evaluation methodology.

\bibliographystyle{abbrv}
\bibliography{CoMP_JT}

\begin{thebibliography}{10}

\bibitem{lte-tr36819}
3GPP.
\newblock {Technical Specification Group Radio Access Network; Coordinated
  multi-point operation for {LTE} physical layer aspects (Release 11), TR
  36.819}, Sept. 2013.

\bibitem{andrews2004scheduling}
M.~Andrews, K.~Kumaran, K.~Ramanan, A.~Stolyar, R.~Vijayakumar, and P.~Whiting.
\newblock Scheduling in a queuing system with asynchronously varying service
  rates.
\newblock {\em Probability in the Engineering and Informational Sciences},
  18(02):191--217, 2004.

\bibitem{andrews2007scheduling}
M.~Andrews and L.~Zhang.
\newblock Scheduling algorithms for multi-carrier wireless data systems.
\newblock In {\em Proc.\ ACM MOBICOM'07}, Sept. 2007.

\bibitem{CCJKKRRSS07}
K.~Balachandran, D.~Calin, F.-C. Cheng, N.~Joshi, J.~H. Kang, A.~Kogiantis,
  K.~Rausch, A.~Rudrapatna, J.~P. Seymour, and J.~Sun.
\newblock Design and analysis of an {IEEE} 802.16e-based {OFDMA} communication
  system.
\newblock {\em BLTJ}, 11(4), 2007.

\bibitem{balanRMPC12}
H.~V. Balan, R.~Rogalin, A.~Michaloliakos, K.~Psounis, and G.~Caire.
\newblock Achieving high data rates in a distributed {MIMO} system.
\newblock In {\em Proc.\ ACM MOBICOM'12}, 2012.

\bibitem{SZJM10}
S.~Brueck, L.~Zhao, J.~Giese, and M.~A. Amin.
\newblock Centralized scheduling for joint transmission coordinated multi-point
  in {LTE}-advanced.
\newblock In {\em Proc.\ IEEE WSA'10}, Feb. 2010.

\bibitem{cadambe2008interference}
V.~Cadambe, S.~A. Jafar, and S.~Shamai.
\newblock Interference alignment on the deterministic channel and application
  to fully connected {AWGN} interference networks.
\newblock In {\em Proc.\ ITW'08}, May 2008.

\bibitem{CG13-2}
R.~Cohen and G.~Grebla.
\newblock Multi-dimensional {OFDMA} scheduling in a wireless network with relay
  nodes.
\newblock In {\em Proc.\ {IEEE} INFOCOM'14}, pages 2427--2435, Apr. 2014.

\bibitem{CG13-1}
R.~Cohen and G.~Grebla.
\newblock Joint scheduling and fast cell selection in {OFDMA} wireless
  networks.
\newblock {\em {IEEE/ACM} Trans. Netw.}, 23(1):114--125, 2015.

\bibitem{COS01}
R.~Cole, K.~Ost, and S.~Schirra.
\newblock Edge-coloring bipartite multigraphs in ${O}({E} \log {D})$ time.
\newblock {\em Combinat.}, 21(1):5--12, 2001.

\bibitem{CYXTL11}
Q.~Cui, S.~Yang, Y.~Xu, X.~Tao, and B.~Liu.
\newblock An effective inter-cell interference coordination scheme for downlink
  {CoMP} in {LTE-A} systems.
\newblock In {\em Proc.\ {IEEE} {VTC'11}}, May 2011.

\bibitem{di2014spatial}
M.~Di~Renzo, H.~Haas, A.~Ghrayeb, S.~Sugiura, and L.~Hanzo.
\newblock Spatial modulation for generalized {MIMO}: challenges, opportunities,
  and implementation.
\newblock {\em Proceedings of the IEEE}, 102(1):56--103, 2014.

\bibitem{ESP05}
A.~Eryilmaz, R.~Srikant, and J.~Perkins.
\newblock Stable scheduling policies for fading wireless channels.
\newblock {\em {IEEE/ACM} Trans.\ Netw.}, 13(2):411--424, 2005.

\bibitem{FMM95}
G.~Fayolle, V.~Malyshev, and M.~Menshikov.
\newblock {\em Topics in the Constructive Theory of Countable Markov Chains}.
\newblock Cambridge University Press, Cambridge, UK, 1995.

\bibitem{fu2014transmission}
S.~Fu, B.~Wu, H.~Wen, P.~Ho, and G.~Feng.
\newblock Transmission scheduling and game theoretical power allocation for
  interference coordination in {CoMP}.
\newblock {\em IEEE Trans. Wireless Commun.}, 13(1):112--123, Jan. 2014.

\bibitem{GMG86}
Z.~Galil, S.~Micali, and H.~Gabow.
\newblock An {$O(EV \log V)$} algorithm for finding a maximal weighted matching
  in general graphs.
\newblock {\em SIAM J. Comput.}, 15(1):120--130, 1986.

\bibitem{GHHSSY10}
D.~Gesbert, S.~Hanly, H.~Huang, S.~Shamai, O.~Simeone, and W.~Yu.
\newblock Multi-cell {MIMO} cooperative networks: A new look at interference.
\newblock {\em IEEE~J. Sel. Areas Commun.}, 28(9):1380--1408, 2010.

\bibitem{hatay1980empirical}
M.~Hata.
\newblock Empirical formula for propagation loss in land mobile radio services.
\newblock {\em IEEE Trans. Veh. Technol.}, 29(3):317--325, 1980.

\bibitem{H81}
I.~Holyer.
\newblock The {NP}-completeness of edge-coloring.
\newblock {\em {SIAM} J. Comp.}, 10:718--720, 1981.

\bibitem{irmer2011coordinated}
R.~Irmer, H.~Droste, P.~Marsch, M.~Grieger, G.~Fettweis, S.~Brueck, H.-P.
  Mayer, L.~Thiele, and V.~Jungnickel.
\newblock Coordinated multipoint: Concepts, performance, and field trial
  results.
\newblock {\em IEEE Commun. Mag.}, 49(2):102--111, 2011.

\bibitem{Book:Kellerer04}
H.~Kellerer, U.~Pferschy, and D.~Pisinger.
\newblock {\em Knapsack Problems}.
\newblock Springer, 2004.

\bibitem{KHHJS13}
K.~Kwak, H.~Lee, H.~W. Je, J.~Hong, and S.~Choi.
\newblock Adaptive and distributed {CoMP} scheduling in {LTE-Advanced} systems.
\newblock In {\em Proc.\ {IEEE} {VTC'13}}, June 2013.

\bibitem{JSCTXX11}
J.~Li, T.~Svensson, C.~Botella, T.~Eriksson, X.~Xu, and X.~Chen.
\newblock Joint scheduling and power control in coordinated multi-point
  clusters.
\newblock In {\em Proc.\ {IEEE} {VTC'11}}, May 2011.

\bibitem{marsch2008multicell}
P.~Marsch and G.~Fettweis.
\newblock On multicell cooperative transmission in backhaul-constrained
  cellular systems.
\newblock {\em Ann. Telecommun.}, 63(5-6):253--269, 2008.

\bibitem{MAW96}
N.~McKeown, V.~Anantharam, and J.~Walrand.
\newblock Achieving 100\% throughput in an input-queued switch.
\newblock In {\em Proc.\ IEEE INFOCOM'96}, Mar. 1996.

\bibitem{PR12}
B.~Patt-Shamir and D.~Rawitz.
\newblock Vector bin packing with multiple-choice.
\newblock {\em Discrete Appl. Math.}, 160(10-11):1591--1600, 2012.

\bibitem{YRLZ13}
F.~Ren, Y.~Xu, H.~Yang, J.~Zhang, and C.~Lin.
\newblock Frequency domain packet scheduling with stability analysis for 3{GPP}
  {LTE} uplink.
\newblock {\em IEEE Trans. Mob. Comput}, 12(12):2412--2426, 2013.

\bibitem{S49}
C.~E. Shannon.
\newblock A theorem on coloring the lines of a network.
\newblock {\em J. Math. Physics}, 28:148--151, 1949.

\bibitem{NAOA11}
N.~Tabia, A.~Gondran, O.~Baala, and A.~Caminada.
\newblock Interference model and evaluation in {LTE} networks.
\newblock In {\em Proc.\ IFIP~WMNC'11}, Oct. 2011.

\bibitem{TE92}
L.~Tassiulas and A.~Ephremides.
\newblock Stability properties of constrained queueing systems and scheduling
  policies for maximum throughput in multihop radio networks.
\newblock {\em IEEE Trans.~Automat.~Contr.}, 37(12):1936--1948, 1992.

\bibitem{tipmongkolsilp2011evolution}
O.~Tipmongkolsilp, S.~Zaghloul, and A.~Jukan.
\newblock The evolution of cellular backhaul technologies: current issues and
  future trends.
\newblock {\em IEEE Communications Surveys \& Tutorials}, 13(1):97--113, 2011.

\bibitem{JQPBY13}
J.~Yu, Q.~Zhang, P.~Chen, B.~Cao, and Y.~Zhang.
\newblock Dynamic joint transmission for downlink scheduling scheme in
  clustered {CoMP} cellular.
\newblock In {\em Proc.\ {IEEE} {ICCC'13}}, Aug. 2013.

\bibitem{ZhangSKRS13}
X.~Zhang, K.~Sundaresan, M.~A. Khojastepour, S.~Rangarajan, and K.~G. Shin.
\newblock {NEMOx}: scalable network {MIMO} for wireless networks.
\newblock In {\em Proc.\ ACM MOBICOM'13}, Sept. 2013.

\bibitem{zhang2012joint}
Y.-P. Zhang, L.~Xia, P.~Zhang, S.~Feng, J.~Sun, and X.~Ren.
\newblock Joint transmission for {LTE}-advanced systems with non-full buffer
  traffic.
\newblock In {\em Proc. IEEE VTC'12}, Sept. 2012.

\bibitem{ZMN05}
X.~Zhou, Y.~Matsuo, and T.~Nishizeki.
\newblock List total colorings of series-parallel graphs.
\newblock {\em J. Discrete Algorithms}, 3(1):47--60, 2005.

\bibitem{ZNSN92}
X.~Zhou, S.-I. Nakano, H.~Suzuki, and T.~Nishizeki.
\newblock An efficient algorithm for edge-coloring series-parallel multigraphs.
\newblock In {\em Proc.\ LATIN'92}, volume 583 of {\em LNCS}, pages 516--529,
  Apr. 1992.

\bibitem{zhuang2014backhaul}
F.~Zhuang and V.~K. Lau.
\newblock Backhaul limited partial cooperations for {MIMO} cellular networks
  via semidefinite relaxation.
\newblock {\em IEEE Trans. Signal Process.}, 62(3):684--693, 2014.

\end{thebibliography}


\normalsize

\newpage
\appendices
\section{Proofs}

\vspace{0.5em}

\ifLongversion
\begin{proof}[{\bf Proof of Proposition~\ref{pro:NPhard-Knapsack}}]
  We show that for $B=1$ the formulation of the OJS problem is equivalent to
  that of the well-known NP-hard Multiple-Choice Knapsack Problem
  (MCKP).

  First, note that for $B=1$, $C=0$ and $D=1$. Therefore
  constraint~\eqref{eq:all-capacities} can be written as: $\sum_{i\in
    \mathcal{I}} \sum_{r=0}^R [w(i,r)]_1\cdot z_{ir} \leq
  K_1$. Constraints~\eqref{eq:all-capacities,eq:one-conf} together with
  the maximization expression of the OJS problem, are equivalent to the MCKP
  formulation given in~\cite{Book:Kellerer04}. Therefore, it is sufficient to
  show that for every $z_{ir}$ for which constraints
  ~\eqref{eq:all-capacities,eq:one-conf} hold, we can find $x_{irs}$ such
  that constraints~\eqref{eq:sufficient,eq:used-once} hold. Since for
  every $(i,r)$, $[w(i,r)]_1=\Gamma(i, r)$, such $x_{irs}$ can be easily
  found.
\end{proof}

\begin{proof}[{\bf Proof of Proposition~\ref{pro:noFPTAS}}]
  We reduce the Partition problem, a well-known NP-hard problem, to OJS with
  $B=2$ and $C=1$. The reduction will show that if an FPTAS exists for OJS with 2 BSs
  then the Partition problem can be solved in polynomial time.

  An input to the partition problem is a set of integers $\mathcal{A}$. Let
  $Z=\sum_{a\in \mathcal{A}} a$, the objective is to decide if there exists a
  subset $\mathcal{A}' \subseteq \mathcal{A}$ such that $\sum_{a'\in
    \mathcal{A}'} a' = \sum_{a \in \mathcal{A} \setminus \mathcal{A}'} a =
  Z/2$.

  The reduction is as follows. We set $\mathcal{B}=\{1, 2\}$,
  $\mathcal{R}=\{0, 1\}$, $l_{12}=(Z/2)$, $S=(Z/2)$. Note that for the reduced
  instance $D=3$. Every $a\in \mathcal{A}$ is transformed into a user $n'$ for
  which BS$(n')$ is BS $1$ and $\widehat{\text{BS}}(n')$ is BS $2$, with a
  single pending packet $i$ defined by $n(i)=n'$ and $\beta(i)=0$ for which $u(i,0)=u(i,1)=1$,
  $w(i,0)=(0, 0, a)$, and $w(i,1)=(a, 0, 0)$. We denote $a(i)$ the item
  $a\in\mathcal{A}$ we transformed to $i$.

  Consider a solution $\pmb{z^*}$, $\pmb{x^*}$ to the reduced OJS instance, whose
  total utility is $|\mathcal{A}|$. For every $i$, $z_{i0}^*=1$ or
  $z_{i1}^*=1$. Due to constraint~\eqref{eq:all-capacities}, choosing
  $\mathcal{A}'=\{a(i):\ z_{i0}^*=1\}$ defines the desired partition of
  $\mathcal{A}$. The other direction, namely that if a partition exists then
  there exists a solution to the reduced OJS instance whose total utility is
  $|\mathcal{A}|$ can be shown using similar ideas.

  To summarize, a solution of utility $|\mathcal{A}|$ for the reduced OJS
  problem exists if and only if the desired partition exists for the Partition
  problem. An FPTAS for OJS can be used to distinguish if a solution of utility
  $|\mathcal{A}|$ exists in polynomial time.
\end{proof}

\fi

\begin{proof}[{\bf (Proposition~\ref{pro:NPhard-coloring}})]
To prove Proposition~\ref{pro:NPhard-coloring}, we use the following
definition:
\begin{definition} \label{def:chromatic}
  The \emph{chromatic index of a graph $G$}~\cite{H81}, $\chi'(G)$, is the
  number of colors required to color the edges of $G$ such
  that no two adjacent edges have the same color.
\end{definition}

  It is known by Vizing's theorem that for every simple graph $G$,
  $\chi'(G)=\Delta(G)$ or $\chi'(G)=\Delta(G)+1$, where $\Delta(G)$ is the
  maximum vertex degree of $G$. The Minimum Edge Coloring Problem
  (MECP)~\cite{H81} is to determine whether $\chi'(G)=\Delta(G)$ or
  $\chi'(G)=\Delta(G)+1$. It is well-known that MECP is NP-hard~\cite{H81},
  therefore to complete the proof we present a polynomial-time reduction from
  MECP to OJS.

\ifTechRep
  Given a simple graph $G=(V,E)$ with maximum vertex degree $\Delta(G)$, we
  now describe how to construct an OJS instance. We set $\mathcal{B}=V$,
  $\mathcal{R}=\{0, 1\}$. For each edge $(v_1,v_2)\in E$, we add a user $n$ to
  $\mathcal{N}$ for which BS$(n)$=$v_1$, $\widehat{\text{BS}}(n)=v_2$, and
  there exists only a single pending packet in $\widehat{Q}_n$. Note that as a
  consequence, for every packet $i$ we have $\beta(i)=1$ and
  thus the only relevant pair is $(i,1)$. For every $i$ we set
  $\Gamma(i,1)=1$, $[w(i,1)]_d=\{1 \text{ if } d\in h(i,1);\ 0 \text{
    otherwise}\}$. The utility is defined as $u(i,1)=1$ and $u(i,0)=0$ for
  every $i$. Also, we set $S=\Delta(G)$, and for every $a,b\in \mathcal{B}$
  $l_{ab}=S$, therefore $\pmb{K}=(S,S,\ldots,S)$. Note that for the
  constructed instance (a)-(d) stated in the proposition hold.

  We now show that the optimal solution to the OJS instance has a utility of
  $|E|$ if and only if $\chi'(G)=\Delta(G)$. Let $\pmb{x}^*,\ \pmb{z}^*$ be an
  optimal solution to the OJS instance with a total utility of $|E|$. Due to
  the utility $u()$ used and since the total number of packets is $|E|$,
  $z^*_{i1}=1$, $\forall i$. Consider an edge $(v_1,v_2)\in E$ and the
  corresponding user $n'$ in OJS with a pending packet $i'$ such that
  $n(i)=n'$ and $\beta(i)=1$. Due to
  constraint~\eqref{eq:sufficient}, there is exactly one $s'$ for which
  $x_{i'1s'}=1$. We assign this edge the color $s'$. Since $1\leq s'\leq S$,
  at most $\Delta(G)$ colors are used. Since
  constraint~\eqref{eq:used-once} also hold, no two adjacent edges are
  colored using the same color $s$. Since we showed an edge coloring with at
  most $\Delta(G)$ colors, $\chi'(G)=\Delta(G)$. The other direction, namely
  showing that if $\chi'(G)=\Delta(G)$ then the optimal solution to the OJS
  instance has a utility of $|E|$, can be proved similarly.
\else
  Given a simple graph $G=(V,E)$ with maximum vertex degree $\Delta(G)$, we
  now describe how to construct an OJS instance. We set $\mathcal{B}=V$. For each edge $\{v_1,v_2\}\in E$, we add a user $n$ to
  $\mathcal{N}$ for which BS$(n)$=$v_1$, $\widehat{\text{BS}}(n)=v_2$, and
  there exists only a single pending packet in $\widehat{Q}_n$. Thus, for every packet $i$ we have $\beta(i)=1$.
 The utility is defined as $u(i,1)=1$ and $u(i,0)=1$ for
  every $i$. Also, we set $S=\Delta(G)$, $K=S$, and $\mathcal{C}=\{\{a,b\},\
  a,b\in \mathcal{B} \wedge a\neq b\}$. Note that for the
  constructed instance (a)-(b) stated in the lemma hold.

  We now show that the optimal solution to the OJS instance has a utility of
  $|E|$ if and only if $\chi'(G)=\Delta(G)$. Let $\pmb{x}^*,\ \pmb{y}^*,\
  \pmb{z}^*$ be an optimal solution to the OJS instance with a total utility
  of $|E|$. Due to the utility $u$ used and since the total number of
  packets is $|E|$, $z^*_{i}=1$, $\forall i$. Consider an edge $e=\{v_1,v_2\}\in
  E$ and a pending packet $i'$ such that $h(i')=\{v_1,v_2\}$. Due to
  \eqref{eq:sufficient}, there is exactly one $s'$ for which
  $x_{i's'}=1$. We assign $e$ the color $s'$ and continue the process
  for the remaining edges and packets. Since $1\leq s'\leq S=\Delta(G)$, at most
  $\Delta(G)$ colors are used. Since \eqref{eq:used-once} also
  holds, no two adjacent edges are colored using the same color $s$. We
  showed an edge coloring with at most $\Delta(G)$ colors,
  $\chi'(G)=\Delta(G)$. The other direction, namely showing that if
  $\chi'(G)=\Delta(G)$ then the optimal solution to the OJS instance has a
  utility of $|E|$, can be proved similarly.
\fi
\end{proof}

\begin{proof}[{\bf (Lemma~\ref{lem:basic-framework})}]
The proof immediately follows from the definitions of the JTK and JTC problems.
\end{proof}

\ifLongversion
\begin{proof}[{\bf (Lemma~\ref{lem:JTC-coloring})}]
  Given a solution $\pmb{x'}$ to a JTC instance, we now define a coloring on
  $G_{\text{SB}}$ that uses at most $S$ colors. Observe that by
  Definition~\ref{def:sb-graph} and since constraint~\eqref{eq:sufficient}
  holds, there exists a one-to-one mapping from every triplet $(i,r,s)$ such
  that $x'_{irs}=1$ into an edge in $E_{\text{SB}}$. This mapping defines an edge
  coloring using at most $S$ colors ($1\leq s\leq S$). Since
  constraint~\eqref{eq:used-once} holds, no two edges of the same color touch
  a vertex in $G_{\text{SB}}$. The other direction, namely, finding a solution
  $\pmb{x'}$ to JTC, given an edge coloring on $G_{\text{SB}}$, can be proved
  similarly.
\end{proof}
\else
\begin{proof}[{\bf (Lemma~\ref{lem:JTC-coloring})}]
  Given a solution $\pmb{x'}$ to a JTC instance, we now define a coloring on
  $G_{\text{SB}}$ that uses at most $S$ colors. Observe that by
  Definition~\ref{def:sb-graph} and since constraint~\eqref{eq:sufficient}
  holds, there exists a one-to-one mapping from every pair $(i,s)$ such
  that $x'_{is}=1$ into an edge in $E_{\text{SB}}$. This mapping defines an edge
  coloring using at most $S$ colors ($1\leq s\leq S$). Since
  constraint~\eqref{eq:used-once} holds, no two edges of the same color touch
  a vertex in $G_{\text{SB}}$. The other direction, namely, finding a solution
  $\pmb{x'}$ to JTC, given an edge coloring on $G_{\text{SB}}$, can be proved
  similarly.
\end{proof}
\fi

\begin{proof}[{\bf (Lemma~\ref{lem:bip})}]
  Using Definition \eqref{def:sb-graph}, it is clear that the
  subgraph $G'=(V',E')$ of $G_{\text{SB}}$ defined by $V'=\mathcal{B}$ and
  $E'=\{\{a,b\}\in E_{SB}:a,b\in \mathcal{B}\}$ is bipartite. Since each $b\in
  V_{\text{SB}}\setminus V'$
has at most one neighbor, 
  $G_{\text{SB}}$ is bipartite.
\end{proof}

\begin{proof}[{\bf (Lemma~\ref{lem:bip-exists-x})}]
  Using the result from~\cite{COS01} and since $G_{\text{SB}}$ is bipartite
  and $\Delta(G_{\text{SB}})\leq S$, $G_{\text{SB}}$ has an edge coloring that
  uses at most $S$ colors. By Lemma~\ref{lem:JTC-coloring}, such a coloring
  defines a solution $\pmb{x'}$ such that
  $\eqref{eq:sufficient}, \eqref{eq:used-once} \text{ hold.}$
\end{proof}

\begin{proof}[{\bf (Theorem~\ref{th:framework})}]
We already showed that for bipartite networks \JTKMMK solves JTK and \JTCBIP
solves JTC. Lemma~\ref{lem:basic-framework} concludes the proof.
\end{proof}

\begin{proof}[{\bf (Lemma~\ref{lem:bip-MMK})}]
Let $\pmb{z}^*,\pmb{y}^*$ be an optimal solution for JTK.
Without~\eqref{eq:sufficient} and~\eqref{eq:used-once}, JTK would be equivalent
to MMK with some restrictions on its parameters (unit-size MMK items and two
MMK capacity values).
Therefore, let $\pmb{z}',\pmb{y}'$ be the solution returned by \JTKMMK,
$U(\pmb{z}',\pmb{y}') \geq \alpha U(\pmb{z}^*,\pmb{y}^*)$.
Finally, the solution is feasible due to Lemmas~\ref{lem:bip} and~\ref{lem:bip-exists-x}.
\end{proof}

\begin{proof}[{\bf (Theorem~\ref{th:JTK-MAT-approx})}]
  Let $\mathcal{B}_0$ be the set of BSs with no backhaul links
    (BSs whose degree is 0 in $G_J$). In line~\ref{ln:leaf-node} of \JTKMAT
    the selected transmissions are determined using $A_{\text{MMK}}$.  This
    set of transmission is an $\alpha$-approximation with respect to a JTK
    instance with $\mathcal{B}'' = \mathcal{B}_0$. Therefore, to complete the
    proof we can assume that every BS has a backhaul link and show that the
    transmissions determined in line~\ref{ln:ret-mat-sol} are an
    $(2\alpha/3\Delta(G_J))$-approximation.

  Let $\pmb{z}^*_{\text{OJS}},\pmb{y}^*_{\text{OJS}}$ be an optimal solution
  for OJS.  The sum of weights for all edges in $\mathcal{C}$, as computed in
  line~\ref{ln:edge-utility} of \JTKMAT, is at least $\alpha
  U(\pmb{z}^*_{\text{OJS}},\pmb{y}^*_{\text{OJS}})$.

  Any graph $G$
  has an edge coloring using at most $(\frac{3}{2})\Delta(G)$
  colors~\cite{S49}. Such a coloring for $G_J$ partitions $\mathcal{C}$ into $(\frac{3}{2})
  \Delta(G_J)$ matchings. Since in line~\ref{ln:max-matching} of \JTKMAT a maximum weight
  matching $\mathcal{E}$ is obtained,
the sum of weight for edges
  in $\mathcal{E}$ is at least $\frac {\alpha U(\pmb{z}^*_{\text{OJS}},\pmb{y}^*_{\text{OJS}})}
  {(3/2) \Delta(G_J)} = (2 \alpha)/(3 \Delta(G_J))
  U(\pmb{z}^*_{\text{OJS}},\pmb{y}^*_{\text{OJS}})$. Let $\pmb{z}',\pmb{y}'$ be the solution
  returned by \JTKMAT. Then, $U(\pmb{z}',\pmb{y}') \geq (2 \alpha)/(3 \Delta(G_J))
  U(\pmb{z}^*_{\text{OJS}},\pmb{y}^*_{\text{OJS}})$.

  We now show that $\pmb{z}',\pmb{y}'$ is feasible. Note that \eqref{eq:one-conf} and
  \eqref{eq:all-capacities} hold since the solution associated with each edge
  \{$a$,$b$\} is feasible (line \ref{ln:edge-utility}) and solutions of
  different edges in $\mathcal{E}$ use items whose $h(i)$ is in
  disjoint (no two edges in $\mathcal{E}$ share a vertex).
  Due to line ~\ref{ln:ret-mat-sol} of Algorithm \JTKMAT, in the
  returned solution if $z'_{i}=1$ and $h(i)=\{a,b\}$ then
  $\{a,b\}\in\mathcal{E}$. Since $\mathcal{E}$ is a matching, $G_{\text{SB}}$
  is bipartite and $\Delta(G_{\text{SB}})\leq S$. Using
  Lemma~\ref{lem:bip-exists-x} we conclude that $\pmb{z}',\pmb{y}'$ is feasible.

  Finally, by Lemma~\ref{lem:JTC-coloring} \JTCBIP solves JTC. By applying
  Lemma~\ref{lem:basic-framework}, we complete the proof.
\end{proof}

\ifLongversion
\begin{proof}[{\bf Proof of Theorem~\ref{th:JTK-STA-approx}}]
  Let $\pmb{z}^*_{\text{OJS}}$ be the optimal solution for OJS.  The sum of
  weights for all vertices in $\mathcal{B}$, as computed in
  line~\ref{ln:vertex-utility} of \JTKSTA, is at least $\alpha
  U(\pmb{z}^*_{\text{OJS}})$.  In each iteration of the repeat loop
  (line~\ref{ln:repeat-loop} of \JTKSTA), in line~\ref{ln:set-z} the utility added to the
  solution $\pmb{z''}$ equals $U(b_{\max})$ (note that $z''_{ir}$ gets updated
  to $1$ at most once due to the update of $I''$ in
  line~\ref{ln:update-I''}). In line~\ref{ln:remove-neighbors}, $b_{\max}$ and
  its neighbors are removed from consideration. The utility lost due to this
  removal is at most $\Delta(G_J) U(b_{\max})$. Therefore, the total utility
  of $\pmb{z''}$ returned in line~\ref{ln:return-z''} is at least $(\alpha /
  \Delta(G_J)) U(\pmb{z}^*_{\text{OJS}})$.

  To show that $\pmb{z''}$ returned by \JTKSTA is feasible, it is sufficient
  to note that due to the correctness of $A_{\text{MMK}}$ function SOL-STAR
  returns a feasible instance with respect to $I''$ and that
  $\tilde{\mathcal{I}}$ (line~\ref{ln:find-tilde-I}) in different iterations
  contains packets of disjoint sets of transmitting BSs (therefore having
  positive weight only in disjoint sets of dimensions).

  Finally, by Lemma~\ref{lem:JTC-coloring} \JTCBIP solves JTC. By applying
  Lemma~\ref{lem:basic-framework}, we complete the proof.
\end{proof}
\else
\begin{proof}[{\bf (Theorem~\ref{th:JTK-STA-approx})}]
  Let $\pmb{z}^*_{\text{OJS}}$ be the optimal solution for OJS.  The sum of
  weights for all vertices in $\mathcal{B}$, as computed in
  line~\ref{ln:vertex-utility} of \JTKSTA, is at least $\alpha
  U(\pmb{z}^*_{\text{OJS}})$.  In each iteration of the repeat loop
  (line~\ref{ln:repeat-loop} of \JTKSTA), in line~\ref{ln:set-z} the utility added to the
  solution $\pmb{z''}$ equals $U(b_{\max})$ (note that $z''_{i}$ gets updated
  to $1$ at most once due to the update of $I''$ in
  line~\ref{ln:update-I''}). In line~\ref{ln:remove-neighbors}, $b_{\max}$ and
  its neighbors are removed from consideration. The utility lost due to this
  removal is at most $\Delta(G_J) U(b_{\max})$. Therefore, the total utility
  of $\pmb{z''}$ returned in line~\ref{ln:return-z''} is at least $(\alpha /
  \Delta(G_J)) U(\pmb{z}^*_{\text{OJS}})$.

  To show that $\pmb{z''}$ returned by \JTKSTA is feasible, it is sufficient
  to note that due to the correctness of $A_{\text{MMK}}$ function SOL-STAR
  returns a feasible instance with respect to $I''$ and that
  $\tilde{\mathcal{I}}$ (line~\ref{ln:find-tilde-I}) in different iterations
  contains packets of disjoint sets of transmitting BSs (therefore having
  positive weight only in disjoint sets of dimensions).

  Finally, by Lemma~\ref{lem:JTC-coloring} \JTCBIP solves JTC. By applying
  Lemma~\ref{lem:basic-framework}, we complete the proof.
\end{proof}
\fi

\begin{proof}[{\bf (Lemma~\ref{lem:sp})}]
  It is clear that if $G_J$ is planar, $G_{\text{SB}}$ is also
  planar. Therefore, to complete the proof it is sufficient to show that
  $G_{\text{SB}}$ is series-parallel.

  To show $G_{\text{SB}}$ is series-parallel, we use the following
  definition~\cite{ZMN05}. A multigraph is \emph{series-parallel} if it has no
  subgraph isomorphic to a subdivision of a clique of size $4$. Since $G_J$
  has no subgraph isomorphic to a subdivision of a clique of size $4$, by
  adding parallel edges to $G_J$ such a subgraph cannot be created in $G_{\text{SB}}$.
  Therefore, by the above definition $G_{\text{SB}}$ is series-parallel.
\end{proof}

\begin{proof}[{\bf (Theorem~\ref{th:psp-framework})}]
The following result, mentioned in~\cite{ZNSN92}, is
needed.
\begin{quote}
  Let $E_U\subseteq E$ denote edges in $E$ whose both vertices are in $U$ and
  let $\delta(G)= \max \{ \frac {2|E_U|} {|U|-1}: U\subseteq V,$ $|U|\geq 3\
  \text{and odd} \}$.  If $G$ is planar and series-parallel then
  $\chi'(G)=\max\{\Delta(G),\lceil \delta(G) \rceil\}$ and \JTCPSP
  from~\cite{ZNSN92} finds an edge coloring that uses $\chi'(G)$ colors.
\end{quote}

Recall that in \JTKPSP an instance for MMK with $D +
|\mathcal{B}_{\text{odd}}|$ dimensions is constructed. The weight constraints
for the new $|\mathcal{B}_{\text{odd}}|$ dimensions are equivalent to
requiring that for a feasible solution $\pmb{z'}$, $\delta(G_{\text{SB}})\leq
S$. Therefore, for such $G_{\text{SB}}$ there exists an edge coloring that
uses at most $S$ colors, and by Lemma~\ref{lem:JTC-coloring} such a coloring
defines a solution $\pmb{x'}$ such that
$\eqref{eq:sufficient}, \eqref{eq:used-once} \text{ hold.}$ Since
\JTKPSP invokes $A_{\text{MMK}}$ which returns an $\alpha$-approximation
solution to the constructed MMK problem, \JTKPSP is an $\alpha$-approximation
for JTK.

Finally, by Lemma~\ref{lem:JTC-coloring}, \JTCPSP solves JTC. By applying
Lemma~\ref{lem:basic-framework}, we complete the proof.
\end{proof}

\begin{proof}[{\bf (Theorem~\ref{thm:stability})}]
Let $\bflambda \in \Lambda$. In order to demonstrate positive recurrence of $\{\bfL(t)\}_{t\ge0}$, we define a Lyapunov function, and show that it has negative drift outside some closed set of states. Let $\bfl = (l_1,\hat{l}_1,\dots,l_N,\hat{l}_N)$ and define the quadratic Lyapunov function $V(\bfl) = \sum_{n=1}^N l_n^2 + \hat{l}_n^2$.

Consider the one-slot drift
\begin{equation}\label{eqn: drift}
\Delta V(\bfl) = \expect{V(\bfL(t+1))-V(\bfL(t)) \mid \bfL(t) = \bfl}.
\end{equation}
By~\eqref{eqn:evolution_1} and~\eqref{eqn:evolution_2} we compute
\begin{align*}
L_n(t+1)^2 ={}& L_n(t)^2 + 2L_n(t) \Big(W_n(t) - \mu_n^{(1)}(t) - \mu_n^{(3)}(t)\Big) + \Big(W_n(t) - \mu_n^{(1)}(t) - \mu_n^{(3)}(t)\Big)^2,\\
\widehat{L}_n(t+1)^2 ={}& \widehat{L}_n(t)^2 + 2\widehat{L}_n(t) \Big(\mu_n^{(3)}(t) - \mu_n^{(2)}(t)\Big)+ \Big(\mu_n^{(3)}(t) - \mu_n^{(2)}(t)\Big)^2,
\end{align*}
Substituting this into~\eqref{eqn: drift} we obtain
\begin{align}
\nonumber \Delta V(\bfl) ={}& \sum_{n = 1}^N \expect{\Big(W_n(t) - \mu_n^{(1)}(t) - \mu_n^{(3)}(t)\Big)^2 + \Big(\mu_n^{(3)}(t) - \mu_n^{(2)}(t)\Big)^2 \mid \bfL(t) = \bfl}\\
&+ 2 \sum_{n = 1}^N \expect{L_n(t) \Big(W_n(t) - \mu_n^{(1)}(t) - \mu_n^{(3)}(t)\Big) + \widehat{L}_n(t) \Big(\mu_n^{(3)}(t) - \mu_n^{(2)}(t)\Big) \mid \bfL(t) = \bfl}. \label{eqn: drift2}
\end{align}
Since the $W_n(t)$ have finite second moment, and the $\mu_n^{(j)}(t)$ have finite support, we can bound (for some constant $C < \infty$),
\begin{equation*}
\sum_{n = 1}^N \expect{\Big(W_n(t) - \mu_n^{(1)}(t) - \mu_n^{(3)}(t)\Big)^2 + \Big(\mu_n^{(3)}(t) - \mu_n^{(2)}(t)\Big)^2 \mid \bfL(t) = \bfl} < C.
\end{equation*}

The second part of~\eqref{eqn: drift2} can be written as
\begin{align}
\nonumber &2 \sum_{n = 1}^N \expect{L_n(t) \Big(W_n(t) - \mu_n^{(1)}(t) - \mu_n^{(3)}(t)\Big) +  \widehat{L}_n(t) \Big(\mu_n^{(3)}(t) - \mu_n^{(2)}(t)\Big) \mid \bfL(t) = \bfl}\\
={}& 2\sum_{n=1}^N \Big( l_n \lambda_n  -  l_n \expect{\mu_n^{(1)}(t) + \mu_n^{(3)}(t) \mid \bfL(t) = \bfl} - \hat{l}_n \expect{\mu_n^{(3)}(t) - \mu_n^{(2)}(t)\mid \bfL(t) = \bfl}\Big).
\label{eqn: drift3}
\end{align}
Since $\bflambda \in\Lambda$, we know by the definition of $\Lambda$ that there exists a flow vector $\bff$ such that the conditions in~\eqref{flow_conditions} hold. Moreover, $\bflambda \in\Lambda$ also implies that there exists a $\bfr \in {\rm Conv}(R)$ such that $f_n^{(j)} < r_n^{(j)}$ if $f_n^{(j)} > 0$, $j =1,2,3$, $n = 1,\dots,N$. Since $\bff$ is dominated by $\bfr$, and $\bfr \in {\rm Conv}(R)$, there exist $\sigma_1,\dots,\sigma_R$ such that
\begin{equation}\label{eqn:combination}
\bff = \sum_{i=1}^{|R|} \sigma_i \bfr_i, \quad \sum_{i=1}^{|R|} \sigma_i < 1,
\end{equation}
where $\bfr_i = (r_{i,1}^{(1)},r_{i,1}^{(2)},r_{i,1}^{(3)},\dots,r_{i,N}^{(1)},r_{i,N}^{(2)},r_{i,N}^{(3)})$ represents the $i$-th vector in $R$.

Using~\eqref{eqn:combination} we obtain
\begin{align}
l_n \lambda_n &= l_n (f_n^{(1)} + f_n^{(3)}) = l_n \sum_{i=1}^{|R|} \sigma_i \big(r_{i,n}^{(1)} + r_{i,n}^{(3)}\big), \label{eqn:arrival1}\\
0 &= \hat{l}_n (f_n^{(2)} - f_n^{(3)}) = \hat{l}_n \sum_{i=1}^{|R|} \sigma_i \big(r_{i,n}^{(2)} - r_{i,n}^{(3)}\big). \label{eqn:arrival2}
\end{align}
By combining~\eqref{eqn:arrival1} and~\eqref{eqn:arrival2}, and exploiting the structure of the $\bfmu$ chosen according to $u_Q$ (see~\eqref{eqn:MWS})
\begin{align}
\nonumber 2 \sum_{n=1}^N l_n \lambda_n &= 2 \sum_{n=1}^N \sum_{i=1}^{|R|} \sigma_i \Big( l_n (r_{i,n}^{(1)} + r_{i,n}^{(3)}) + \hat{l}_n(r_{i,n}^{(2)} - r_{i,n}^{(3)})\Big)\\
&\le 2 \sum_{n=1}^N \Big( l_n (\expect{\mu_{n}^{(1)}(t) + \mu_{n}^{(3)}(t) \mid \bfL(t) = \bfl}) + \hat{l}_n(\expect{\mu_{n}^{(2)}(t) - \mu_{n}^{(3)}(t)\mid \bfL(t) = \bfl })\Big) \sum_{i=1}^{|R|} \sigma_i. \label{eqn:arrival3}
\end{align}

Substituting~\eqref{eqn:arrival3} into~\eqref{eqn: drift3} yields
\begin{align}
\nonumber &2 \sum_{n = 1}^N \expect{L_n(t) \Big(W_n(t) - \mu_n^{(1)}(t) - \mu_n^{(3)}(t)\Big) +  \widehat{L}_n(t) \Big(\mu_n^{(3)}(t) - \mu_n^{(2)}(t)\Big) \mid \bfL(t) = \bfl}\\
 \le{}& - 2\Big(1 - \sum_{i=1}^{|R|}\sigma_i \Big)\sum_{n=1}^N \Big(l_n \expect{\mu_n^{(1)}(t) + \mu_n^{(3)}(t) \mid \bfL(t) = \bfl}- \hat{l}_n \expect{\mu_n^{(3)}(t) - \mu_n^{(2)}(t)\mid \bfL(t) = \bfl}\Big) < 0, \label{eqn: drift4}
\end{align}
where the last inquality follows from the choice of $\mu_n^{(j)}$~\eqref{eqn:MWS}. The expression in~\eqref{eqn: drift4} can be made arbitrarily small by increasing the state $\bfl$. Positive recurrence of $\{\bfL(t)\}_{t\ge0}$ then follows from \cite[Theorem 2.2.4]{FMM95}.
\end{proof}


\ifTechRepFinal
\newpage

\section{Extended Results}\label{sec:extension}

In this appendix we relax assumptions (i)-(iii) from Section
  \ref{sec:model} and show the applicability of our algorithms to the case
  where a packet can be transmitted using one of several Modulation and Coding
Schemes (MCSs). This appendix contains three sections. In
Section~\ref{ap:model} we present the extended network model. In
Section~\ref{ap:prob} we introduce the extended OJS Problem and in
Section~\ref{ap:algs} we describe the decomposition framework and extend our algorithms.

\subsection{Network Model} \label{ap:model}

 Each wireless transmission requires an MCS, which is selected from a finite set $\mathcal{M}=\{1,\ldots, M\}$ of supported MCSs.
We define the configuration set $\mathcal{R}=\{0\} \cup
\mathcal{M}$ such that $r\ge 1$ indicates wireless transmission using MCS $r$, and $r=0$
indicates a packet forwarded to another BS over the backhaul. 
Thus, a pair $(i, r)$ ($i\in \mathcal{I},\ r\in
\mathcal{R}$) defines a wireless transmission of packet $i$ with MCS $r$ ($r \ge 1$) or packet $i$ forwarded over the backhaul ($r=0$). Note that the pair $(i,0)$ is not feasible if $\beta(i)=1$.

For each packet-configuration pair $(i, r)$, we define two properties: its size $\Gamma(i,r)$ and
its success probability $p(i,r)$. In case of forwarding over the backhaul, $\Gamma(i, 0)$ represents the size of
packet $i$ in bytes. For $r \ge 1$, the $\Gamma(r, c)$ represents the number
of scheduled blocks required for transmission with MCS $r$, which
depends on
the packet size in bytes $\Gamma(i,0)$ and the MCS $r$.

The success probability $p(i,r)$ represents the probability that packet $i$ will be successfully received by the user $n(i)$ ($r\ge 1$) or that packet $i$ forwarded over the backhaul is successfully received by $\widehat{\text{BS}}(n)$ ($r=0$). 
We assume that $p(i,m)$ is higher if $\beta(i) = 1$ compared to $\beta(i) =
0$, since the SINR of a user is greater when a packet is joint-transmitted.
The capacity of a
backhaul link between BSs $a$ and $b$ is $l_{ab}$
bytes.

\subsection{The OFDMA Joint Scheduling (OJS) Problem} \label{ap:prob}

We now formulate the extended version of the joint scheduling problem presented in Section~\ref{sec:scheduling_problem}. 
Capacity constraints apply both to the subframes of the $B$ BSs
as well as the $C$ backhaul links, and we denote the total number of such constraints by
$D=B+C$. We order these constraints such that the constraint $b$
corresponds to the BS $b$, $b = 1,\dots,B$,
and constraints $B+1,\dots,D$
correspond to the backhaul. Define the $D$-dimensional capacity vector
$\pmb{K}=(K_1,\ldots, K_D)$ such that for $1\leq d\leq B$, $K_d=S$ (number of
scheduled blocks), and $K_d=l_{ab}$ for
$d\ge B+1$.

In order to describe the capacity used by the wireless transmission of packet $i$, we introduce
\begin{equation*}
h(i) = \left\{
\begin{array}{ll}
\{\text{BS}(n(i))\}    & {\rm ~if~} \beta(i) = 0,\\
\{\text{BS}(n(i)),\widehat{\text{BS}}(n(i))\}    & {\rm ~if~} \beta(i) = 1.\\
\end{array}\right.
\end{equation*}
If $\beta(i) = 0$ then $h$ returns only the serving BS, and if $\beta(i) = 1$ it returns both the serving and secondary BS. 
We define the function $w:\mathcal{I}\times \mathcal{R}\rightarrow
(\mathbb{N}_0)^D$, where $w(i, r)$ denotes the $D$-dimensional vector that
represents the capacity used for $(i,r)$. If $r \ge 1$, then $w(i, r)$ is the
all-zero vector except for $[w(i, r)]_{b} = \Gamma(i, r)$ for $b \in h(i)$. If
$r=0$, then only the entry corresponding to the appropriate backhaul link is
positive, and equal to the length in bytes of packet $i$. The function $u:\mathcal{I} \times \mathcal{R} \mapsto \mathbb{R}_+$
represents the utility of scheduling packet $i$ according to configuration
$r$, and is essentially an extension of the utility function presented in Section~\ref{sec:scheduling_problem}.

Based on the set of packets $\mathcal{I}$ and the utility function $u$, the centralized scheduler determines
the set of transmissions to take place in
the upcoming subframe, by maximizing the aggregate utility. The scheduler must also determine which scheduled
blocks will be used for each packet transmission, such that for JT the
scheduled blocks of the serving BS and secondary BS are aligned (i.e., have
identical index).
The
scheduling decisions are represented using indicator variables $z_{ir} \in \{0,1\}$ and
$x_{irs}\in \{0,1\}$, where $z_{ir}$ indicates if a transmission $(i,r)$
takes place, and $x_{irs}$ indicates if scheduled block $s$ is used by a
transmission $(i, r)$.
The scheduler needs to solve the
following integer programming problem (with $\pmb{z} = (z_{ir})_{i\in\mathcal{I},r\in\mathcal{R}}$ and $\pmb{x} = (x_{irs})_{i\in\mathcal{I},r\in\mathcal{R},s \in \mathcal{S}}$).

\noindent{\bf OFDMA Joint Scheduling (OJS) Problem}:
\begin{alignat}{2}
\underset{\pmb{x},\pmb{z}}{\text{max}}&
&\:\: & \sum_{i=1}^I \sum_{r=1}^R z_{ir}
  u(i,r) =: U(\pmb{z}) \notag
\\
\text{s.t.}&
&&  \sum_{r=0}^R z_{ir} \leq 1, \quad \forall\  i\in
\mathcal{I}, \label{eq:one-conf_e} \\
&
&&   \sum_{i=1}^I \sum_{r=0}^R z_{ir} [w(i,r)]_d \leq K_d,
 \quad \forall
  1\leq d\leq D, \label{eq:all-capacities_e} \\
&
&&  \sum_{s=1}^S x_{irs}=z_{ir} \Gamma(i, r),  \quad \forall
  i \in \mathcal{I}\  \forall r\ge 1, \label{eq:sufficient_e} \\
&
&&   \sum_{r=1}^R \sum_{\{i:\ b\in h(i)\}} x_{irs} \leq 1, \quad \forall b\in
  \mathcal{B}\ \forall s\in \mathcal{S}, \label{eq:used-once_e}\\
&
&&   z_{ir} \in \{0,1\}, \quad \forall i \in \mathcal{I}\ \forall r \in \mathcal{R}, \label{eq:integer_z_e}\\
&
&&   x_{irs} \in \{0,1\}, \quad \forall i \in \mathcal{I}\ \forall r \in \mathcal{R}\ \forall s \in \mathcal{S}. \label{eq:integer_x_e}
\end{alignat}
Constraint \eqref{eq:one-conf_e} ensures a packet is scheduled only
once; \eqref{eq:all-capacities_e} ensures capacities in each subframe and backhaul link are not exceeded; \eqref{eq:sufficient_e} ensures
sufficient scheduled blocks are allocated for each $(i,r)$ transmission;
and \eqref{eq:used-once_e} ensures that each scheduled block is used at most
once in the subframe of each BS.

\subsection{OJS Problem - Algorithms} \label{ap:algs}

\addtolength{\tabcolsep}{-.35em}
\begin{table*}
\scriptsize
\begin{minipage}[t]{0.6\linewidth}
\centering
\caption{The performance and input required for different algorithms within the decomposition
  framework
  \DF{$A_{\text{JTK}}$}{$A_{\text{JTC}}$}
  \label{table:algorithms_summary_e}}
\begin{tabular}{|l|l|l|l|p{1.6cm}|} \hline 
{\bf $A_{\text{JTK}}$} & {\bf $A_{\text{JTC}}$} & {\bf Ratio} & {\bf Running time} & {\bf Input $G_J$}
\\
\hline
\JTKMMK & \JTCBIP & $\alpha$ & $O(T_{\text{MMK}}(I,R,D,S))$ & bipartite  \\ \hline
\JTKMAT & \JTCBIP & $\frac {2  \alpha}
  {3  \Delta(G_J)}$ & $O(C \cdot T_{\text{MMK}}(I,R,3,S))$ & any \\ \hline
\JTKSTA & \JTCBIP & $\frac{\alpha}{\Delta(G_J)}$ & $O(B^2
T_{\text{MMK}}(I,R,2\Delta(G_J)+1,S))$ & any \\ \hline
\JTKPSP & \JTCPSP &$\alpha$ & $O(2^{B}+T_{\text{MMK}}(I,R,2^{B},S))$ &
planar ser.paral.\\
\hline
\end{tabular}
\end{minipage}
\hfill
\begin{minipage}[t]{0.34\linewidth}
\centering
\caption{Algorithms for MMK \label{table:MMK_e}}
\begin{tabular}{|l|l|p{1.9cm}|} \hline 
{\bf $A_{\text{MMK}}$} & {\bf Ratio} & {\bf $T_{\text{MMK}}(I,R,D,S)$} \\
\hline
DP~\cite{Book:Kellerer04} & Optimal & $O(S^D I R D)$ \\
\hline
PTAS~\cite{PR12} & $1/(1+\epsilon)$ & $O((I R)^{(D/\epsilon)})$ \\ \hline
Greedy~\cite{Book:Kellerer04} & $\infty$ & $O(I R \log(IR))$ \\ \hline
\end{tabular}
\end{minipage}
\normalsize
\end{table*}
\addtolength{\tabcolsep}{.35em}

\subsubsection{Decomposition Framework}

Similar to what was done in Section~\ref{sec:scheduling_problem} for the basic problem, we can decompose OJS into two parts as follows.

\noindent \textbf{Joint Transmission Knapsack (JTK) Problem:}
\begin{alignat*}{2}
\underset{\pmb{z}}{\text{max}} & & \:\: & U(\pmb{z})\\
\text{ s.t.} &&& \eqref{eq:one-conf_e},
\eqref{eq:all-capacities_e}, \eqref{eq:integer_z_e},\  \exists
\pmb{x}:\ \eqref{eq:sufficient_e}, \eqref{eq:used-once_e}, \eqref{eq:integer_x_e} \text{ hold}
\end{alignat*}
\noindent \textbf{Joint Transmission Coloring (JTC) Problem:}
\begin{alignat*}{2}
\text{given } \pmb{z} \text{, find}& &\:\: & \pmb{x} \text{
  s.t. \eqref{eq:sufficient_e}, \eqref{eq:used-once_e}, \eqref{eq:integer_x_e}
hold}
\end{alignat*}

Solving JTK to obtain
$\pmb{z}$
and then solving JTC to obtain
$\pmb{x}$, provides a solution to OJS. 
%

The scheduled block graph for this version of JTC is the same as what was presented in Definition~\ref{def:sb-graph}, except that every pair
  $(i,r)$ such that $r\ge 1$ and $z_{ir}=1$ contributes $\Gamma(r,c)$ edges to
  $E_{\text{SB}}$ instead of 1.

\subsubsection{Algorithms for General Graphs} \label{ap:alg-general}

We now describe the algorithms \JTKMAT and \JTKSTA for the
  extended model. The performance ratio of \JTKMAT and \JTKSTA in the
  extended model are the same as those in Section~\ref{sec:algorithms}. The
  proofs use similar ideas to those used in Section~\ref{sec:algorithms} and
  are omitted for brevity. Tables \ref{table:algorithms_summary_e} and \ref{table:MMK_e} summarize the different options for solving OJS using the
decomposition framework. Finally, note that the stability result presented in Theorem~\ref{thm:stability} can also be extended to this setting.

\begin{varalgorithm}{\JTKMAT}
\caption{Based on matching}
\label{alg:JTK-MAT_e}
\begin{algorithmic}[1]
\For{$b\in \mathcal{B}$}
\State\parbox[t]{\dimexpr\linewidth-\algorithmicindent} {If $b$ has no
  backhaul link, run $A_{\text{MMK}}$ to solve a JTK instance with
$\mathcal{B}'=\{b\}$, $\mathcal{I}'=\{i\in\mathcal{I}:\ h(i)=\mathcal{B}'\}$,
$\mathcal{C}'=\{\}$ and set $z_{ir}$ as determined by $A_{\text{MMK}}$.} \label{ln:leaf-node_e}
\EndFor
  \For{$e=\{a,b\} \in \mathcal{C}$}
  \State {Run
    $A_{\text{MMK}}$ to solve a JTK instance with $\mathcal{B}'=\{a,b\}$, $\mathcal{I}'=
    \{i\in\mathcal{I}:\ \exists r\in\mathcal{R},\ h(i,r)\subseteq\mathcal{B}'\}$} \label{ln:MMK_e}
  \State \parbox[t]{\dimexpr\linewidth-\algorithmicindent}{Assign
    $U(\pmb{z})$ found by $A_{\text{MMK}}$ as a weight for $e$ \label{ln:edge-utility_e}}
\EndFor
\State{Compute maximum weight matching on $G_J$ and store the result in the edge set
  $\mathcal{E}$ \label{ln:max-matching_e}}
\State{Set $z_{ir}=1$ only if $\exists e\in\mathcal{E}$
  such that $z_{ir}=1$ in the solution returned in line~\ref{ln:MMK_e} for
  edge $e$} \label{ln:ret-mat-sol_e}
\end{algorithmic}
\end{varalgorithm}

\begin{varalgorithm}{\JTKSTA}
\caption{Based on star subgraphs}
\label{alg:JTK-STA_e}
\begin{algorithmic}[1]
\Function{SOL-STAR}{$b$, $\mathcal{B}'$, $\mathcal{C}'$, $\mathcal{I}'$}
  \State \parbox[t]{\dimexpr\linewidth-\algorithmicindent}{Run
    $A_{\text{MMK}}$ to solve a JTK instance defined with
    $\widetilde{\mathcal{B}'}=\{b\}\cup\{a:\ \{a,b\}\in \mathcal{C}'\}$, $\widetilde{\mathcal{C}'}=\{
    \{a,b\}:\ \{a,b\}\in \mathcal{C}',\ a\in \widetilde{\mathcal{B}'},\ b\in \widetilde{\mathcal{B}'}\}$,
$\widetilde{\mathcal{I}'}=
    \{i\in\mathcal{I}':\ h(i)\subseteq\widetilde{\mathcal{B}'}\}$} \label{ln:STA-DMCKP_e}
  \Return $\pmb{z}$ as determined by $A_{\text{MMK}}$
\EndFunction

  \For{$b \in \mathcal{B}$}
  \State {$\pmb{Z}[b] \gets$ SOL-STAR($b$, $\mathcal{B}$,
    $\mathcal{C}$, $\mathcal{I}$)}
  \State \parbox[t]{\dimexpr\linewidth-\algorithmicindent}{Assign
    $U(\pmb{Z}[b])$ as a weight for $b$ in
    $\mathcal{B}$ \label{ln:vertex-utility_e}}
\EndFor
\State{Initialize $\mathcal{B}''\gets \mathcal{B}$; $\mathcal{C}''\gets
  \mathcal{C}$; $\mathcal{I}''\gets
\mathcal{I}$}
\Repeat \label{ln:repeat-loop_e}
\State \parbox[t]{\dimexpr\linewidth-\algorithmicindent}{Find the vertex $b_{\max}$ in $\mathcal{B}''$ with maximum weight}

\State {$\widetilde{\mathcal{I}}\gets
  \{i\in\mathcal{I}'': \ \exists r,\  \pmb{Z}[b_{\max}]_{ir}=1\}$} \label{ln:find-tilde-I_e}
\State{for all $i\in \tilde{\mathcal{I}}$, $z''_{ir}\gets \pmb{Z}[b_{\max}]_{ir}$} \label{ln:set-z_e}
\State\parbox[t]{\dimexpr\linewidth-\algorithmicindent}{$\mathcal{J}\gets \{
  a\in \mathcal{B}'':\exists b\in \mathcal{B}'',\{b_{\max},b\}\in \mathcal{C}'',\{a,b\}\in \mathcal{C}''\}$}
\State{$\mathcal{I}'' \gets \mathcal{I}''\setminus \tilde{\mathcal{I}}$;
  $\mathcal{B}'' \gets \mathcal{B}''\setminus \{a:\ \{a,b_{\max}\}\in
  \mathcal{C}''\}$} \label{ln:update-I''_e} \label{ln:remove-neighbors_e}
\State{Remove from $\mathcal{C}''$ edges with an endpoint not in $\mathcal{B}''$}
\For{$b\in \mathcal{J}$} \label{ln:update_weight_e}
\State{$\pmb{Z}[b] \gets$ SOL-STAR($b$, $\mathcal{B}''$,
  $\mathcal{C}''$, $\mathcal{I}''$)}
\State{Update $U(\pmb{Z}[b])$ as a weight for $b$ in
  $\mathcal{B}''$}
\EndFor

\Until{$\mathcal{B}''$ is empty \label{ln:empty-vtag_e}}
\State{\text{{\bf return} }$\pmb{z}''$} \label{ln:return-z''_e}
\end{algorithmic}
\end{varalgorithm}


\fi










\end{document}